%% file: main.tex
\documentclass[sigconf,nonacm,dvipsnames,balance=false]{acmart}
\usepackage{enumitem}
\usepackage{multirow}
\usepackage{placeins}
\usepackage{subcaption}
\usepackage{listings}

\definecolor{ForestGreen}{RGB}{34,139,34}
\definecolor{BurntOrange}{RGB}{204,85,0}
\definecolor{RoyalBlue}{RGB}{65,105,225}
\theoremstyle{remark}
\newtheorem{example}{Example}[section]
\newcommand{\myparagraph}[1]{\par\smallskip\noindent\textbf{#1.}\ }
\newcommand{\ours}{query-centric system }
\setcopyright{none}
\AtBeginEnvironment{thebibliography}{\interlinepenalty=10000}
\hypersetup{hidelinks}
\begin{document}
\title[Query-Centric Optimization of AI Workflows]{Query-Centric Optimization of AI Workflows via Approximate Query Processing and Proxy Models}
\author{Huayi Wang}
\affiliation{\institution{Georgia Institute of Technology}\city{Atlanta}\country{USA}}
\email{hwang762@gatech.edu}
\author{Jun Xu}
\affiliation{\institution{Georgia Institute of Technology}\city{Atlanta}\country{USA}}
\email{jx@cc.gatech.edu}
\author{Gromit Yeuk-Yin Chan}
\affiliation{\institution{Adobe Research}\city{San Jose}\country{USA}}
\email{ychan@adobe.com}
\renewcommand{\shortauthors}{Wang, Xu, and Chan}
\input{sections/abstract}
\maketitle
\pagestyle{plain}
\begingroup\small\noindent\textbf{Artifact availability.}
The source code, data, and other artifacts are available at
\url{https://github.com/Why1221/Query-Centric-AI-optimization}.
\endgroup
\input{sections/introduction}
\input{sections/problem}
\input{sections/related_work}
\input{sections/method}
\input{sections/evaluation_tpc}
\FloatBarrier
\input{sections/evaluation_llm}

\input{sections/conclusion}
\FloatBarrier
\bibliographystyle{ACM-Reference-Format}
\bibliography{references}
\end{document}

%% file: sections/abstract.tex
\begin{abstract}
Many modern AI workflows---ranging from LLM post-training pipelines to
agentic reasoning tasks---can be expressed as declarative queries whose
expensive predicate is evaluated by a large model or reward function.
We propose a \emph{query-centric} formulation of these workflows and
show that classical database techniques, namely approximate query
processing (AQP) and proxy-model (PM) based filtering, can substantially reduce the number of expensive model invocations by identifying calls that can be omitted while retaining useful aggregate or selected-output quality, without changing the database, model, or pipeline.
AQP treats the workflow as an online aggregation problem:
it progressively samples records, maintains a running aggregate estimate
with {an approximate confidence interval and an empirical stopping rule based on interval narrowing and stabilization.}
PM trains a lightweight, CPU-resident decision tree on a small set of oracle-labeled examples and uses it to pre-filter records
using structural features, with oracle confirmation and sampled correction for aggregates and a gate-and-filter cascade for response selection.
{A shared label cache supports both AQP estimation and proxy training, and the plan requiring fewer remaining oracle calls completes first.}
We evaluate {on TPC-DS aggregate queries and real LLM}
post-training pipelines including math reasoning, general
instruction following, and code generation.
On TPC-DS, AQP keeps aggregate error under 10\% {using about 10\% to 15\% of oracle calls in the balanced-distribution experiments}, and PM reduces oracle calls
by 60--70\% on {several} natural-label workloads,
{although natural-label small-table queries can retain substantial error.}
On LLM pipelines, {fixed-budget sampling provides a cost-quality
reference, and PM uses up to 19 times fewer RM calls than full scoring with accuracy of 91.16\% vs 92.95\% on LLaMA-3.1-8B Math.}
\end{abstract}

%% file: sections/introduction.tex
\section{Introduction}
\label{sec:introduction}

\begin{figure}[t]
    \centering
    \includegraphics[width=\linewidth]{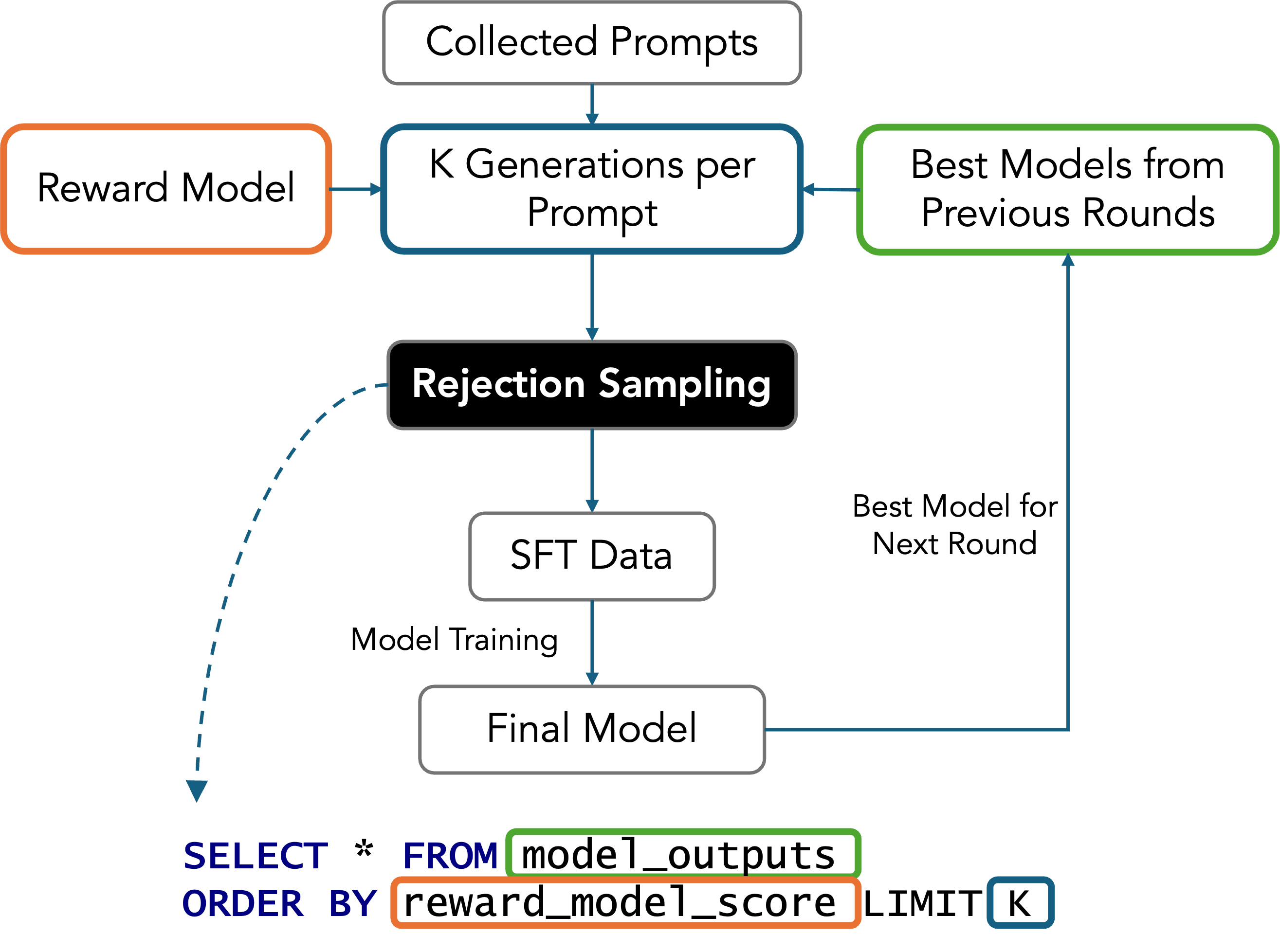}
    \caption{Illustration of rejection sampling as a SQL query in the LLM post-training recipe.}
\label{fig:eg1}
\end{figure}

The rapid growth of expensive Large Language Model (LLM) computation in AI workflows~\cite{aero2025,chen2023frugalgpt}---from training foundation models to invoking them hundreds of times in agentic task pipelines~\cite{biswal2025tag,patel2025lotus}---has created new opportunities to optimize latency and cost. More importantly, many of the current AI-centric workflows can be expressed as database queries. Consider the following two examples:
\begin{example}
    A recurring pattern in modern LLM post-training pipelines — explicitly documented in both the Llama 2 and Llama 3 technical reports~\cite{grattafiori2024llama,touvron2023llama2} — is what can be understood as the following SQL query operation:
    \begin{lstlisting}
    SELECT * FROM model_outputs 
    ORDER BY reward_model_score DESC LIMIT K
    \end{lstlisting}
    Demonstrated in Figure~\ref{fig:eg1}, for each training prompt, {$N$ candidate responses} are sampled from the current model, scored by a reward model, and only the top-scoring response is retained as supervised fine-tuning data for the next training round. {Here, $N$ denotes the generated pool size and $K$ the number retained with descending reward order in the query.} This generate-rank-select loop is a central mechanism by which the model bootstraps its own training data quality without additional human annotation. 
\end{example}

\begin{figure}[t]
    \centering
    \includegraphics[width=.9\linewidth]{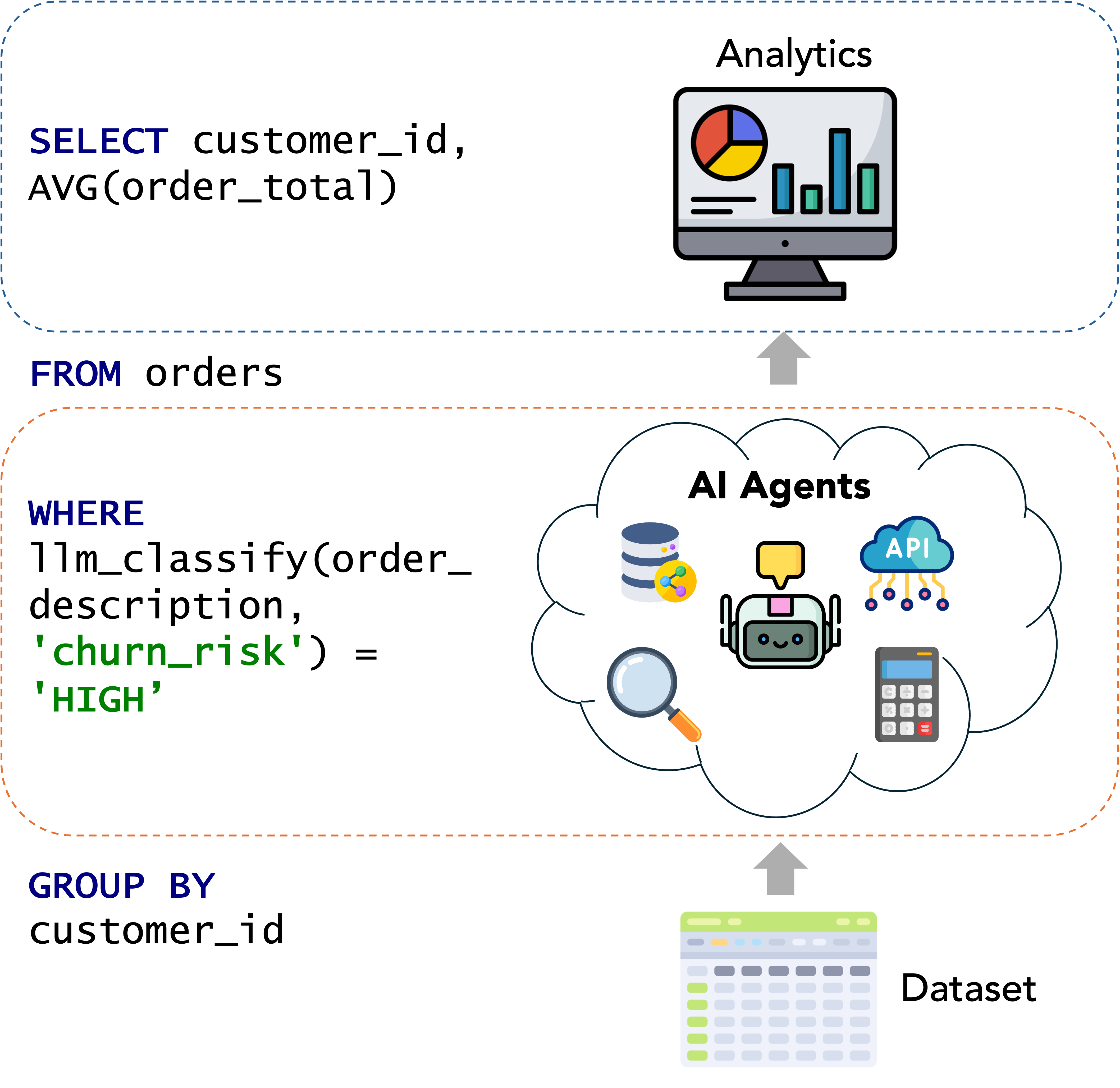}
    \caption{Illustration of an analytical query involving AI agents as a User Defined Function (UDF).}
\label{fig:eg2}
\end{figure}

\begin{example}
    Agentic reasoning tasks---workflows in which an AI model must gather context, apply multi-step inference, and produce a decision or action---are typically treated as orchestration problems, implemented through bespoke application logic spread across pipelines, queues, and LLM API calls. Yet consider a canonical example: \textit{``computing the average order value among customers whose recent orders suggest churn risk''}, a task that today is commonly delegated to an agent framework with hand-written retrieval, prompt assembly, and filtering logic~\cite{shankar2025docetl,biswal2025tag}. As illustrated in Figure~\ref{fig:eg2}, this workflow can be expressed as the following single declarative query:
    \begin{lstlisting}
    SELECT customer_id, AVG(order_total)
    FROM orders WHERE llm_classify(order_description, 
    'churn_risk') = 'HIGH' GROUP BY customer_id
    \end{lstlisting}
    where \texttt{llm\_classify} itself may encapsulate a multi-step agentic pipeline, yet appears in the query simply as a semantic predicate that filters records before a standard aggregation is applied. This observation suggests that agentic reasoning tasks can be naturally viewed as declarative queries with semantic predicates.
\end{example}

{ These examples illustrate query-centric AI workflows, where the final result determines how much model evaluation is useful: an aggregate summarizes many records, while a selection retains only a few. This exposes oracle evaluations that contribute little to the requested result, and we avoid them through aggregate sampling and structural proxy filtering, leaving the oracle unchanged.}

\noindent\textbf{Our Approach.}
{
We use standard database and statistics tools: progressive uniform sampling estimates aggregates, with an approximate interval guiding the sampling budget. A CPU decision tree provides a complementary proxy from structural features, such as response length and formatting. Oracle confirmation and sampled correction support aggregate queries, while a gate and filter reduce RM calls for response selection. The query formulation connects these database techniques to the cost and quality of the workflow's final output.}
{Our shared execution reuses uniform oracle labels and original data representations for both estimation and proxy training, avoiding a second labeling budget and further data transformations. It completes the plan with fewer remaining oracle calls first and reuses cached labels if the other plan continues. Unlike optimizing agreement with individual oracle labels, we evaluate the numerical aggregate or selected response required by the query. Related Work distinguishes this execution design from declarative AI systems~\cite{rev_palimpzest,rev_quest,shankar2025docetl}, prediction-based approximation~\cite{rev_bargain,rev_crossprediction} and embedding-based query optimization approaches~\cite{rev_chung,rev_uqe,tasti2021}.}

We conduct a comprehensive experimental evaluation on two classes of query-centric AI workloads. The first is a set of TPC-DS queries {that replay known predicate labels to measure oracle-row demand;} these experiments demonstrate the effectiveness of the AQP-based approach, showing that progressive sampling achieves $<$10\% aggregate error while invoking the oracle on only around 10\% of tuples {under adjusted balanced label distributions. Natural small-table queries can remain difficult at a roughly 20\% budget. An additional shared-execution experiment measures the cost of an early PM output followed by AQP, without requiring full evaluation.} The second is a suite of real LLM post-training pipelines spanning math reasoning (GSM8K), general instruction following (UltraFeedback), and code generation (HumanEval+, MBPP), {where we evaluate PM selection against fixed-budget reward-model sampling. The proxy-based filter uses up to 19 times fewer RM calls than full scoring, including training; for LLaMA-3.1-8B Math, accuracy changes from 92.95\% to 91.16\%.}

\noindent\textbf{Contributions.} Our main contributions are as follows:
\begin{itemize}
    \item {We formulate AI workflows as aggregate or selected-output queries and use their result requirements to identify and avoid unnecessary oracle evaluations through AQP and structural proxy filtering.}
    \item {We design shared AQP and PM execution that reuses oracle labels, preserves both estimates, and returns the plan with fewer remaining oracle calls first.}
    \item {We evaluate both aggregate strategies on TPC-DS and PM selection on real LLM candidate pools,} {using up to 19 times fewer RM calls than full scoring on the evaluated structured LLM tasks, including training, and oracle-call reductions of 85\% to 90\% with $<$10\% aggregate error in the balanced TPC-DS experiments.} {Additional experiments compare CPU BARGAIN and embedding proxies and measure query-result agreement and variability.}
\end{itemize}

%% file: sections/problem.tex
\section{Background and Problem}

In this section, we describe the background, as well as the inputs, outputs, and goals of our query-centric AI workflow.
To begin with, we explain the characteristics of common AI workflows that can be expressed as queries. Consider the following query examples:
\begin{lstlisting}
-- Q1
SELECT AVG(churn_rate), predicted_segment
FROM customer_analytics_db
GROUP BY predicted_segment
-- Q2
SELECT genre
FROM movie_db
WHERE predicted_rating > 4
-- Q3
SELECT *
FROM loan_application_db
WHERE predicted_default = True
-- Q4
SELECT t1.user, t2.movie
FROM user_db as t1
LEFT JOIN movie_db as t2
ON t1.predicted_preferred_genre = t2.genre
--Q5
SELECT *
FROM census_db
ORDER BY predicted_income ASC
LIMIT 50000
\end{lstlisting}

The queries \textbf{Q1-5} demonstrate various needs of predictive AI workflows to output designated tables. In a straightforward data science pipeline, we first fetch all the rows into memory and input them into the workflows. Then, after acquiring the predictions, we store them in a separate column and aggregate or filter the rows to output the results. Yet from a data management perspective, the above workflow causes a lot of data processing on all rows in the database, but only a small fraction of rows appear in the final output. For example, in \textbf{Q3}, if the default rate is only 10\%, a reliable proxy or early filter may reject many likely-negative entries before they reach the full AI pipeline. To sum up, a query-centric AI workflow contains two types of speed-up opportunities. First, there is a \textit{filter} condition (\textbf{Q2-4}) where we only need to focus on a subset of labels. We could derive a simpler pipeline that optimizes the subset's task. Query clauses like \texttt{WHERE}, \texttt{JOIN}, and \texttt{LIMIT} could trigger such a condition. Secondly, there is also an \textit{aggregate} condition (\textbf{Q1}) where we could consider different weights on different labels. For example, for conditions like \texttt{GROUP BY}, rows with a particular label might contain similar value distribution on the numerical column so that we do not need to query many of them to get the average. 

Furthermore, given the needs of feature engineering, the AI pipeline might need to query lots of columns to produce the predictions. Many predictive systems are powered by expensive AI pipelines that contain complex (1) feature engineering and (2) model architecture. We refer to these inference steps as \textit{targeted AI pipelines}. The pipeline queries the data needing predictions from sources and acquires the data in a tabular data format. Then, it transforms the input data into a target schema. For example, records are transformed into labels by a classification pipeline, numbers by a regression pipeline, and clusters by a clustering pipeline. Yet, as inference pipelines grow more sophisticated, they become increasingly expensive to execute.
{We call the supplied pipeline an \textit{oracle}, whose labels or scores define the query result to approximate.} 

We summarize these problems into two goals. Given a database, a query, and an AI pipeline, we want to (1) derive an estimator of the query result based on samples to reduce the number of records sent through the AI pipeline (\textbf{G1}); and (2) build a simpler proxy model that estimates the AI workflow's result on the query efficiently and accurately (\textbf{G2}). 
While many query optimization techniques for ML inference have been proposed to speed up the process~\cite{lu2018pp, yang2022core, kang2017noscope, aero2025}, {our design also aims to operate without changing the database architecture.} In the industrial setting, it is preferred to have minimal changes on the query processor and the layers underneath due to extra costs incurred on database migration and maintenance. Besides, the database already contains several optimization features like caching and indexing that we can leverage. 

\subsection{Inputs, Outputs, and Goals}
Our strategies take a query, a dataset, and an AI workflow as input, and output an approximated query result. {Our goal is to reduce the number of records sent through the AI pipeline and the cost of its model invocations.}
{The application identifies the expensive predicate or score function, and ordinary relational operations produce its input relation. Our prototype does not discover arbitrary model calls or search join orders. Aggregate error, selected-response quality, and exact winner agreement are separate evaluation objectives.}

To be specific, for the targeted ML pipeline, we expect the output to be either a \textit{classification} label or a \textit{regression} value.
Our scope of the queries is as follows:
\begin{enumerate}[leftmargin=6mm]
    \item \textit{Aggregate Queries} on a numerical column with the group-by clause based on ML pipeline.
    \item \textit{Selection Queries} on particular rows with the filter clause based on ML pipeline.
    \item \textit{Limit Queries} that control the size of output with sorting criteria based on the ML pipeline.
\end{enumerate}

From these queries, we can observe that some predicted labels or values are more important than others. For example, we only need to focus on those data points whose predicted values are in a certain range. Thus, a query-centric AI workflow, unlike typical Machine Learning (ML) pipelines that assign equal importance to each row’s predictions, results in different preferences on the pipeline output.
{Our strategies first collect a uniform sample of oracle outputs. The same labels support an aggregate estimate and training a proxy for the query's required output as we will elaborate in Section~\ref{sec:method}}.

%% file: sections/related_work.tex
\section{Related Work}
\label{sec:related_work}
Our work draws on and departs from three bodies of
literature. We first review classical approximate query
processing in Section~\ref{subsec:relate_aqp}, which provides the statistical foundation for
our AQP formulation. We then discuss query optimization
systems that handle ML predicates inside a database system in Section~\ref{subsec:query_opt_ml}, which
share our goal of reducing oracle invocations {through sampling or proxy models.}
Finally, we review LLM pipeline optimization work in Section~\ref{subsec:llm_opt},
{including declarative systems that exploit query structure. Our focus is reducing calls to a supplied oracle according to aggregate and selected-output requirements.}

\subsection{Approximate Query Processing}\label{subsec:relate_aqp}
Online aggregation~\cite{hellerstein1997online} introduced the
idea of computing aggregate results progressively as rows are
scanned: after each batch of samples, the system updates the
aggregate estimate and its confidence interval, {allowing users to inspect the current precision.}
WanderJoin~\cite{li2016wander} extends this framework to
queries with joins. The core difficulty is that uniform
sampling from each table independently produces few join
matches when join selectivity is low, making the estimator
high-variance. WanderJoin resolves this by performing random
walks over the join graph {and weighting their contributions to obtain an unbiased aggregate estimate.}

Our AQP formulation applies this framework to AI workflows,
{where each sampled value requires an expensive oracle or reward-model call. Our experiments start from a materialized relation and focus on reducing oracle invocations. The approximate intervals guide the sampling budget, while empirical error is reported separately.}
{BlinkDB~\cite{rev_blinkdb} uses stratified samples, while VerdictDB~\cite{rev_verdictdb} provides AQP through middleware without changing the database. Our unchanged-engine design is compatible with such integration.}

\subsection{Query Optimization with ML Predicates}\label{subsec:query_opt_ml}
Prior work on optimizing queries with expensive ML predicates
falls into {several related settings.}
{NoScope~\cite{kang2017noscope} uses specialized cascades for video queries, while BlazeIt~\cite{kang2019blazeit} uses proxy models for video aggregation and LIMIT queries. TASTI~\cite{tasti2021} constructs semantic indexes for unstructured data, including video, text, and speech. Its embeddings supply proxy scores without training a separate proxy for each query.}

{For general ML predicates expressed as UDFs,}
PP~\cite{lu2018pp} and CORE~\cite{yang2022core}
rewrite the query plan to insert proxy filters ahead of
costly ML UDFs inside the DBMS.
CORE relaxes PP's independence assumption across predicates using
branch-and-bound search, achieving up to 63\% higher
throughput than PP.
{ABae~\cite{kang2021abae} uses a learned proxy to allocate sampling effort across strata. Aero~\cite{aero2025} monitors UDF statistics at runtime and adapts predicate evaluation.}

{Our proxy deliberately relies on classical database and data-science tools rather than neural encoders. Cheap, explicit features (length, symbol density, formatting, lexical statistics) feed a single decision tree that is inexpensive to fit, easy to inspect, and whose splits are comparisons expressible as SQL predicates. It evaluates extraction, fitting, and prediction on CPU outside the engine.}

\subsection{LLM Pipeline Optimization}\label{subsec:llm_opt}
A parallel line of work optimizes LLM-powered workflows
from the machine learning and systems perspective. These
efforts fall into three categories: designing a declarative structure
of LLM pipelines, routing queries to cheaper models, and
reducing generation cost in Best-of-$N$ sampling.

First, LOTUS~\cite{patel2025lotus} introduces semantic operators,
a declarative programming model that extends relational
algebra with LLM-based operations such as semantic
filtering, joining, and ranking over unstructured text.
{Abacus~\cite{abacus2025} selects implementations of semantic operators under quality, cost, or latency constraints.}
TAG~\cite{biswal2025tag} frames natural language questions as table-augmented generation
tasks, enabling accurate analytical query answering without
Text-to-SQL. DocETL~\cite{shankar2025docetl} takes an
agentic approach to pipeline design, rewriting
user-specified LLM pipelines into decomposed plans {for document processing.}
{These systems optimize the pipeline itself. We treat it as a black box and reduce oracle calls through sampling and filtering, so the two lines of work are complementary.}

For optimization through model routing, FrugalGPT~\cite{chen2023frugalgpt} routes each query to
the cheapest LLM that can answer it accurately.
Task Cascades~\cite{shankar2026cascades} decomposes a task into a cascade
of cheaper sub-operations, escalating only uncertain records to the
expensive oracle.
{The approaches can reduce expensive model calls through choices of models, document portions, or tasks. Our evaluation instead keeps the oracle unchanged.}

{Best-of-$N$ selection is used in LLM post-training~\cite{touvron2023llama2,grattafiori2024llama}. Speculative Rejection~\cite{zanette2024specrejection} and the Pandora's Box formulation~\cite{kalayci2025pandora} reduce candidate-generation work. Our comparison holds generated candidates fixed but reduces RM scoring. Generation and fine-tuning costs are outside the reported savings.}
\myparagraph{Closest approximation methods}
BARGAIN~\cite{rev_bargain} targets agreement with oracle labels or precision and recall for filtering only. Our objective also covers the numerical aggregate or selected response requested by the query. Thus, in Section~\ref{sec:eval:tpcds}, we compare the authors' AT implementation on a subset of TPC-DS queries using the same CPU proxy and training rows.

Another recent study from Chung \textit{et al.}~\cite{rev_chung} uses lightweight CPU classifiers over embeddings from neural networks for semantic filtering and ranking. Its gains depend on representation quality and the amortization of embedding and training costs. Our RM cascade uses structural features to predict continuous scores and reduce scoring through a gate and candidate filter. As we will show in Section~\ref{sec:eval:llm}, surface features used in our proxy model retain competitive selection quality while avoiding neural feature-extraction cost on our workloads. For aggregates, our shared execution preserves a uniform-sample AQP estimate and uses the same labels to train PM. The remaining oracle requests determine which configured plan completes first, with label reuse if the other plan continues as we will elaborate in Section~\ref{sec:strategy_choice}.

\myparagraph{Approximate and declarative AI queries}
ThalamusDB~\cite{rev_thalamus} uses model scores and user calibration to bound multimodal query results. UQE~\cite{rev_uqe} supports aggregate and retrieval queries, using embedding-based stratification for aggregation and active learning for retrieval. Palimpzest~\cite{rev_palimpzest} chooses AI-operator implementations, while DocETL~\cite{shankar2025docetl} rewrites document-processing tasks. ZenDB~\cite{rev_zendb} uses document hierarchies, and QUEST~\cite{rev_quest} reduces extraction work through retrieval and document-specific plans. SEED~\cite{rev_seed} compiles data-curation cascades using generated code and small models. The systems establish the value of declarative AI optimization. Our query-centric design uses aggregate and selected-output requirements to coordinate uniform AQP with a structural CPU proxy for a supplied oracle.

Cross-prediction~\cite{rev_crossprediction} combines fitted predictions and sampled labels to obtain corrected statistical estimates. Our shared execution instead preserves a proxy-independent AQP estimate alongside the PM result. Uniform labels serve both estimation and proxy training, and a shared cache eliminates duplicate oracle calls. Comparing the remaining requests determines which configured plan completes first, making its result available before the other plan finishes.

Our experiments fix the input relation or candidate pool and oracle to isolate call reduction at comparable aggregate or selection quality. In contrast, ThalamusDB and UQE use calibration or embedding-based strata; Palimpzest and DocETL optimize operator implementations or task decompositions; QUEST and ZenDB optimize document extraction through retrieval or document hierarchies, whereas our queries operate on materialized records or generated response candidates. Their full-pipeline results would thus include effects beyond our estimation and selection strategies. We therefore discuss these systems qualitatively rather than use them as direct experimental baselines.

%% file: sections/method.tex
\section{Query-Centric AI Workflow Inference}
\label{sec:method}
The core idea of an inference optimization from a query-centric AI workflow is that we do not need to fetch all the data points and invoke the AI pipeline on each of them to obtain accurate results that require predictions.

For example, if the query aims at collecting the \texttt{AVG(c1)} where the predicted values at \texttt{c2} are equal to \texttt{"label 1"} while all values in \texttt{c1} are equal to 0,
{the average is zero whenever at least one row qualifies. No further labels are needed to compute that value.}

On the other hand, if we do not aggregate the rows in the query result (e.g. \texttt{SELECT}),
we still do not need to invoke the AI pipeline on much of the database if the query has a filter by prediction labels.
Therefore, there exists an execution plan that decides how many AI-pipeline invocations we need, as well as how they should be constructed.
In the following sections, we describe two main strategies to process the queries without invoking the AI pipeline on all data from the database, their tradeoffs, and how to interplay them together to optimize query performance.
\subsection{Strategy AQP: Treating the Workflow as an Approximate Query Processing Problem}
\label{sec:strategy_aqp}
The first strategy treats these workflows as an Approximate Query Processing (AQP) problem and estimates aggregate results from samples.
We use an online aggregation approach~\cite{hellerstein1997online} to provide approximate results with confidence intervals and gradually improve with an increasing number of samples.
Given an \texttt{AGG(expression)} and any samples where \texttt{AGG} could be \texttt{SUM}, \texttt{AVG}, \texttt{COUNT} or \texttt{VAR} and \texttt{expression} could involve any columns in the database,
{the desired accuracy target for the estimator $\hat Y$ is}
\begin{equation}\label{eq:conf_interval}
    Pr[| \hat{Y} - \texttt{AGG(expression)}| \leq \epsilon] \geq \alpha.
\end{equation}
{Here $\epsilon$ is the requested half-width and $\alpha$ the confidence level. Each round draws a uniform batch with replacement and updates the estimate and approximate half-width. The width guides the sampling budget, and a stalled width is reported with its consumed budget.}
{Equation~\ref{eq:conf_interval} states the desired coverage. The normal interval below is a fixed-budget approximation, not a simultaneous guarantee after repeated inspection. Repeated draws retain their multiplicities, while cached labels avoid repeated oracle calls.}

\subsubsection{Illustration: AQP with \texttt{\upshape SUM}}
For concreteness, we first address the following \texttt{SUM} query:
\begin{lstlisting}
SELECT SUM(c1) WHERE c2 = "label 1"
\end{lstlisting}

{Let $z_\gamma$ indicate whether row $\gamma$ satisfies the oracle predicate, and let $a_\gamma$ be its value in \texttt{c1}. A uniform draw has probability $p(\gamma)=1/N$, so $z_\gamma a_\gamma/p(\gamma)$ estimates the full \texttt{SUM} without bias. For a batch of $b$ draws, the round estimate is $v_i=(N/b)\sum_{\gamma\text{ drawn}}z_\gamma a_\gamma$, with repeated draws counted repeatedly. Averaging independent round estimates reduces variance as more samples are collected. The interval below summarizes the uncertainty of the estimate and guides the sampling budget.}

\subsubsection{AGG Functions}
\label{sec:strategy_1_CI}
{SUM and COUNT use scaled sample totals, with $a_\gamma=1$ for COUNT. AVG uses the ratio of the sampled qualifying sum to the qualifying count. A round with no qualifying row has an undefined AVG and is excluded from the finite round estimates. VAR requires an estimator based on the appropriate sample moments.}
Let the {finite aggregate estimate on the $i$-th round} be $v_i$. 
Then the estimator of $\hat{Y}$ with $n$ trials is: 
$\hat{Y} = \frac{1}{n} \sum_{i=1}^n {v_i}$.
{To compute the approximate half-width in Equation~\ref{eq:halfwidth}, we estimate the sample standard deviation:}
$s = \sqrt{\frac{1}{n-1} \sum_{i=1}^n {(v_i-\hat{Y})}^2} $.
{With the estimator $\hat{Y}$ and the sample standard deviation $s$, the approximate half-width is}

\begin{equation}\label{eq:halfwidth}
 h_n=\frac{z_{(1+\alpha)/2}s}{\sqrt n}.
\end{equation}

{Here $n$ counts finite round estimates and $z_{0.975}=1.96$ for $\alpha=0.95$. The reported half-width describes empirical precision at the current budget, rather than certifying coverage after repeated inspection.}

For the relative-error experiments, let $r$ count sampling rounds and let $n(r)$ count their finite estimates. The relative half-width is $H_r=100h_{n(r)}/|\hat Y_{n(r)}|$, with a target of $10\%$. A plateau requires finite current and earlier widths, and is detected when
\begin{equation}\label{eq:plateau}
 H_r>10,\quad H_{r-5}>0,\quad
 \frac{H_{r-5}-H_r}{H_{r-5}}<0.05,\quad r\geq7.
\end{equation}
The patience window avoids reacting to a single round. When a plateau is detected, the system reports the current interval and the oracle cost consumed so far, and the user can stop or escalate to full evaluation (Section~\ref{sec:plateau_results}).

Other AGG functions, such as MIN and MAX, are difficult for the sampling method. They usually do not follow a certain distribution like Gaussian.
Assuming these values depend on one unique row only, {and each independent round samples $b$ distinct rows uniformly,} the expected number and standard deviation of sampling trials needed will be $\frac{N}{b}$ and $\sqrt{(1-\frac{b}{N})/(\frac{b}{N})^2}$ where $N$ is the number of rows and $b$ is the sample size.
{Each round hits the extreme row with probability $b/N$, so the waiting time is geometric. With $b$ draws with replacement, the hit probability instead is $1-(1-1/N)^b$.}
If $b \ll N$, {the waiting-time variation is large.} If $b$ is not much smaller than $N$, sampling becomes comparable to a full table scan. In both cases, a full scan is preferable.
To address the limitation on these AGG functions, we introduce the second strategy that we should construct queries to select points with target labels with small cost, in the next section.

\subsubsection{LIMIT Queries}
\label{sec:strategy_1_limit}

Beyond AGG functions, Strategy~AQP extends naturally to \texttt{LIMIT} queries of the form:
\begin{lstlisting}
SELECT * ORDER BY score DESC LIMIT k
\end{lstlisting}

This pattern appears directly in LLM post-training pipelines: for each prompt, $N$ candidate responses are generated and scored by a reward model, and only the top-$k$ responses are retained~\cite{grattafiori2024llama}.
Rather than invoking the oracle on all $N$ candidates, we ask: how many candidates must we sample so that the global top-$k$ are likely to be included?

\myparagraph{{Candidate coverage}}
{
For a uniform sample of fixed size $m\geq k$ without replacement, the probability of including every global top-$k$ candidate is}
\begin{equation}\label{eq:limit_stopping}
    P_k(m, N) = \prod_{i=0}^{k-1} \frac{m - i}{N - i}.
\end{equation}
{The identity counts subsets containing the designated top-$k$, without assumptions about score independence. A fixed budget can be chosen to satisfy $P_k(m,N)\geq\alpha$. For $k=1$, the probability is $m/N$. For large $N$ and fixed $k$, the required budget is approximately}

\begin{equation}\label{eq:limit_approx}
m\approx N\alpha^{1/k}.
\end{equation}

{The fraction increases with $k$, because every designated candidate must be present. Winner recovery and task quality therefore have different sampling requirements.}

\myparagraph{{Adaptive Stopping in Practice}}
{Equation~\ref{eq:limit_stopping} is exact for fixed-size uniform sampling. For applications that accept an empirical quality tradeoff, a score-stabilization heuristic instead uses a tolerance $\delta>0$ and checks whether the worst score among the current top-$k$ changes little:}

\begin{equation}\label{eq:adaptive_stop}
    |r_{(k)}^{(m)} - r_{(k)}^{(m-1)}| \leq \delta,
\end{equation}
where $r_{(k)}^{(m)}$ denotes the $k$-th highest reward score observed after drawing $m$ ($m \geq k+1$) candidates.
{Equation~\ref{eq:adaptive_stop} records score stabilization rather than a probability of recovering unseen candidates. An unchanged score can occur before a better candidate is sampled. We retain the rule as an empirical option, separate from the fixed-size identity in Equation~\ref{eq:limit_stopping}.
Our evaluation (Section~\ref{sec:eval:llm}) shows that scoring a small fraction of candidates already gets close to full-RM quality, at 1.9$\times$ to 8.0$\times$ lower estimated oracle scoring time at the offline stabilization points.
}

\subsection{Strategy PM: Proxy Model to Increase Throughput on Model Inference}
\label{sec:strategy_pm}

The second strategy to avoid feeding all data to the AI workflow is to develop a predictive model that \textbf{handles the ``easy'' input} before passing the remaining data to the AI pipelines.
We define easy data as rows whose pipeline outcome can be predicted with high confidence by a rule-based model, or whose prediction error does not affect the \textit{final query result}.
{For example, errors on candidates outside the returned set need not change a selection query's output.}
Thus, the idea is to train an efficient proxy model with limited ground truth samples and process the majority of the data {while retaining useful result quality}, thus increasing the throughput of the query.

\subsubsection{Illustration: SELECT rows with label ``1''}
Let us consider a simple query:
\texttt{SELECT * FROM table where label = '1'}
{A lightweight proxy predicts the predicate label from inexpensive row features. First, a uniform sample obtains oracle labels, or reuses those already collected by AQP. Then, the sampled labels train a query-specific decision tree. Afterward, its predictions determine which remaining rows receive oracle evaluation. For aggregate queries, a random audit also accounts for positives among excluded rows. The proxy's purpose is to reduce work for the query result, rather than reproduce every oracle prediction.}
\begin{figure}[t]
    \centering
    \includegraphics[width=\linewidth]{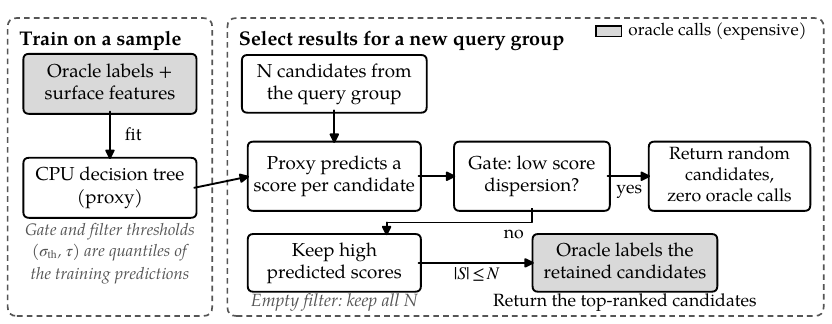}
    \caption{{Strategy PM. A CPU proxy trained on a small oracle-labeled sample predicts a score for each candidate in a query group, and only retained candidates receive oracle labels. Reward-model scoring of LLM responses is one instance; Table~\ref{tab:cascade_presets} gives the cascade presets used there.}}

\label{fig:strategy-pm-overview}
\end{figure}

\subsubsection{Overview of Proxy Model}
{We use a Decision Tree (DT) because it is inexpensive to fit and its feature comparisons can be expressed as SQL predicates. The proxy learns from a small uniform set of oracle-labeled examples, including samples already collected by AQP. Figure~\ref{fig:strategy-pm-overview} illustrates the workflow: a sampled subset provides oracle labels for a regression tree, which then predicts scores for the candidates of unseen query groups from surface features. In the RM workload, the oracle is the reward model and a query group is the set of candidate responses for one prompt. Our evaluated training budgets are 5\% or 10\% of prompts in the LLM workload and a sweep of row fractions in TPC-DS. The budgets are workload settings rather than a universal minimum.}
In our prototype, feature extraction, fitting, and prediction run on CPU outside the database.
\myparagraph{Aggregate correction}\phantomsection\label{sec:pm_agg}
For TPC-DS aggregates, PM includes known training labels exactly and obtains oracle labels for all remaining rows that the proxy does not confidently exclude. 
The remaining predicted-negative pool $U$ receives a uniform random audit at a configurable fraction; Section~\ref{sec:eval:tpcds} gives the fractions used in our experiments.
Let $C$ contain the training and confirmed rows, and let $A\subseteq U$ be the audit sample. The corrected SUM is
\begin{equation}\label{eq:audit_sum}
 \hat Y_{\mathrm{PM}}=\sum_{i\in C}z_i a_i+
 \frac{|U|}{|A|}\sum_{i\in A}z_i a_i .
\end{equation}
An empty pool contributes zero, and a non-empty pool receives at least one audit row. COUNT uses $a_i=1$. AVG divides the corrected SUM by the corrected positive count. The sampled correction accounts for positives among excluded rows. It is separate from the per-prompt selection cascade below.

\subsubsection{Feature Engineering}
\label{sec:pm_features}
{The proxy uses cheap features from the input record, including length, symbol density, formatting, and lexical statistics. For LLM responses, answer markers and numeric structure can help predict oracle scores without a neural encoder. Smaller models and embeddings can also observe such properties, so our claim concerns extraction cost and usefulness. Section~\ref{sec:proxy_ablation} compares representations separately from the regression model.}

\subsubsection{Cascade Gating}\label{sec:pm_cascade} We now expand the rejection-sampling cascade, which is the form used throughout our LLM evaluation (Section~\ref{sec:eval:llm}). The query is \texttt{SELECT * ORDER BY rm\_score DESC LIMIT k} over the $N$ candidates of each prompt, and the proxy is a decision tree that predicts the oracle score from surface features. {We specialize to top-1 ($k=1$) because our evaluated workflow selects one of 32 responses per prompt.} {Each prompt forms an independent query group for the cascade's two coupled tests.}

\myparagraph{Variance gate (group-level skip)} The tree first scores all $N$ candidates of a prompt and we measure the within-prompt dispersion of these predictions, $\sigma_{\mathrm{pred}} = \mathrm{std}(\hat{s}_1,\dots,\hat{s}_N)$. {When $\sigma_{\mathrm{pred}}$ falls below the threshold $\sigma_{\mathrm{th}}$, the gate returns one candidate using a fixed random seed and issues zero oracle calls for the prompt. Low predicted dispersion identifies groups for which the proxy provides little ranking information. The gate exploits the empirical quality of candidates in structured tasks, rather than certifying equality of their true oracle scores.}

\myparagraph{Candidate filter (within-group pruning)} If a prompt survives the variance gate, we avoid scoring all $N$ candidates by {pruning candidates with low predicted scores.} We retain the survivor set $S = \{\, c : \hat{s}(c) \ge \tau \,\}$, where $\tau$ is a score threshold, forward only $S$ to the oracle, and select $\arg\max_{c \in S}$ over the true oracle scores. If $S$ is empty, meaning no candidate clears $\tau$, we fall back to forwarding all $N$ candidates {to obtain a full-oracle selection for that prompt.} {This costs $|S|$ calls for nonempty $S$ ($N$ when empty).}

{Both thresholds are computed on the training sample alone, after fitting the tree to oracle labels. Let $q_s$ denote the score quantile and $q_\sigma$ the dispersion quantile. We set $\tau=Q_{q_s}(\{\hat s(c):c\text{ in training}\})$ and $\sigma_{\mathrm{th}}=Q_{q_\sigma}(\{\sigma_{\mathrm{pred}}(p):p\text{ in training}\})$. The standard deviation uses divisor $N$. No held-out labels enter threshold selection.}
{The filter is one-sided, and the oracle ranks only the retained candidates. A nonempty survivor set does not establish exact top-$k$ recovery. Section~\ref{sec:selection_agreement} measures agreement with full-RM selection and the effect of scoring five candidates on prompts skipped by the gate.}

\myparagraph{Two cascade presets} {Preset A disables the gate and applies only the candidate filter. Preset B enables both tests. General instruction following uses A, while Math and Code use B when the mean training oracle score is at least 6.5 and A otherwise. Table~\ref{tab:cascade_presets} separates the preset parameters from the workflow figure. The policy uses the observed quality of training candidates to adjust the scoring budget.}
\begin{table}[!tp]

\centering\small
\caption{Cascade presets used in the LLM experiments.}
\label{tab:cascade_presets}
\setlength{\tabcolsep}{4pt}
\begin{tabular}{llll}
\toprule
Task and training condition & Preset & $q_\sigma$ & $q_s$\\
\midrule
Math/Code, mean RM $<6.5$ & A & off & 0.60\\
Math/Code, mean RM $\geq6.5$ & B & 0.99 & 0.50\\
General, Qwen3-8B & A & off & 0.90\\
General, LLaMA-3.1-8B & A & off & 0.97\\
General, Mistral-7B-V0.3 & A & off & 0.30\\
General, Qwen3-32B & A & off & 0.88\\
\bottomrule
\end{tabular}

\end{table}

{
\myparagraph{Scope} Overall, PM targets selection queries where any sufficiently good candidate is an
acceptable answer, as in Math and Code, where a pool often holds several
correct responses (Section~\ref{sec:eval:llm}).
It does not guarantee recovery of the exact top-scoring candidate. When that candidate matters, as on the
open-ended General task, it saves little and loses quality. The fallback behavior and the requirement for exact winner recovery are discussed below.
}
\subsubsection{Fallback or Early Termination}
\label{sec:pm_fallback}
{The implemented cascade returns a candidate without oracle calls when the variance gate fires. If the gate does not fire, an empty filtered set triggers full-oracle scoring. The mean training oracle score selects the preset, and there is no separate reliability test on the gate. Users requiring the exact full-oracle winner can disable approximate pruning and score all candidates. For approximate selection, the measured task quality and agreement describe the resulting tradeoff.}

\subsection{Deciding Which Strategy to Use}
\label{sec:strategy_choice}
\begin{figure}[t]

\centering
\includegraphics[width=\linewidth]{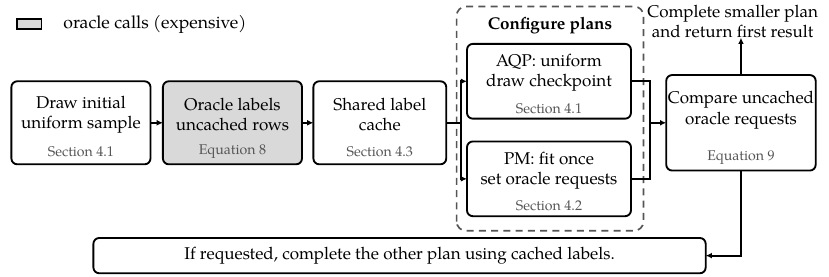}

\caption{Shared AQP and PM execution. Initial uniform labels support AQP and train PM. Comparing remaining uncached requests determines which configured plan completes first; the other can continue using cached labels.}
\label{fig:query_execution}

\end{figure}

{Query structure determines the output objective. Aggregates can use either estimator, while PM supports approximate selection and LIMIT admits fixed-size sampling. Low selectivity and variable qualifying values can make AQP less precise. PM helps when inexpensive features identify useful records, but class imbalance alone does not establish a good proxy. The strategies share labels rather than requiring a second sampling phase. Our 5\% to 10\% initial budgets are workload settings, not universal minimums.}
\myparagraph{Shared samples and oracle cost}
Shared execution avoids paying twice for oracle labels needed by AQP and PM. First, ordinary joins and filters produce the input relation for the supplied expensive function. Second, uniform oracle sampling gives an initial AQP estimate. Third, the already labeled sample trains a CPU proxy without a second training-label budget. Figure~\ref{fig:query_execution} shows how both plans reuse the labels throughout execution.

If $L$ contains cached row labels and a plan requests rows $D$, only $D\setminus L$ needs oracle evaluation. Total cost is
\begin{equation}\label{eq:cost_accounting}
 C=C_{\mathrm{features}}+C_{\mathrm{fit}}+
 C_{\mathrm{predict}}+C_{\mathrm{route}}+
 \sum_{i\in L_{\mathrm{final}}}c_f(i).
\end{equation}
Each cached oracle row is charged once, while the CPU terms account for feature extraction, fitting, prediction, and routing.

\myparagraph{Output order and continuation}
After fitting PM, its confirmation set and configured audit determine requests $D_{\mathrm{PM}}$. A chosen AQP checkpoint has a uniform draw schedule with distinct row set $D_{\mathrm{AQP}}$. Before observing their unknown labels, compare
\begin{equation}\label{eq:remaining_oracle_cost}
 c_{\mathrm{PM}}=|D_{\mathrm{PM}}\setminus L|,
 \qquad c_{\mathrm{AQP}}=|D_{\mathrm{AQP}}\setminus L|.
\end{equation}
If $c_{\mathrm{PM}}<c_{\mathrm{AQP}}$, we complete PM and return its estimate first. Otherwise, we complete AQP first, with ties also favoring AQP. 
The other plan can then continue on the cached labels if the user wants its result too. Until either plan completes, the AQP estimate from the initial sample serves as a running answer.

Overall, the plans share labels while preserving their respective estimators: AQP averages only its own uniform draws, and PM's proxy and audit weights are fixed once fitted, so either plan can finish first without changing the other's estimate. The coordination also adds little CPU work. In Section~\ref{sec:eval:llm}, on the LLM workloads, feature extraction,
tree fitting, and prediction together take a few seconds, roughly 1\% of the
estimated time for full reward-model scoring (167--10,891\,s) and at most about 5\% in the worst case.

%% file: sections/evaluation_tpc.tex
\section{Evaluation}
\label{sec:eval}
In this section, we evaluate our query-centric optimization on two kinds of AI workflows. Our goals are:
\begin{itemize}
    \item Using AI workflows derived from TPC-DS queries, we demonstrate that under balanced distributions AGG results can be approximated within $10\%$ error {at the original stabilization reference of 10\% to 15\% of tuples, corresponding to 85\% to 90\% fewer oracle calls.}
    \item Using LLM post-training tasks with Reward Models as an example, {we evaluate Strategy PM against fixed-budget sampling and use up to 19 times fewer RM calls than full scoring, including training. We report task-specific quality gaps and their variability.}
\end{itemize}

\myparagraph{Implementation Details}
We run the CPU experiments on a workstation with an AMD Ryzen~9 9950X CPU and 96\,GB of DRAM. All proxy feature extraction, fitting, and prediction run on CPU. We use a single NVIDIA GeForce RTX~5090 GPU for Qwen3-8B, LLaMA-3.1-8B, and Mistral-7B-V0.3 generation. Qwen3-32B generation uses eight A100 80\,GB GPUs.

\subsection{Evaluating AGG Queries with TPC-DS}
\label{sec:eval:tpcds}
\myparagraph{Experimental Setup}
\label{sec:eval:setup}
We choose TPC-DS as the evaluation workload for two reasons.
First, real-world AI-agentic query workflows are largely black-box:
their internal predicates and data distributions are opaque,
making controlled experimentation difficult.
Second, TPC-DS is a widely adopted benchmark that covers
representative SQL query patterns with diverse aggregate functions,
providing a well-understood setting in which we can precisely
control label distributions and isolate the effect of each strategy.

Specifically, we select eight representative TPC-DS queries (Q7, 13, 50, 52, 53, 54, 55, 56) spanning three aggregate functions (SUM, AVG, and COUNT).
For each query, we split the original WHERE predicate into
two parts: the \emph{predicted} predicate that we could predict using columns in the table and the other structural filters and joins.
We construct a data table by joining the relevant fact and
dimension tables according to the original query's join
conditions, materializing the filters explicitly.
The predictive predicate is converted into a binary row label.
We then evaluate a single scalar aggregate (SUM, AVG, or COUNT)
over all rows satisfying the learned predicate, stripping the
original query's GROUP~BY, ORDER~BY, and LIMIT clauses to
isolate the core problem---accurately estimating an aggregate
under a predictive predicate with limited oracle calls that mimics an AI workflow.

Since the value distributions of TPC-DS are referenced from real world data tables, applying the original WHERE predicate to
dimension columns (e.g., \texttt{i\_category},
\texttt{i\_class} for Q53) to produce a natural label distribution limits our opportunities to stress test our strategies with a comprehensive set of ranges of label distributions.
Thus, we also set the positive-label rate to 30\% and 50\% to simulate more balanced distributions
that may arise in practice.
We define the oracle budget as the number of calls to retrieve these labels from the column.
{Using known predicate labels, we measure oracle calls and relative aggregate error against full evaluation.}

{We repeat each configuration multiple times. We report mean aggregate error, with shading showing the 25th to 75th percentiles in the error curves.}
Both strategies use uniform random sampling to select the data for calling the oracle initially.

\myparagraph{Compared methods and settings}
We compare the following methods using the same natural predicate labels and full-oracle aggregate.
\begin{itemize}
\item \textbf{AQP} estimates the aggregate from uniform 1\% batches. AQP (stop) in Table~\ref{tab:bargain} returns at the first plateau in Equation~\ref{eq:plateau}, or at the 25-round cap. It does not stop at the first width below 10\% or assume that every returned estimate meets that target.
\item \textbf{PM} in Table~\ref{tab:bargain} uses oracle confirmation and random auditing, as described in Section~\ref{sec:pm_agg}. Training uses 5\% of rows, except 10\% for Q54. The audit fraction is 10\% by default, 5\% for Q50, and 2\% for Q54.
\item \textbf{BARGAIN AT} uses the authors' adaptive-threshold implementation~\cite{rev_bargain}. AT-90, AT-95, and AT-99 target row accuracies of 90\%, 95\%, and 99\%, respectively. They use failure probability 0.05 and $M=20$, with the target adjusted for known training labels.
\end{itemize}
PM and AT share training rows and the same CPU decision tree, using one CPU thread for fitting and prediction. We count every distinct oracle row, including training, calibration, confirmation, and audit. We repeat the comparison multiple times with paired seeds and report mean relative aggregate error. We also repeat shared execution multiple times per query to measure label reuse and output order.

\begin{figure*}[!t]
  \centering
  \begin{minipage}{\textwidth}
    \centering
    {\small
      \textcolor{ForestGreen}{\rule{8pt}{8pt}}~adjusted $p{=}50\%$
      \qquad
      \textcolor{BurntOrange}{\rule{8pt}{8pt}}~adjusted $p{=}30\%$
      \qquad
      \textcolor{RoyalBlue}{\rule{8pt}{8pt}}~true label (natural)
    }
  \end{minipage}\\[6pt]
  \includegraphics[width=0.24\linewidth]{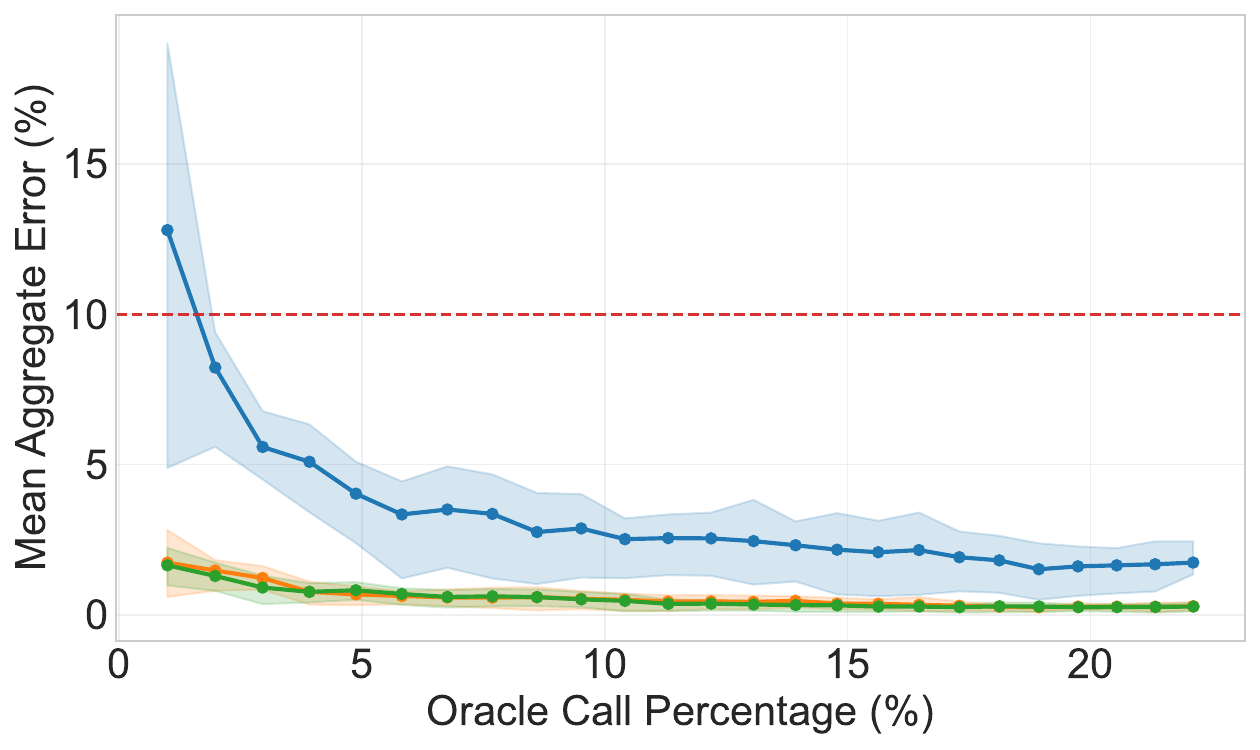}\hfill
  \includegraphics[width=0.24\linewidth]{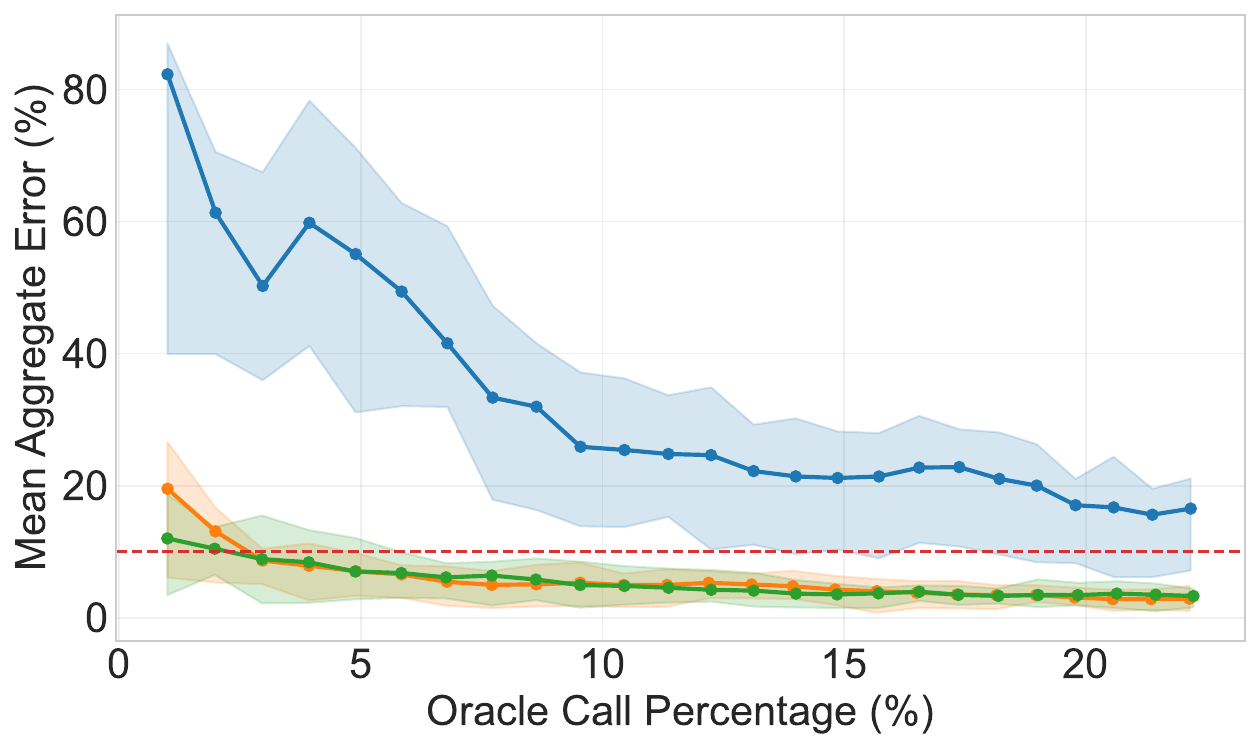}\hfill
  \includegraphics[width=0.24\linewidth]{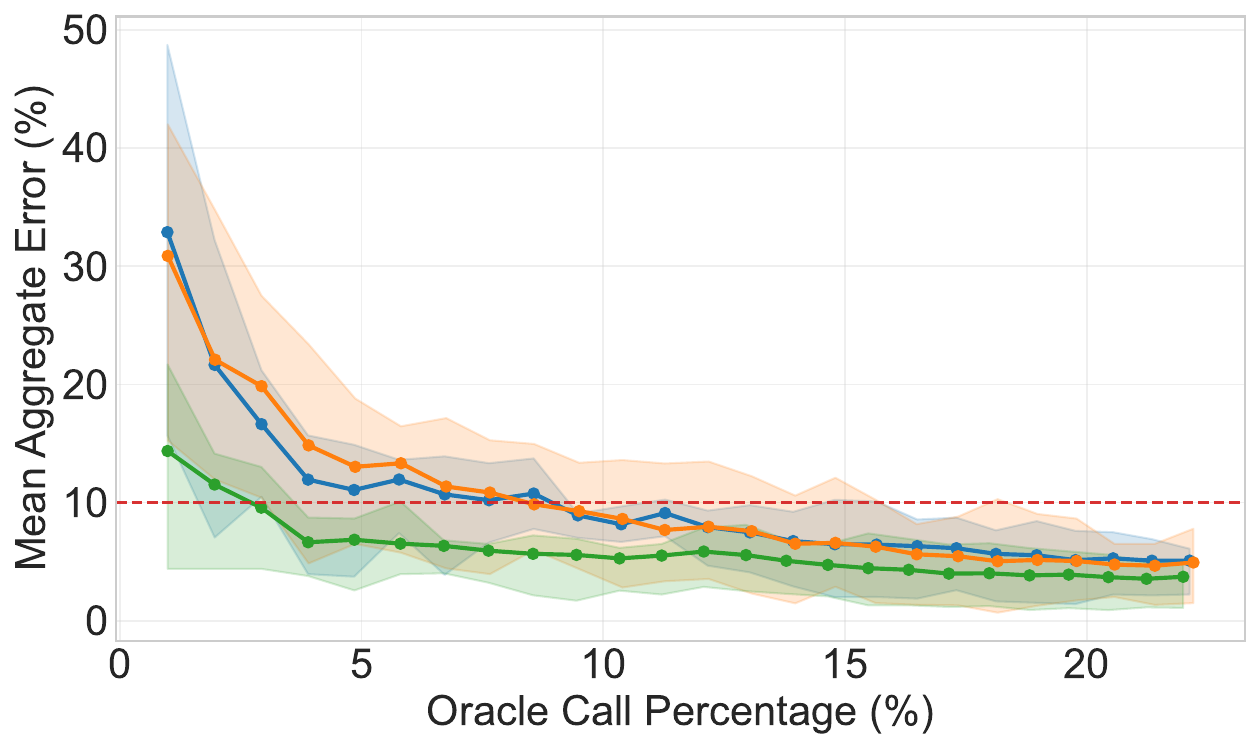}\hfill
  \includegraphics[width=0.24\linewidth]{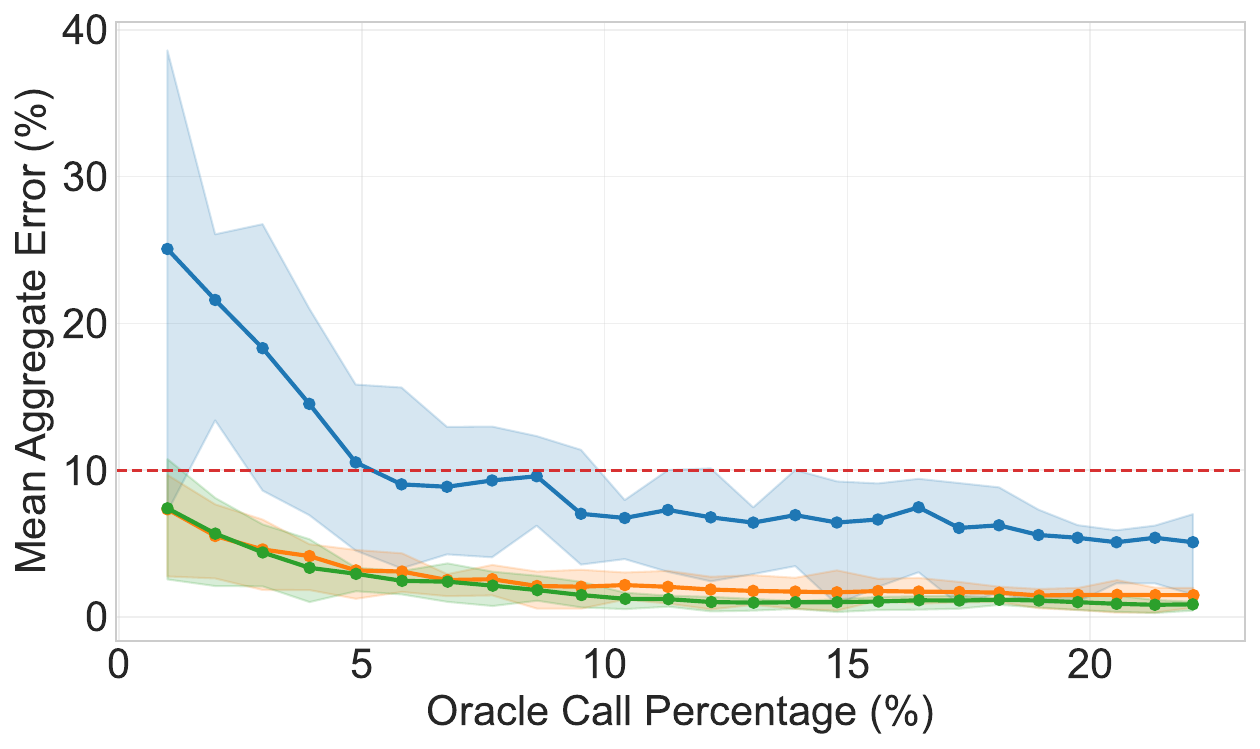}\\[-2pt]
  {\small (a) Q7 ($p{=}1.4\%$,$n=529,437$) \hfill
         (b) Q13 ($p{=}1.0\%$, $n=6,477$) \hfill
         (c) Q50 ($p{=}26.3\%$, $n=2,311$) \hfill
         (d) Q52 ($p{=}1.7\%$,$n=86,313$)}\\[6pt]
  \includegraphics[width=0.24\linewidth]{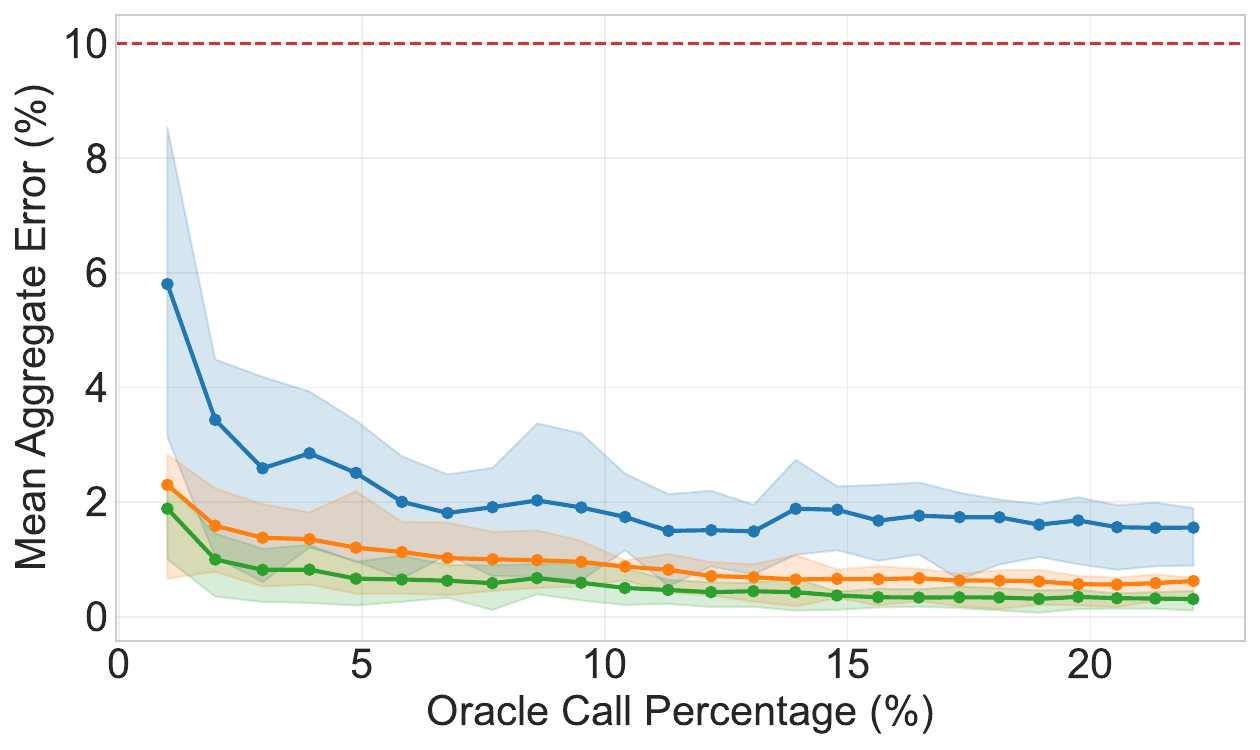}\hfill
  \includegraphics[width=0.24\linewidth]{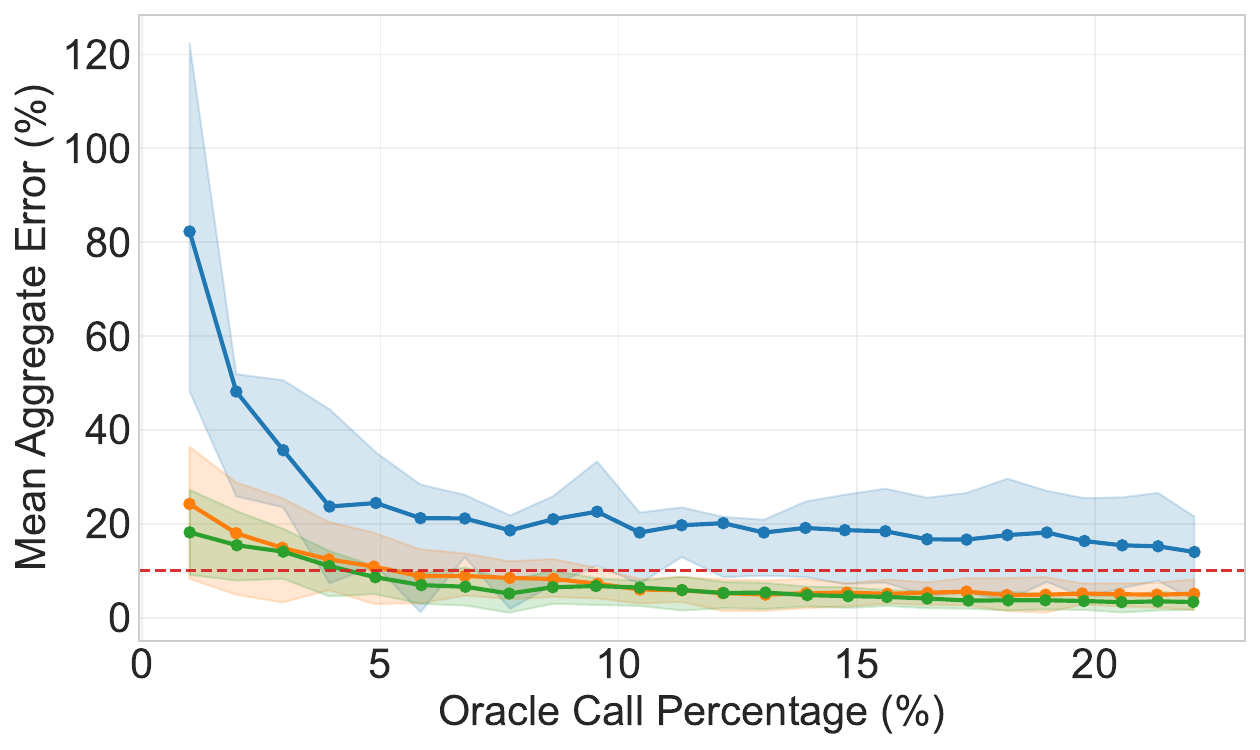}\hfill
  \includegraphics[width=0.24\linewidth]{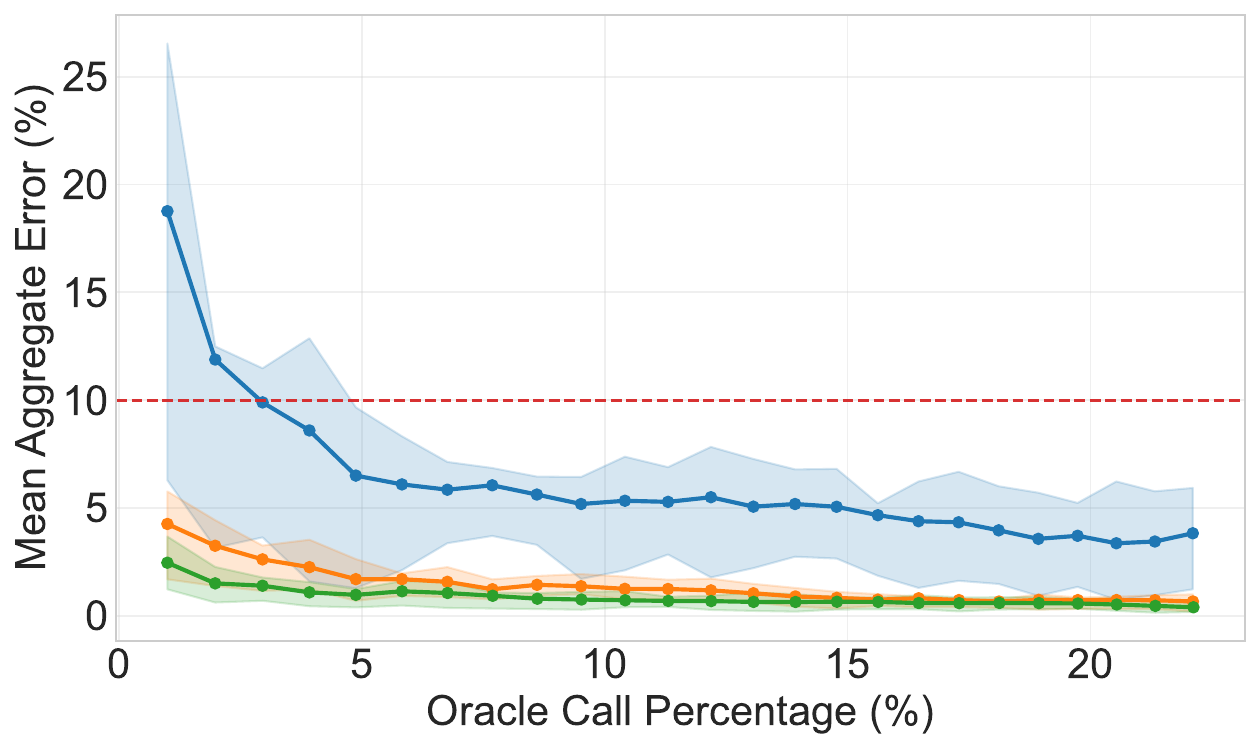}\hfill
  \includegraphics[width=0.24\linewidth]{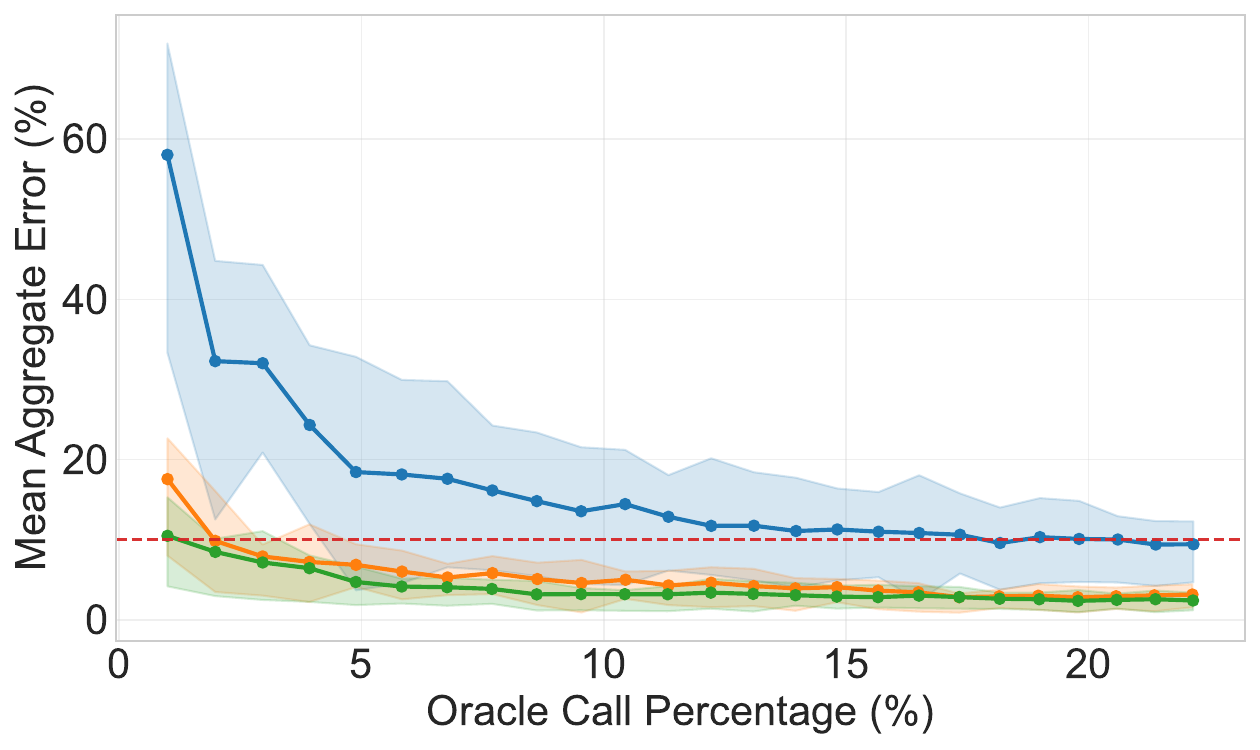}\\[-2pt]
  {\small(e) Q53 ($p{=}4.1\%$,$n=541,763$) \hfill (f) Q54 ($p{=}6.1\%$,$n=2,201$) \hfill
         (g) Q55 ($p{=}1.7\%$,$n=94,025$) \hfill
         (h) Q56 ($p{=}4.5\%$,$n=20,165$)}
  \caption{
    \textbf{Strategy~AQP: mean aggregate error vs.\ oracle-call
    percentage} for all eight TPC-DS queries.
    Shaded regions show the 25th--75th percentile range;
    the dashed red line marks 10\% error.
    $p$ denotes the natural positive-label rate and $n$ denotes the table size.
    Under balanced distributions (green, orange), sampling converges
below 10\% error within 10\% of oracle calls.
{Under the natural distribution (blue), Q13 and Q54 remain above
10\% error at a 20\% oracle budget. Their achieved error is reported
without requiring full evaluation. 
}}
  \label{fig:tpcds_s1_dist}
\end{figure*}

\begin{figure*}[!t]
  \centering
  \begin{minipage}{\textwidth}
    \centering
    {\small
      \textcolor{ForestGreen}{\rule{8pt}{8pt}}~adjusted $p{=}50\%$
      \qquad
      \textcolor{BurntOrange}{\rule{8pt}{8pt}}~adjusted $p{=}30\%$
      \qquad
      \textcolor{RoyalBlue}{\rule{8pt}{8pt}}~true label (natural)
    }
  \end{minipage}\\[6pt]
  \includegraphics[width=0.24\linewidth]{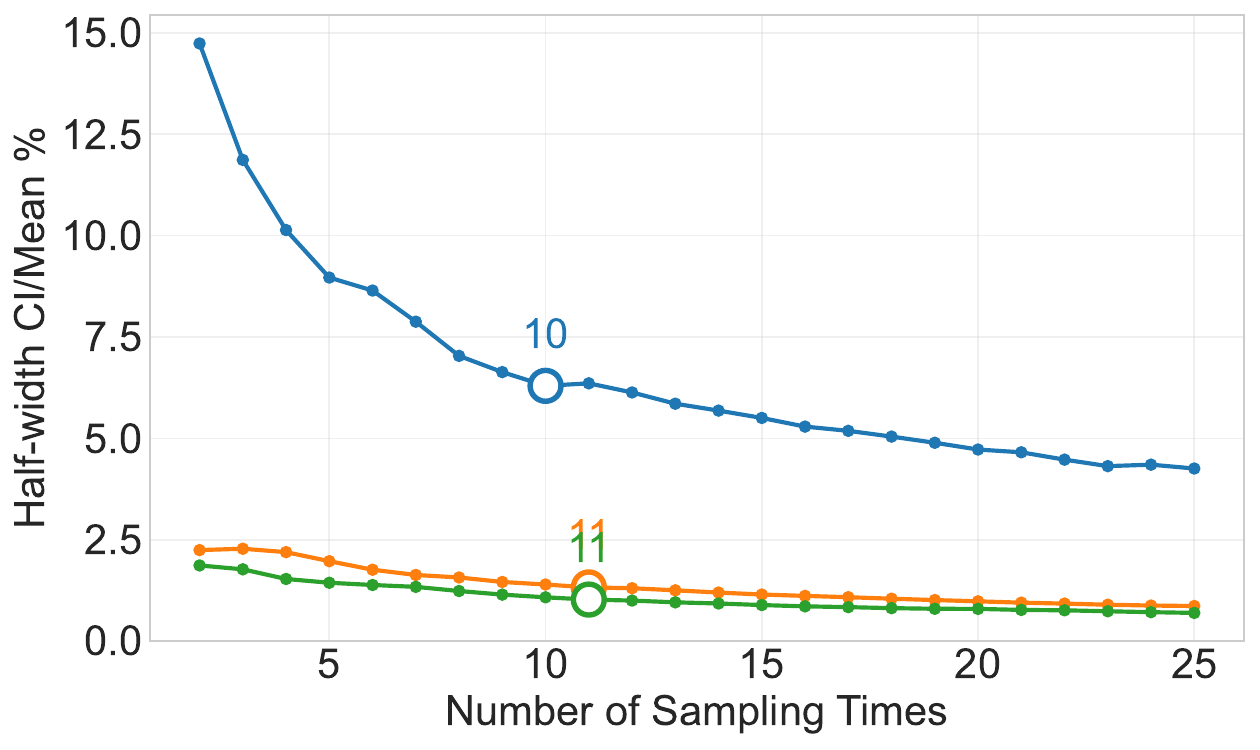}\hfill
  \includegraphics[width=0.24\linewidth]{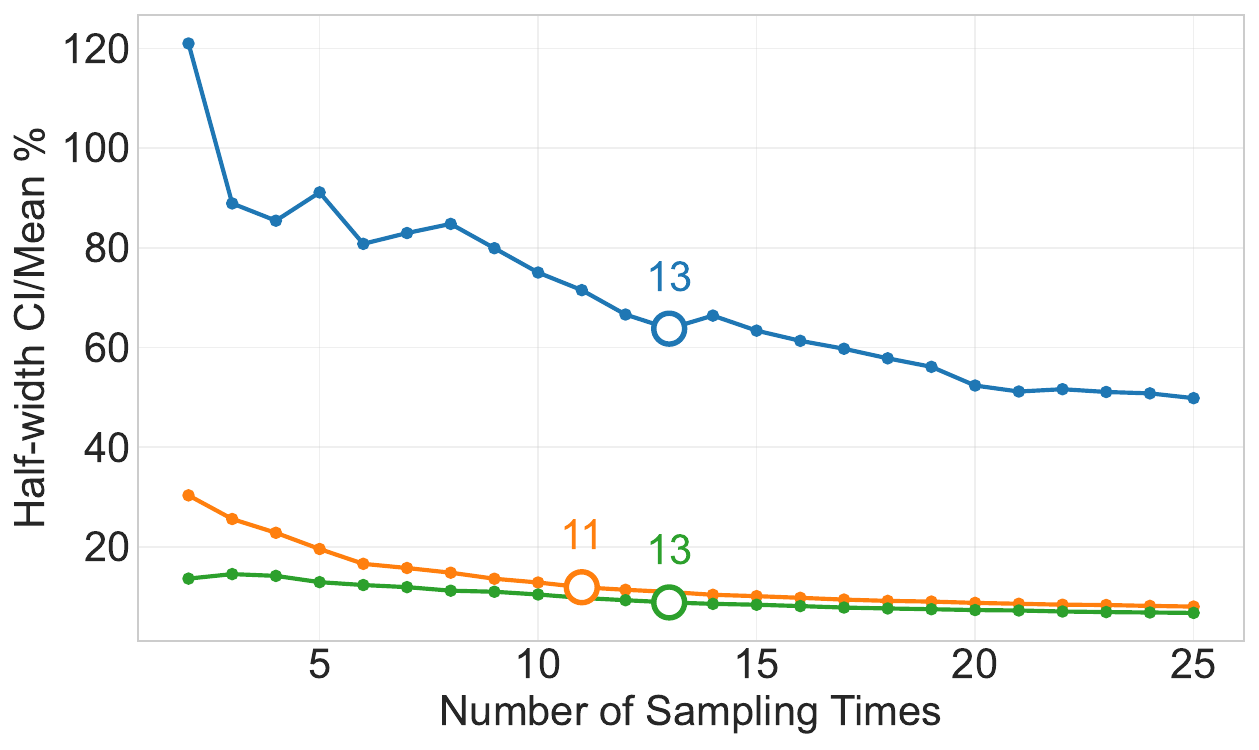}\hfill
  \includegraphics[width=0.24\linewidth]{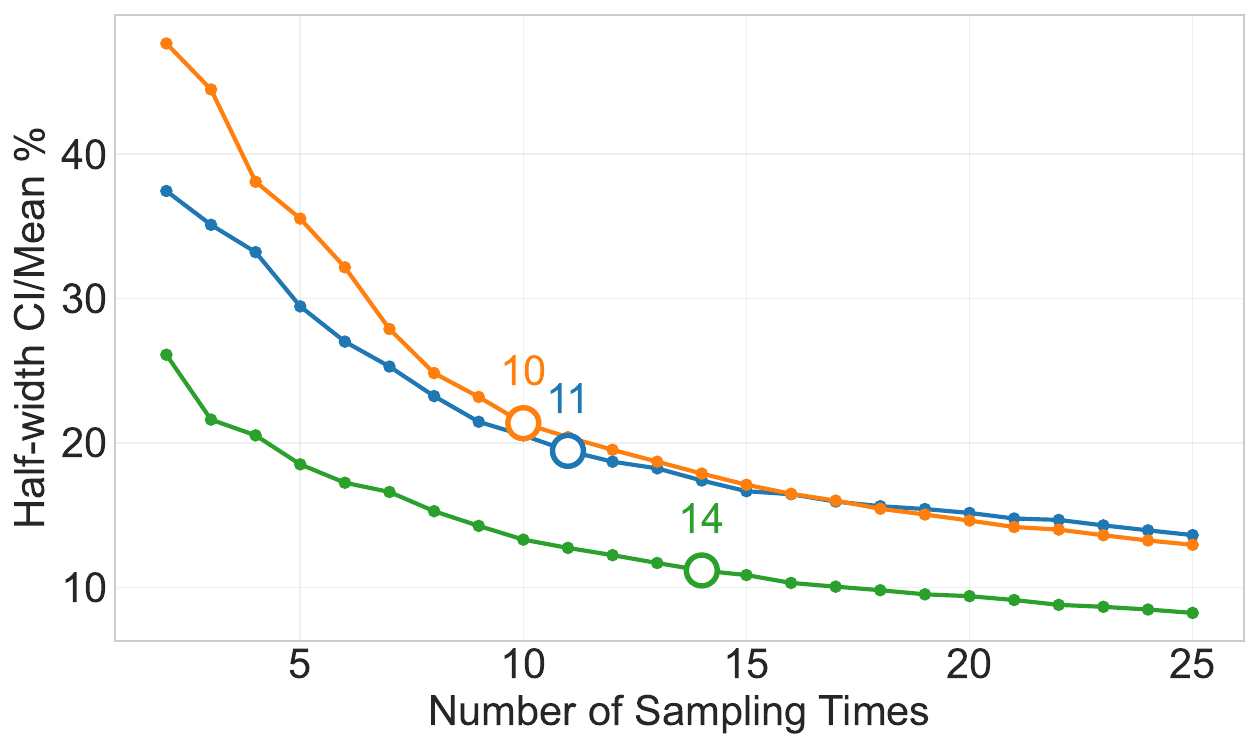}\hfill
  \includegraphics[width=0.24\linewidth]{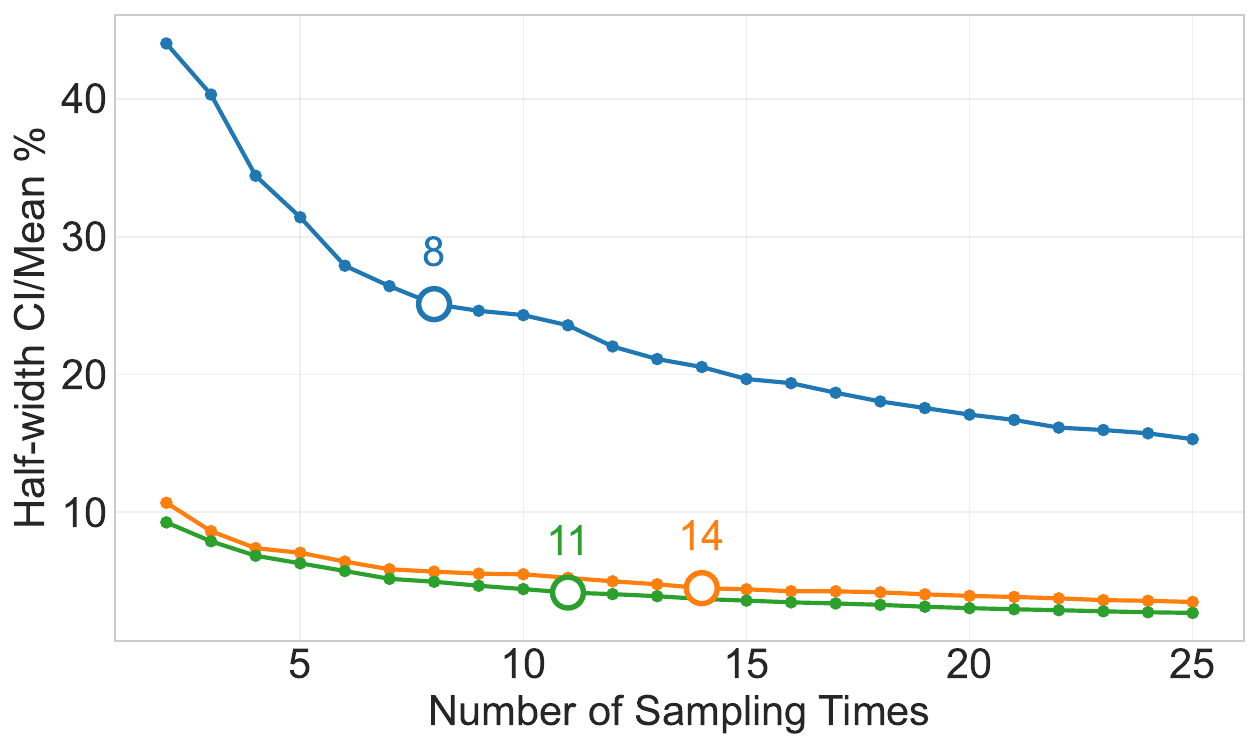}\\[-2pt]
  {\small (a) Q7 (AVG, $p{=}1.4\%$) \hfill
         (b) Q13 (AVG, $p{=}1.0\%$) \hfill
         (c) Q50 (COUNT, $p{=}26.3\%$) \hfill
         (d) Q52 (SUM, $p{=}1.7\%$)}\\[6pt]
  \includegraphics[width=0.24\linewidth]{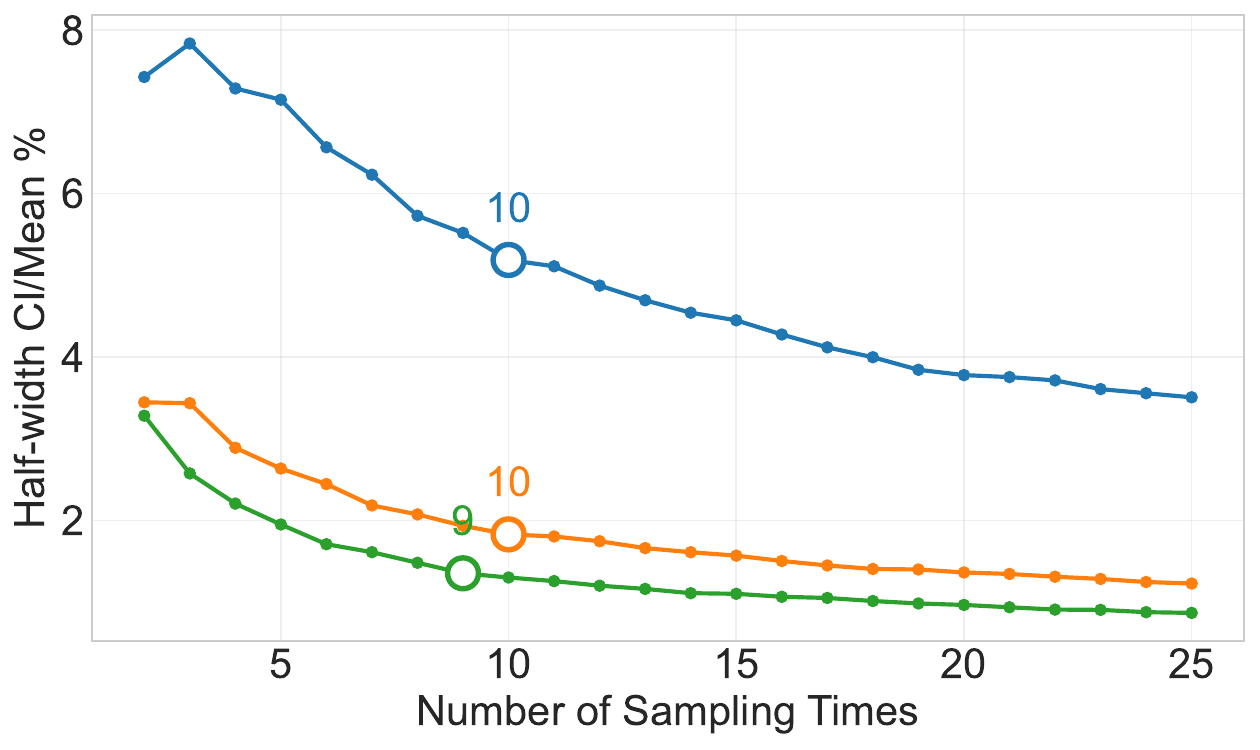}\hfill
  \includegraphics[width=0.24\linewidth]{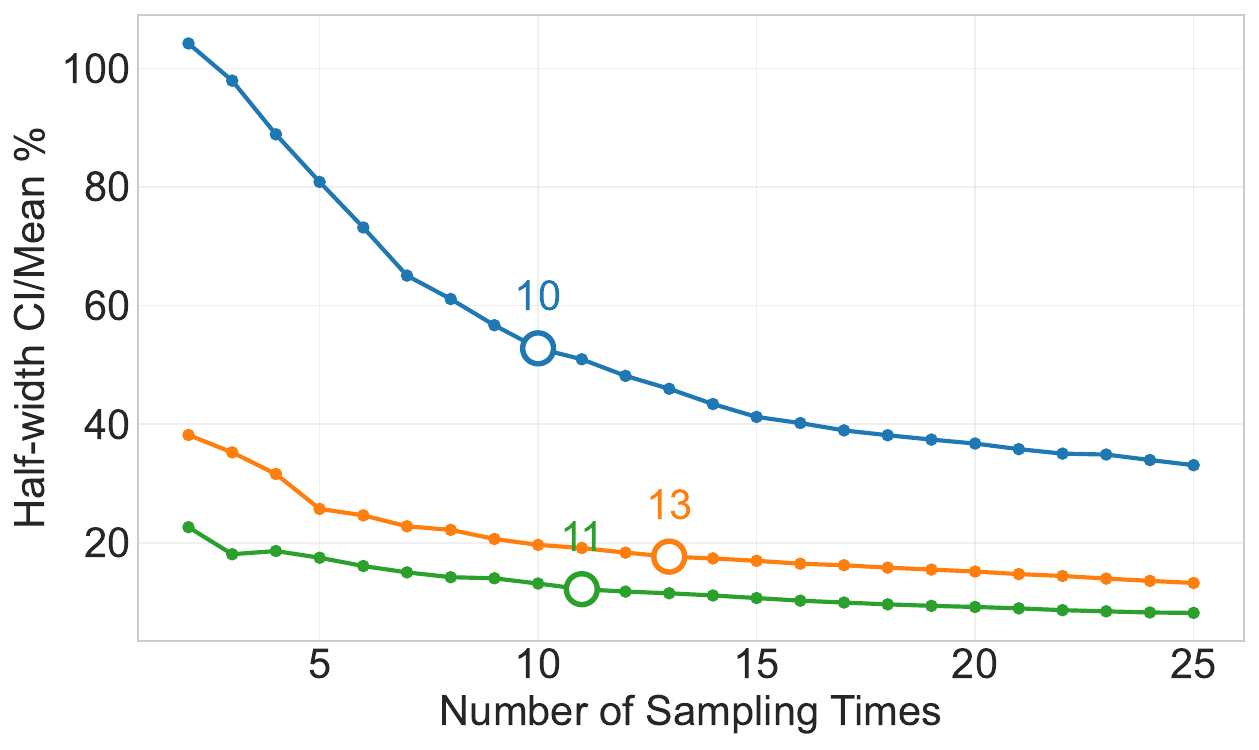}\hfill
  \includegraphics[width=0.24\linewidth]{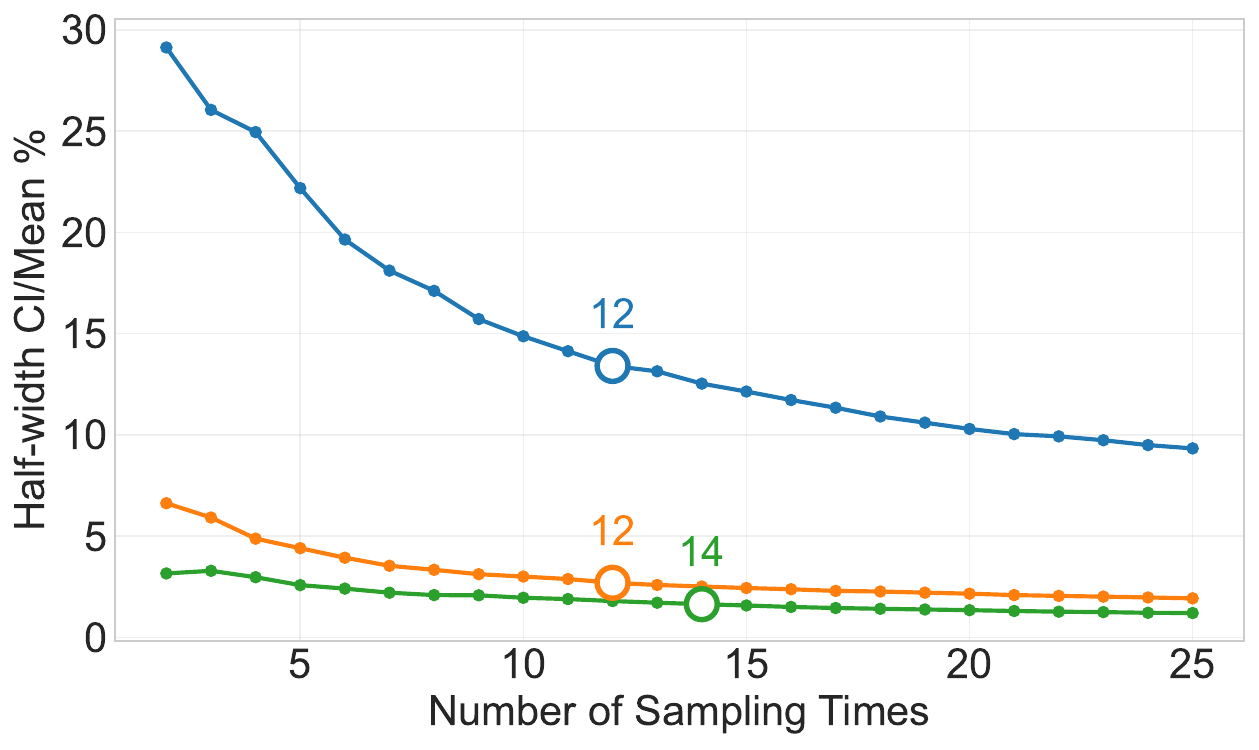}\hfill
  \includegraphics[width=0.24\linewidth]{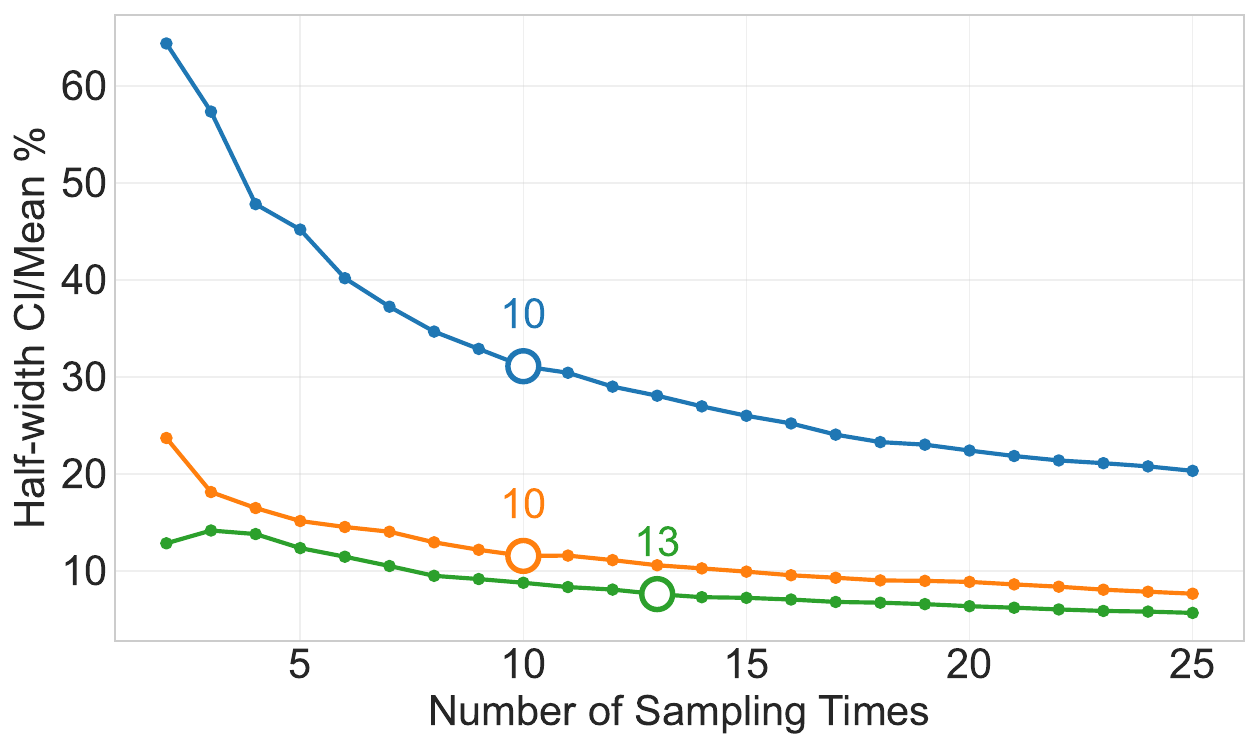}\\[-2pt]
  {\small (e) Q53 (SUM, $p{=}4.1\%$) \hfill
         (f) Q54 (COUNT, $p{=}6.1\%$) \hfill
         (g) Q55 (COUNT, $p{=}1.7\%$) \hfill
         (h) Q56 (SUM, $p{=}4.5\%$)}
  \caption{
  {Strategy AQP: relative half-width of the approximate 95\% interval
(half-width divided by the current sample mean) vs.\ number of sampling
rounds, for all eight queries. $p$ denotes the natural positive-label rate.
Circles mark the offline stabilization reference points. Under balanced
distributions the half-width is below 10\% at these points; under natural
labels, Q13 and Q54 remain above 10\%.}
    }
  \label{fig:tpcds_s1_req}
\end{figure*}

\smallskip\noindent\textbf{Results for Strategy~AQP.}
We present the results of Strategy AQP under the settings described above in Figures~\ref{fig:tpcds_s1_dist} and~\ref{fig:tpcds_s1_req}.
Figure~\ref{fig:tpcds_s1_dist} measures \textit{the real errors of the queries under different percentages of oracle calls} and Figure~\ref{fig:tpcds_s1_req} measures \textit{the estimated CI with increasing numbers of sampling rounds}.
For Figure~\ref{fig:tpcds_s1_dist}, the dots along the line represent the increasing number of sampling trials (e.g., the 4th dot means we have done four trials of sampling). Each trial {draws 1\% of the data uniformly with replacement} so that the total of oracle calls might be less than the total number of samples due to overlapping.
{Under the adjusted positive-label rates of 30\% and 50\%, aggregate error falls below 10\% with less than 10\% of oracle calls.}
{Under natural TPC-DS labels, several larger relations retain low error, while Q13 and Q54 (small tables with fewer than 10,000
rows) can remain above the target after evaluating more than 20\% of their rows. 
}
{In the batch-size ablation (Table~\ref{tab:sample_size}), 1\% offers a near-optimal aggregate-error and oracle-call trade-off among the tested settings at a moderate oracle budget. Smaller batches generally use fewer oracle calls but have higher error; larger batches can reduce error at substantially higher cost.}
We hypothesize that our sampling strategy obeys the Law of Large Numbers and it might be beneficial for small data to just go through the original AI workflows.

Figure~\ref{fig:tpcds_s1_req} {shows the original stabilization reference points at around 10 to 13 rounds. The markers describe the plotted interval trends and correspond to roughly 85\% to 90\% fewer oracle calls.}
{Under balanced distributions, the approximate interval has narrowed below 10\% at those points. Under natural labels, the small-table cases retain wide intervals. 
}

\begin{figure*}[!t]
  \centering
  \includegraphics[width=0.24\linewidth]{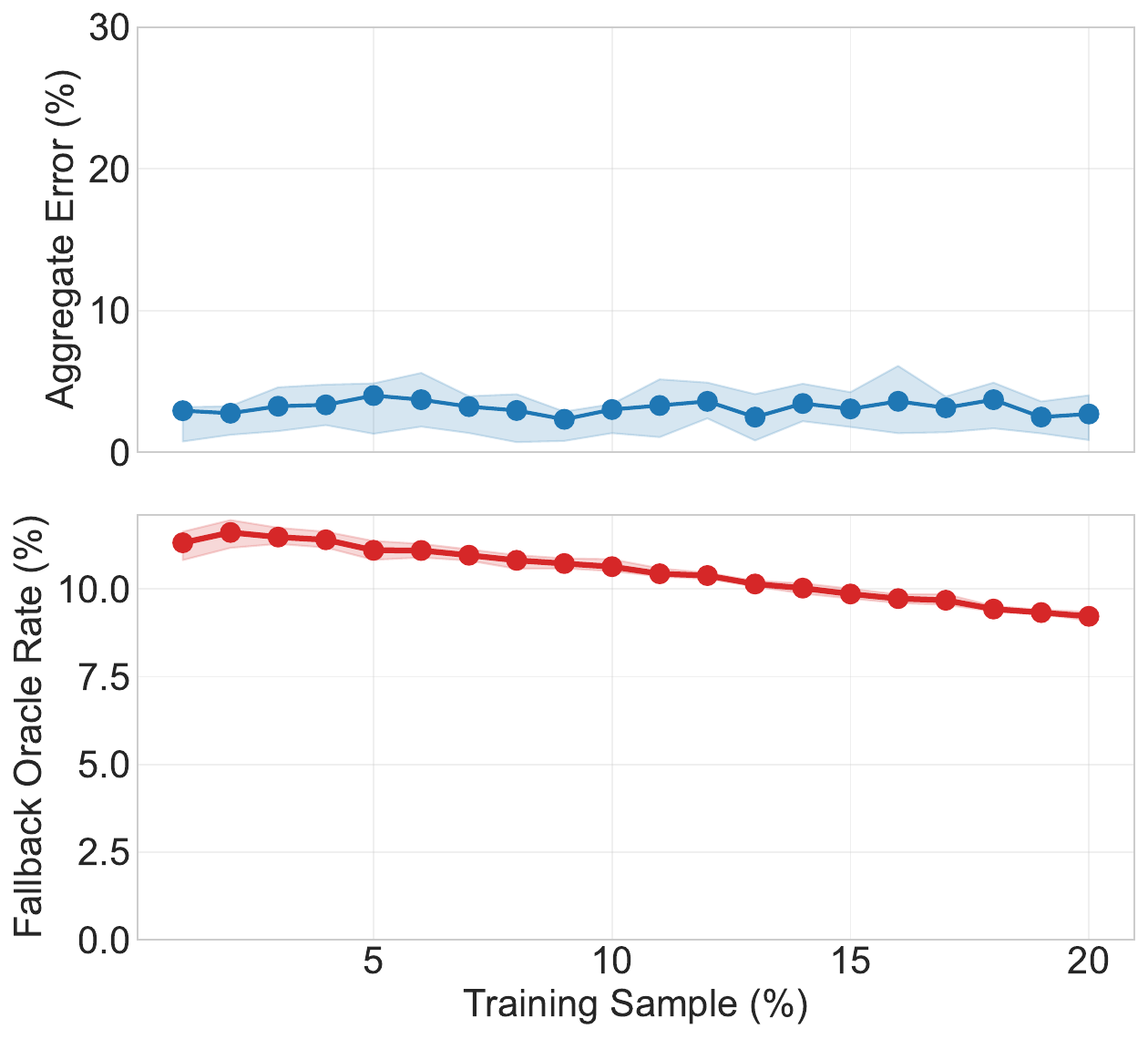}\hfill
  \includegraphics[width=0.24\linewidth]{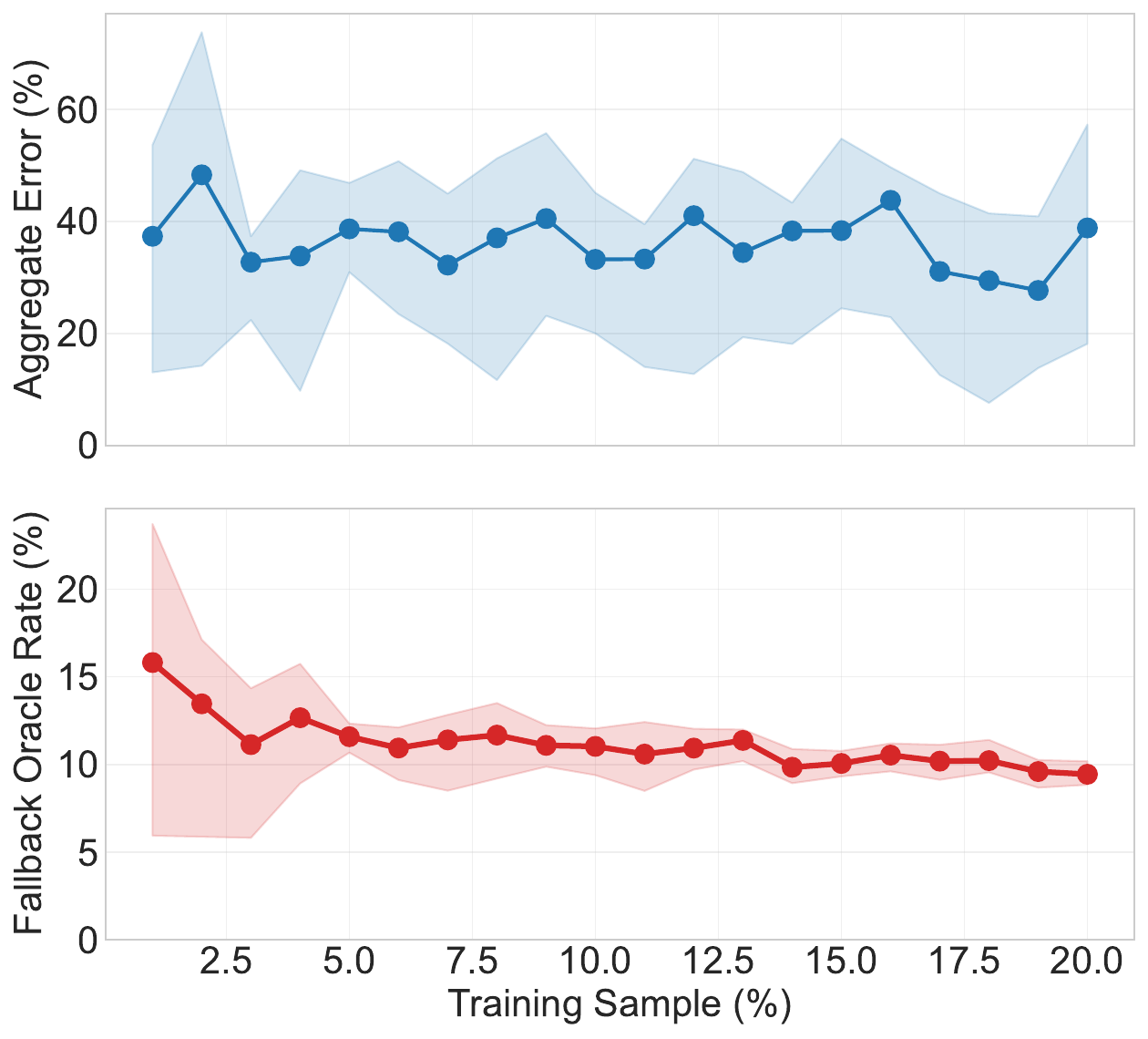}\hfill
  \includegraphics[width=0.24\linewidth]{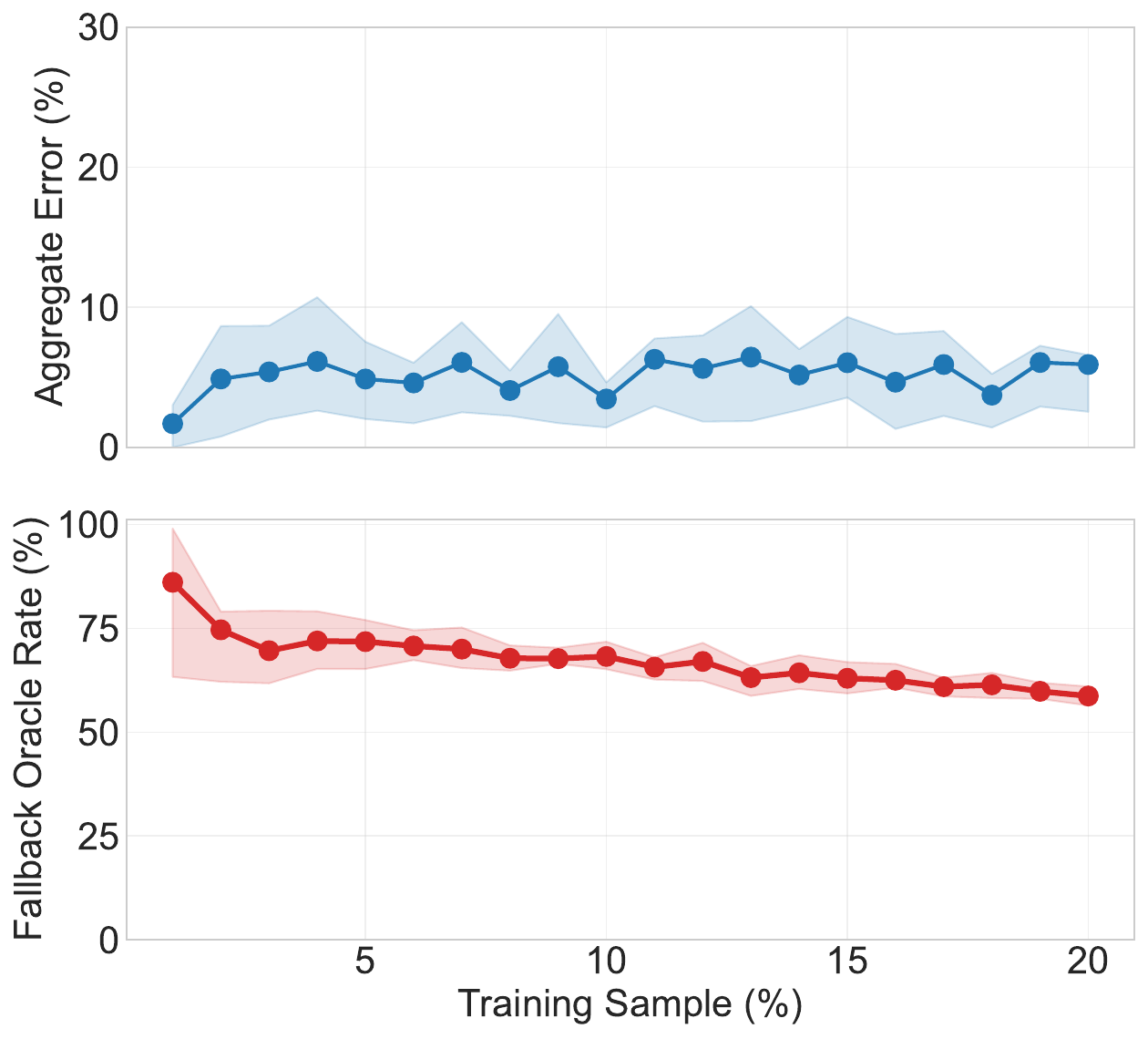}\hfill
  \includegraphics[width=0.24\linewidth]{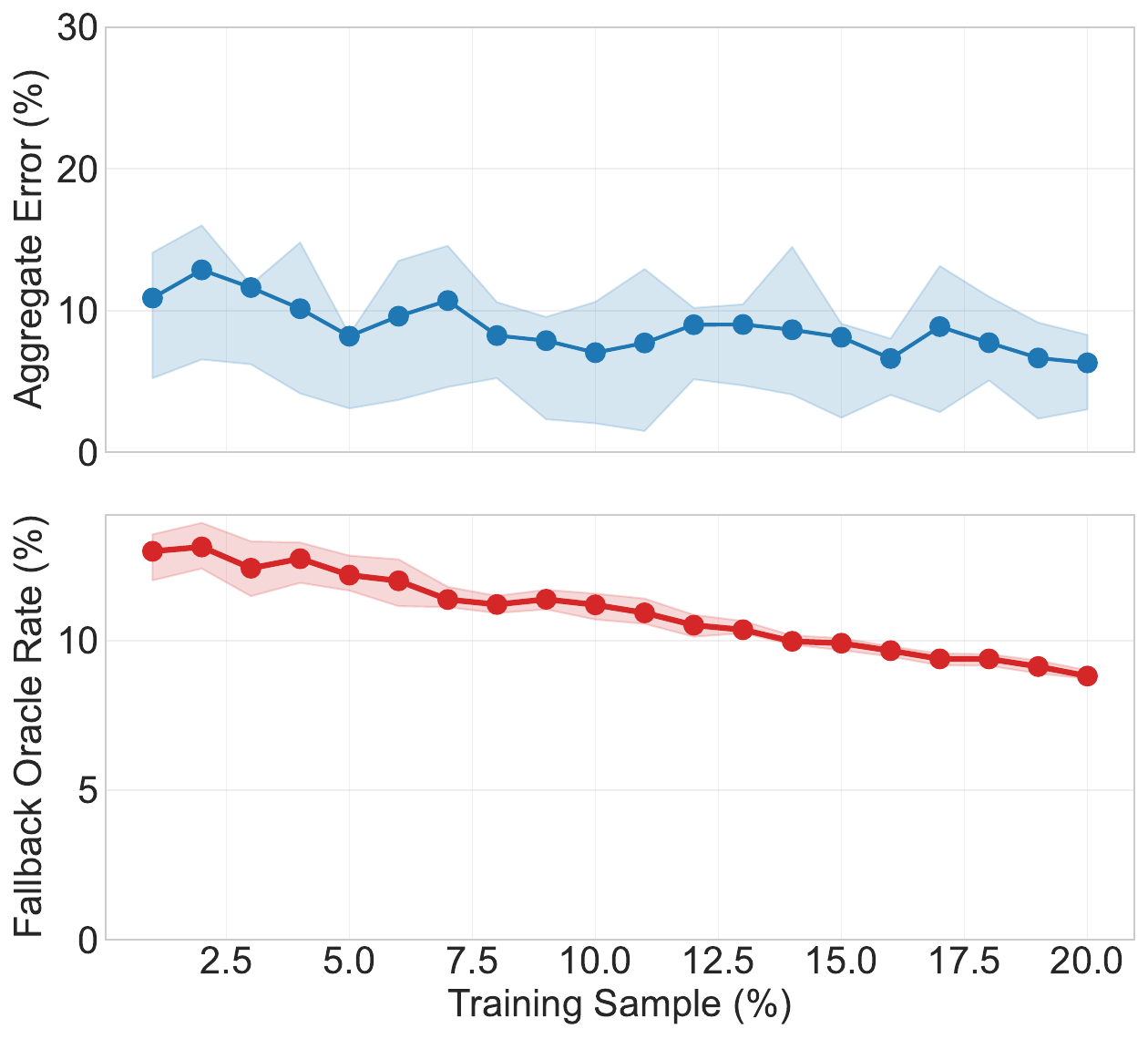}\\[-2pt]
  {\small (a) Q7 (AVG, $p{=}1.4\%$) \hfill
         (b) Q13 (AVG, $p{=}1.0\%$) \hfill
         (c) Q50 (COUNT, $p{=}26.3\%$) \hfill
         (d) Q52 (SUM, $p{=}1.7\%$)}\\[6pt]
  \includegraphics[width=0.24\linewidth]{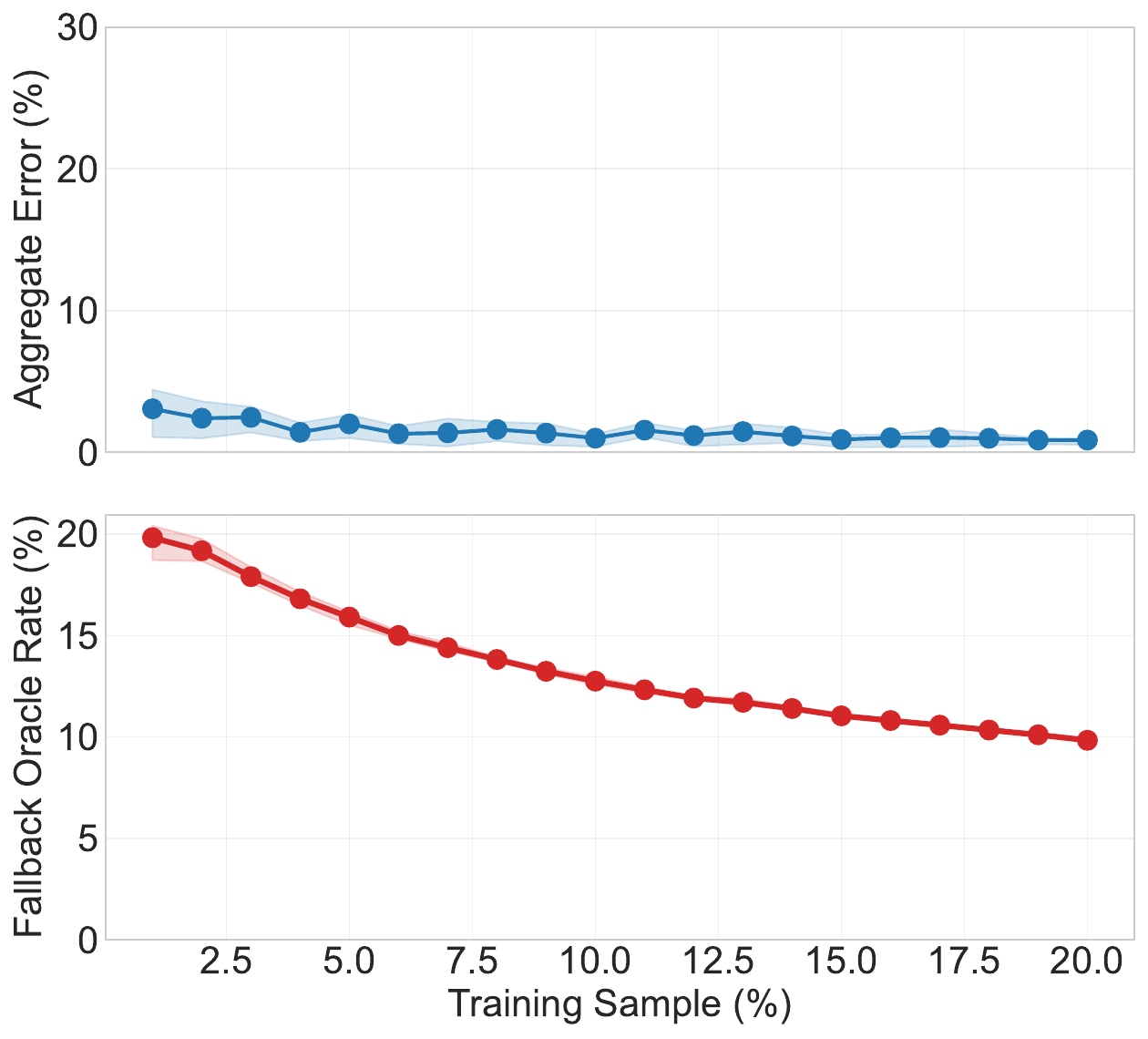}\hfill
  \includegraphics[width=0.24\linewidth]{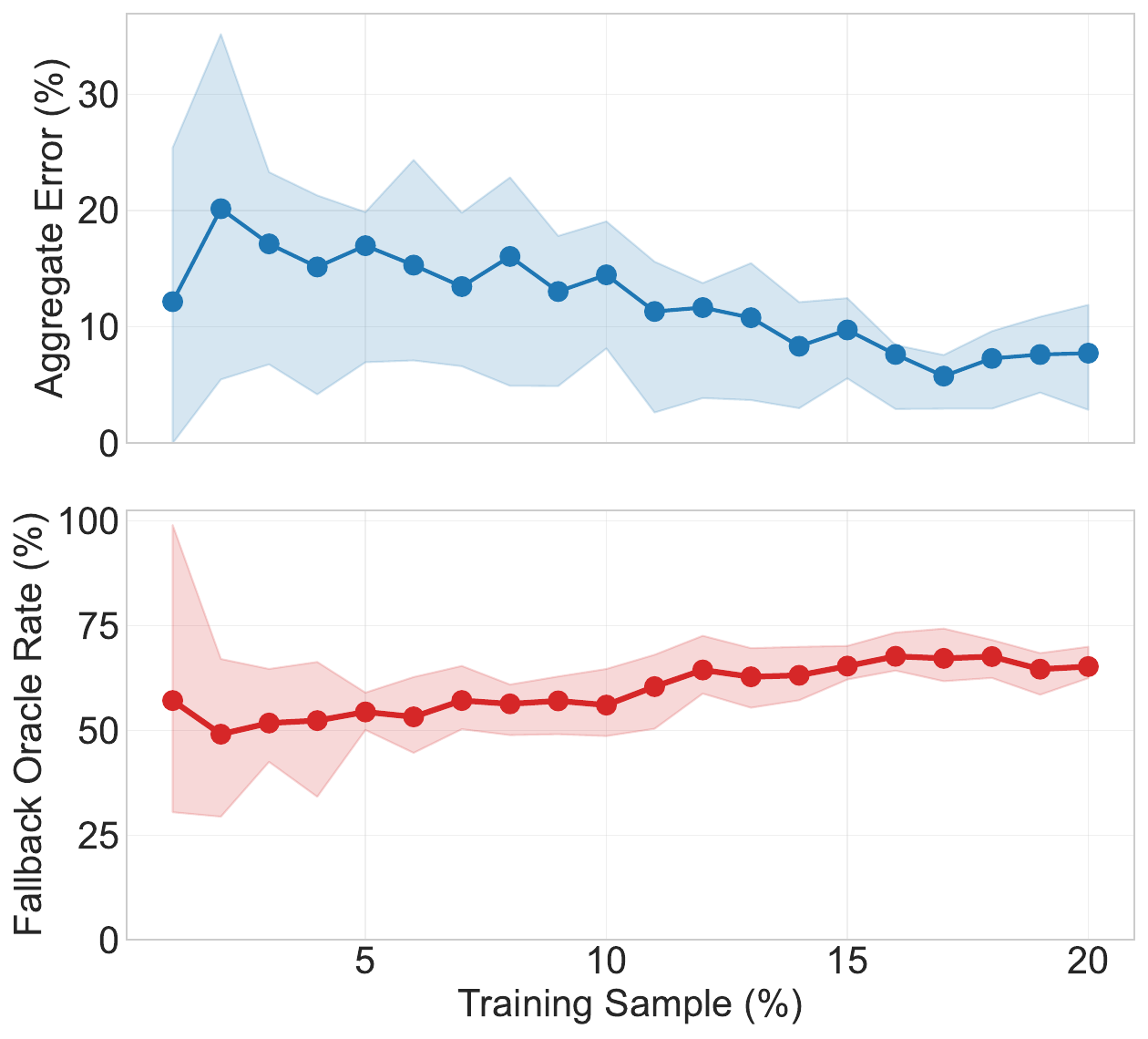}\hfill
  \includegraphics[width=0.24\linewidth]{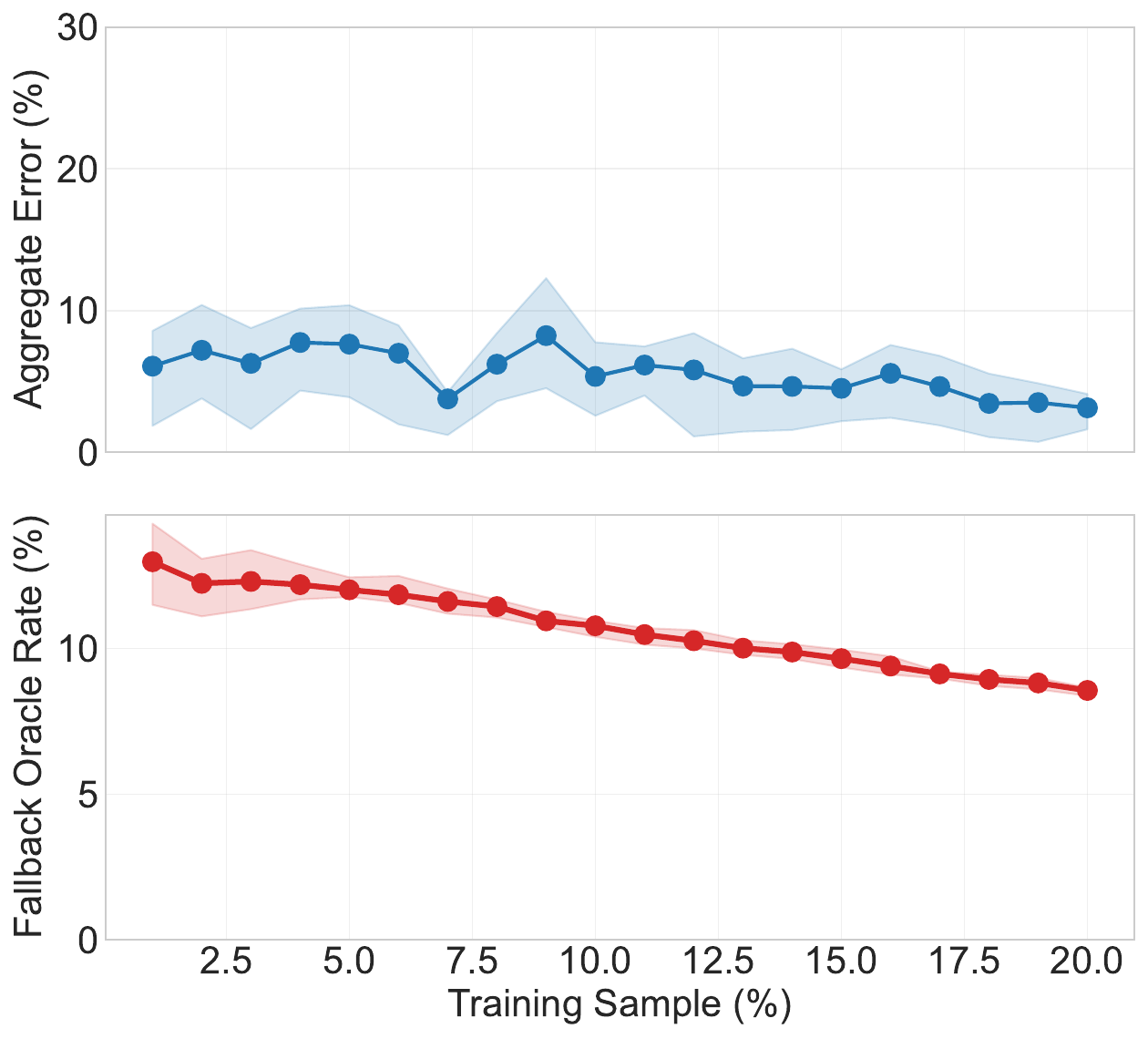}\hfill
  \includegraphics[width=0.24\linewidth]{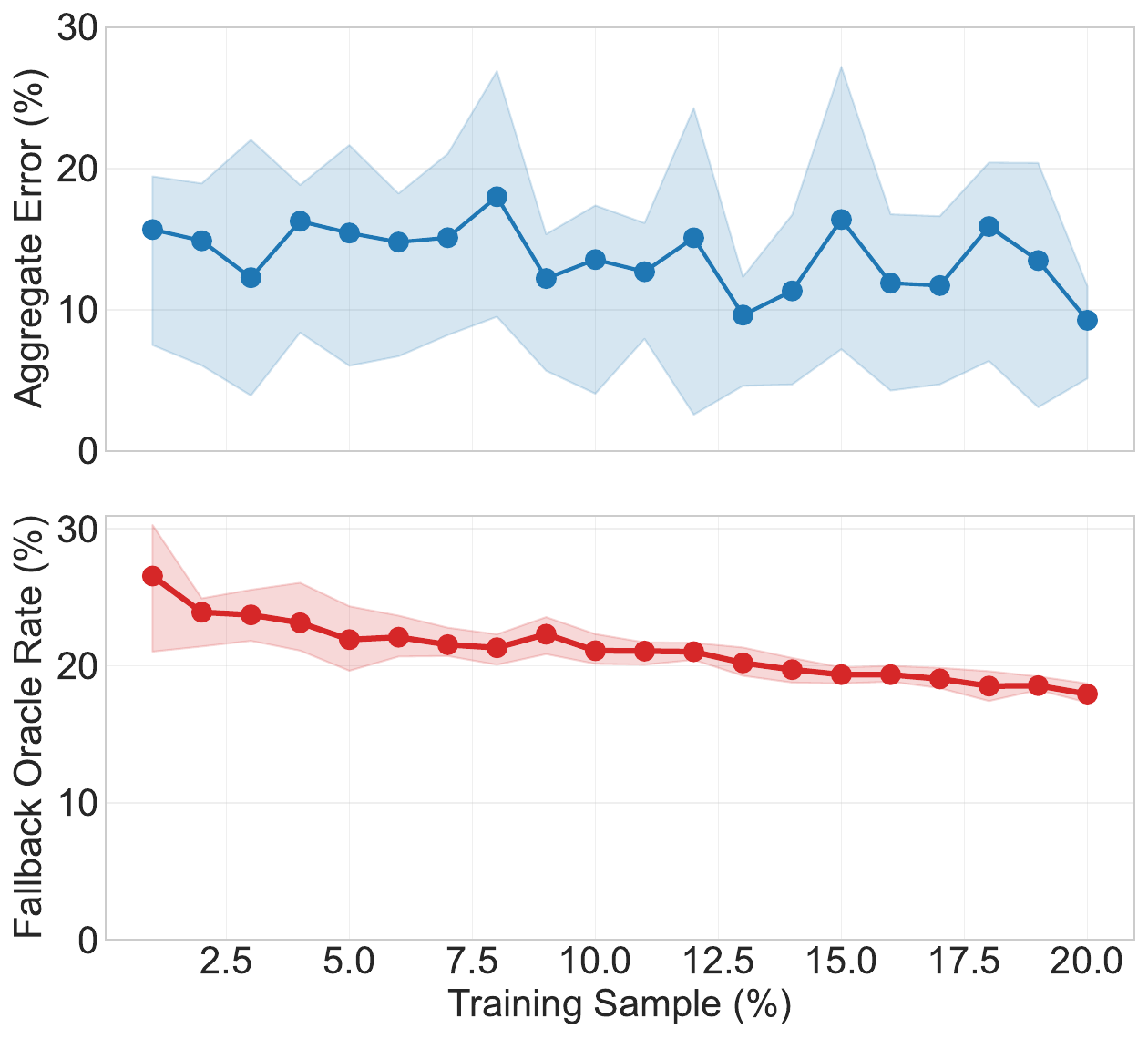}\\[-2pt]
  {\small (e) Q53 (SUM, $p{=}4.1\%$) \hfill
         (f) Q54 (COUNT, $p{=}6.1\%$) \hfill
         (g) Q55 (COUNT, $p{=}1.7\%$) \hfill
         (h) Q56 (SUM, $p{=}4.5\%$)}
\caption{
    \textbf{Strategy~PM aggregate error vs.\ training sample percentage}
    under the natural label distribution.
    {Each configuration is evaluated multiple times, and shaded regions show the 25th to 75th percentiles.}
    {The plotted fallback rate counts confirmation and audit calls,
    excluding training. It does not indicate mandatory full evaluation.}}
  \label{fig:tpcds_s2}
\end{figure*}

\smallskip\noindent\textbf{Results for Strategy~PM.}
For Strategy~PM, we use the natural TPC-DS labels rather than randomized labels, because randomized labels would destroy learnable structure for the decision tree. {Figure~\ref{fig:tpcds_s2} retains the original aggregate errors and additional oracle-call rates. The latter count confirmation and audit after training, rather than a separate full-evaluation policy. Training calls are added when reporting total oracle cost.} {Its original audit settings differ from the final comparison above.}
Overall, 5 out of 8 queries (Q7, 52, 53, 55, 56) performed as expected, in the sense that the errors reached or stayed at a low margin ($<10\%$) as the training samples increased to at most 20\%, with additional oracle calls decreasing along the way. In total, the oracle calls are around 30--40\%, implying a 60--70\% reduction in oracle calls to the expensive AI pipeline {for these configurations.}
{Q50 and Q54 require many oracle confirmations because the tree confidently excludes fewer rows. On Q13, aggregate error remains high even when PM excludes more rows. The results identify where proxy filtering saves oracle work and where retaining the AQP estimate is valuable.}
{The training-size ablation under the final comparison settings (Table~\ref{tab:sample_size}) shows that 5\% training for most queries (10\% for Q54) gives a near-optimal aggregate-error and oracle-call trade-off when oracle cost is prioritized. Larger training samples can lower error on some queries, but usually increase total calls and do not improve error consistently.}

\subsubsection{Comparison with BARGAIN}\phantomsection\label{sec:bargain}
Table~\ref{tab:bargain} compares AQP at its stopping point, PM, and BARGAIN AT~\cite{rev_bargain} on natural labels. Query-result accuracy exposes savings that row-accuracy targets alone do not capture. On Q7, PM gives 2.04\% aggregate error at 19.75\% oracle cost, compared with 4.12\% at 18.07\% for AT-90. On Q53, AQP gives 1.33\% error at 22.12\% cost, while AT-95 gives 21.61\% error at 37.55\% cost. The differences show why the aggregate objective belongs in the execution decision. AQP here uses the online plateau test in Equation~\ref{eq:plateau}, or the 25-round cap. The stabilization markers in Figure~\ref{fig:tpcds_s1_req} are separate offline reference points.

AT-99 further reduces aggregate error but uses 75.89\% to 100\% of oracle rows. For Q7, it gives 0.20\% error at 87.42\% cost. Table~\ref{tab:bargain_at99} reports the complete AT-99 results. On Q55, AT-99 still has 12.43\% error at 83.93\% calls, compared with AQP's 4.12\% error at 12.94\% calls. Higher row agreement does not ensure a fixed aggregate-error bound, and the methods offer different cost-quality tradeoffs.

\begin{table}[!tp]

\centering\small
\caption{Oracle calls (\% of rows) / aggregate error (\%) on natural TPC-DS labels. Lower is better for both; bold marks the fewest in each row. 
}
\label{tab:bargain}
\setlength{\tabcolsep}{2pt}
\begin{tabular}{@{}lrrrr@{}}
\toprule
Query & AQP (stop) & PM & AT-90 & AT-95 \\
\midrule
Q7  & 21.35 / 2.16 & 19.75 / 2.04 & \textbf{18.07} / 4.12 & 42.54 / \textbf{0.90} \\
Q13 & 12.35 / 30.74 & 20.15 / 26.45 & \textbf{12.22} / 35.15 & 28.39 / \textbf{23.48} \\
Q50 & \textbf{13.67} / 8.47 & 77.05 / \textbf{5.23} & 81.55 / 15.00 & 93.04 / 6.49 \\
Q52 & \textbf{14.60} / \textbf{10.88} & 20.64 / 11.37 & 16.24 / 69.75 & 40.71 / 46.61 \\
Q53 & 22.12 / \textbf{1.33} & 24.23 / 1.54 & \textbf{17.36} / 28.44 & 37.55 / 21.61 \\
Q54 & \textbf{15.42} / 19.55 & 72.77 / 22.36 & 71.60 / 32.11 & 75.97 / \textbf{15.37} \\
Q55 & \textbf{12.94} / \textbf{4.12} & 20.49 / 5.84 & 13.14 / 60.04 & 42.60 / 46.27 \\
Q56 & \textbf{12.92} / 13.65 & 29.89 / \textbf{11.27} & 32.40 / 60.93 & 57.58 / 41.76 \\
\bottomrule
\end{tabular}

\end{table}

\begin{table}[!tp]

\centering\small
\caption{BARGAIN AT-99 oracle calls and aggregate error.}
\label{tab:bargain_at99}
\begin{tabular}{lrr}
\toprule
Query & \shortstack{Oracle calls\\(\%)} & \shortstack{Aggregate error\\(\%)} \\
\midrule
Q7 & 87.42 & 0.20 \\
Q13 & 75.89 & 8.39 \\
Q50 & 100.00 & 0.00 \\
Q52 & 85.36 & 10.94 \\
Q53 & 86.00 & 5.46 \\
Q54 & 94.61 & 3.96 \\
Q55 & 83.93 & 12.43 \\
Q56 & 94.94 & 5.07 \\
\bottomrule
\end{tabular}

\end{table}

\subsubsection{Sample-Size Sensitivity}
\label{sec:sample_size}
Table~\ref{tab:sample_size} reports aggregate error together with distinct
oracle calls as the sampling budget changes. AQP draws 0.25\%--5\% of rows
per round and applies the same plateau rule at each batch size. PM trains
on 5\%--20\% of rows using the final confirmation and audit settings.
Unlike the original PM curves in Figure~\ref{fig:tpcds_s2}, these PM call
totals include training, confirmation, and audit under the settings used
for the BARGAIN comparison.

\myparagraph{AQP batch size}
Smaller batches generally save calls but yield less accurate aggregates.
On Q55, 0.5\% batches give 8.53\% error at 6.44\% calls, whereas 1\%
batches give 4.12\% error at 12.94\% calls. Q7 offers a useful lower-cost
option: 0.5\% batches give 3.39\% error at 8.25\% calls, compared with
2.16\% error at 21.35\% calls for 1\% batches. Increasing the batch size
can help difficult queries, at higher cost. On Q54, increasing it from
1\% to 5\% lowers error from 19.55\% to 8.02\% while calls rise from
15.42\% to 45.14\%. Q13 still has 16.03\% error at 47.94\% calls with
5\% batches. Across the eight queries, the 1\% setting uses
12.35\%--22.12\% of oracle rows and provides a moderate-budget tradeoff;
no batch size is optimal for every query. A fixed-draw control on Q52
holds sampled rows and oracle cost constant and gives 8.73\%--8.74\%
error across batch sizes. Most stopping-point differences therefore
reflect the number of labels consumed by the round-based rule rather
than how the same draws are grouped.

\myparagraph{PM training size}
The default training budget favors oracle savings rather than minimum
error. On Q7, increasing training from 5\% to 20\% changes error from
2.04\% to 1.77\% while total calls rise from 19.75\% to 32.21\%.
Q52 gains more accuracy, from 11.37\% to 5.35\% error, but calls rise
from 20.64\% to 31.89\%. For Q54, increasing the default 10\% training
budget to 15\% reduces error from 22.36\% to 11.60\% while calls rise
from 72.77\% to 80.28\%. At 20\% training its error rises again to
16.62\%. More training thus does not consistently improve the final
aggregate. Decision-tree fitting takes 0.0006--2.0556 seconds in these
natural-label settings, excluding oracle latency; this CPU cost is
separate from the table's oracle-call totals.
\begin{table*}[!tp]

\centering\small
\caption{Sample-size sensitivity on natural-label TPC-DS queries. Cells give mean aggregate error / mean oracle calls (\%).}
\label{tab:sample_size}
\setlength{\tabcolsep}{5pt}
\begin{tabular}{lrrrrr}
\toprule
\multicolumn{6}{l}{\textbf{AQP: fraction of rows drawn per round}} \\
Query & 0.25\% & 0.5\% & 1\% & 2\% & 5\% \\
\midrule
Q7  & 5.96 / 3.20 & 3.39 / 8.25 & 2.16 / 21.35 & 1.22 / 39.35 & 0.87 / 71.34 \\
Q13 & 35.68 / 4.39 & 30.39 / 7.09 & 30.74 / 12.35 & 17.16 / 21.48 & 16.03 / 47.94 \\
Q50 & 16.33 / 2.73 & 9.17 / 5.82 & 8.47 / 13.67 & 3.98 / 25.53 & 2.98 / 57.28 \\
Q52 & 21.91 / 3.12 & 12.18 / 5.68 & 10.88 / 14.60 & 7.01 / 25.57 & 5.53 / 53.08 \\
Q53 & 3.74 / 4.89 & 1.93 / 10.51 & 1.33 / 22.12 & 0.97 / 39.35 & 0.75 / 71.37 \\
Q54 & 43.83 / 4.20 & 26.07 / 7.31 & 19.55 / 15.42 & 13.89 / 28.02 & 8.02 / 45.14 \\
Q55 & 10.20 / 3.11 & 8.53 / 6.44 & 4.12 / 12.94 & 3.70 / 31.36 & 1.31 / 67.45 \\
Q56 & 30.12 / 3.49 & 18.95 / 5.92 & 13.65 / 12.92 & 10.93 / 19.29 & 4.64 / 45.06 \\
\bottomrule
\end{tabular}
\par\smallskip
\begin{tabular}{lrrrr}
\toprule
\multicolumn{5}{l}{\textbf{PM: fraction of rows used for training}} \\
Query & 5\% & 10\% & 15\% & 20\% \\
\midrule
Q7  & 2.04 / 19.75 & 2.69 / 24.01 & 2.38 / 28.05 & 1.77 / 32.21 \\
Q13 & 26.45 / 20.15 & 29.05 / 24.23 & 21.42 / 28.34 & 24.76 / 32.56 \\
Q50 & 5.23 / 77.05 & 7.06 / 75.20 & 6.16 / 77.99 & 3.55 / 79.74 \\
Q52 & 11.37 / 20.64 & 7.43 / 24.74 & 7.05 / 28.10 & 5.35 / 31.89 \\
Q53 & 1.54 / 24.23 & 0.91 / 26.03 & 0.65 / 29.22 & 0.80 / 32.84 \\
Q54 & --- & 22.36 / 72.77 & 11.60 / 80.28 & 16.62 / 82.27 \\
Q55 & 5.84 / 20.49 & 5.08 / 24.03 & 3.52 / 27.77 & 3.67 / 31.50 \\
Q56 & 11.27 / 29.89 & 12.40 / 34.56 & 10.01 / 37.55 & 10.40 / 40.38 \\
\bottomrule
\end{tabular}
\par\smallskip
{\footnotesize AQP uses the same plateau rule for every batch size and stops at its own first plateau or at 25 rounds. Calls count distinct oracle rows. PM calls include training, confirmation, and audit; Q54 was not tested at 5\% training.}

\end{table*}

\subsubsection{Shared Labels and Output Order}
\phantomsection\label{sec:shared_sampling_results}
We compare shared and separate execution of AQP and PM on the eight
natural-label queries. In the shared setting, PM trains on distinct rows from
five 1\% AQP rounds (ten for Q54). The nominal 5\%/10\% budgets correspond to
Table~\ref{tab:bargain}, but repeated draws give slightly fewer training rows. PM uses the same
thresholds and audit fractions. Before any further labels
are collected, we compare the uncached rows needed by PM and by a 25-round AQP run (Equation~\ref{eq:remaining_oracle_cost}) and finish the smaller one first.

Table~\ref{tab:shared_sampling} reports both completion orders and reuse savings for every query. PM completes first in 75 of 160 (8 queries $\times$ 20 trials) runs. On Q7 it gives 1.98\% error at 19.65\% oracle cost. Continuing to the AQP checkpoint gives 1.66\% error at 34.22\% combined cost, saving 7.55 percentage points over separate execution. Q50 completes AQP first at 22.08\% cost and 4.21\% error. Every replay preserves both estimates under either completion order. Q13 remains difficult, with 31.25\% PM error and 18.41\% AQP error. Overall, we show that label reuse reduces total oracle work while allowing the caller to use the first completed result without paying for both plans.

Repeated draws are removed from the PM training set; no extra rows are
sampled to fill the nominal budget. The mean training fractions are
4.853\%--4.893\% on the seven queries using five initial rounds and
9.493\% on Q54. Table~\ref{tab:bargain} instead uses the specified
training size sampled without replacement. Shared and separate execution
use the same trained PM and AQP requests, so the comparison isolates
label reuse. Savings range from 7.55 to 17.94 percentage points of full
oracle evaluation when both plans complete.
\begin{table*}[!tp]

\centering\small
\caption{First completed result and oracle-call savings from shared execution.}
\label{tab:shared_sampling}
\setlength{\tabcolsep}{4pt}
\begin{tabular}{@{}lrrrrrr@{}}
\toprule
Query & \shortstack{PM\\calls / error (\%)} & \shortstack{AQP, 25 rounds\\calls / error (\%)} & \shortstack{PM first\\runs (\%)} & \shortstack{First result\\calls / error (\%)} & \shortstack{Both plans\\calls (\%)} & \shortstack{Calls saved\\by reuse (pp)} \\
\midrule
Q7 & 19.65 / 1.98 & 22.12 / 1.66 & 100 & 19.65 / 1.98 & 34.22 & 7.55 \\
Q13 & 20.99 / 31.25 & 22.19 / 18.41 & 75 & 20.66 / 27.80 & 35.42 & 7.76 \\
Q50 & 76.79 / 5.48 & 22.08 / 4.21 & 0 & 22.08 / 4.21 & 80.93 & 17.94 \\
Q52 & 20.59 / 7.74 & 22.11 / 7.10 & 100 & 20.59 / 7.74 & 34.99 & 7.71 \\
Q53 & 24.28 / 1.38 & 22.12 / 2.05 & 0 & 22.12 / 2.05 & 38.00 & 8.40 \\
Q54 & 64.73 / 22.59 & 22.16 / 11.78 & 0 & 22.16 / 11.78 & 69.70 & 17.20 \\
Q55 & 20.64 / 6.29 & 22.12 / 3.88 & 100 & 20.64 / 6.29 & 35.02 & 7.74 \\
Q56 & 30.70 / 11.57 & 22.14 / 9.58 & 0 & 22.14 / 9.58 & 43.25 & 9.58 \\
\bottomrule
\end{tabular}

\par\smallskip
{\footnotesize Calls are percentages of full oracle evaluation; errors are relative aggregate errors.
Reuse savings are percentage points (pp), relative to completing the two plans separately.}

\end{table*}

\subsubsection{Stalled Intervals and Continuation Policies}
\label{sec:plateau_results}
Table~\ref{tab:plateau_cost} reports the plateau frequency, stopping round,
oracle calls, relative interval half-width, and aggregate error for all
eight queries. The user specifies the desired half-width and confidence
level, while AQP recomputes the estimate and approximate interval each
round. A plateau flags a width that remains too large; it does not certify
the accuracy of the returned result.

On Q13 and Q54, the interval half-width remains above the 10\% target.
Table~\ref{tab:plateau_policy} compares stopping, continuing to the
25-round cap, and evaluating all remaining rows after a plateau.
Stopping uses 12.35\% of rows on Q13 and 15.42\% on Q54 but leaves
30.74\% and 19.55\% error. Continuing to 25 rounds costs about 22\% of
rows and lowers the errors to 18.11\% and 14.32\%. The full-on-plateau
policy removes approximation error on Q13 at 100\% cost. On Q54 it
averages 3.79\% error at 76.62\% cost because only 70\% of runs trigger
a plateau; the other runs return at the 25-round cap. Every fully
evaluated run has zero approximation error. These call fractions measure
expensive model work; wall-clock throughput also depends on oracle latency.
\begin{table*}[!tp]

\centering\small
\caption{AQP cost and error at the plateau or sampling cap.}
\label{tab:plateau_cost}
\begin{tabular}{lrrrrr}
\toprule
Query & \shortstack{Plateau detected\\(\% of runs)} & \shortstack{Stopping\\round} & \shortstack{Oracle calls\\(\%)} & \shortstack{Relative CI\\half-width (\%)} & \shortstack{Aggregate error\\(\%)} \\
\midrule
Q7 & 5 & 24.1 & 21.35 & 4.63 & 2.16 \\
Q13 & 100 & 13.2 & 12.35 & 63.27 & 30.74 \\
Q50 & 85 & 14.9 & 13.67 & 19.25 & 8.47 \\
Q52 & 70 & 16.1 & 14.60 & 21.79 & 10.88 \\
Q53 & 0 & 25.0 & 22.12 & 3.47 & 1.33 \\
Q54 & 70 & 17.0 & 15.42 & 39.57 & 19.55 \\
Q55 & 70 & 14.2 & 12.94 & 13.83 & 4.12 \\
Q56 & 75 & 14.1 & 12.92 & 26.98 & 13.65 \\
\bottomrule
\end{tabular}

\end{table*}
\begin{table}[!tp]

\centering\small
\caption{Continuing, stopping, or fully evaluating a stalled query.}
\label{tab:plateau_policy}
\begin{tabular}{llrr}
\toprule
Query & Policy & \shortstack{Oracle\\calls (\%)} & \shortstack{Aggregate\\error (\%)} \\
\midrule
\multirow{3}{*}{Q13} & Fixed 25 rounds & 22.17 & 18.11 \\
 & Stop at plateau & 12.35 & 30.74 \\
 & Full on plateau & 100.00 & 0.00 \\
\midrule
\multirow{3}{*}{Q54} & Fixed 25 rounds & 22.09 & 14.32 \\
 & Stop at plateau & 15.42 & 19.55 \\
 & Full on plateau & 76.62 & 3.79 \\
\bottomrule
\end{tabular}

\end{table}

\subsubsection{Alternative Predicate Classifiers}
\label{sec:tpc_predictors}
Table~\ref{tab:tpc_predictors} compares the decision tree (DT) with
Logistic Regression (LR) and XGBoost for binary TPC-DS predicate labels.
Training rows, features, confirmation thresholds, and audit rules remain
fixed. LR uses standardization fitted only on the training rows.
The table uses 5\% training, except 10\% on Q54.

Across all 31 tested query/training-budget settings, LR lowers mean
aggregate error but increases oracle calls in every setting; 27 settings
evaluate at least 97\% of oracle rows. On Q7 at the default budget, LR
reduces error from DT's 2.04\% to 0.24\% while calls rise from 19.75\%
to 99.38\%. Under the unchanged confirmation threshold, LR confirms
another 94.31\% of rows after training, compared with 5.83\% for DT.
Q13 is the exception to nearly full evaluation at the initial budgets,
but its error changes only from 26.45\% to 24.96\% as calls rise from
20.15\% to 62.41\%.

XGBoost also trades error against oracle work. At the initial budgets,
every reduction in aggregate error uses more calls, and every call saving
increases error. On Q7 it changes error from 2.04\% to 1.88\% while
calls rise from 19.75\% to 69.36\%. Neither alternative consistently
improves both aggregate error and oracle savings over DT under these
features and execution rules. The comparison supports DT as an inexpensive
default, without claiming it is the most accurate classifier.
\begin{table}[!tp]

\centering\small
\setlength{\tabcolsep}{3pt}
\caption{Decision tree versus Logistic Regression and XGBoost on TPC-DS.}
\label{tab:tpc_predictors}
\begin{tabular}{lrrr}
\toprule
Query & \shortstack{DT\\calls / error\\(\%)} & \shortstack{LR\\calls / error\\(\%)} & \shortstack{XGBoost\\calls / error\\(\%)} \\
\midrule
Q7 & 19.75 / 2.04 & 99.38 / 0.24 & 69.36 / 1.88 \\
Q13 & 20.15 / 26.45 & 62.41 / 24.96 & 16.09 / 34.13 \\
Q50 & 77.05 / 5.23 & 98.66 / 1.04 & 52.84 / 7.88 \\
Q52 & 20.64 / 11.37 & 98.88 / 1.22 & 21.97 / 8.04 \\
Q53 & 24.23 / 1.54 & 99.79 / 0.01 & 72.90 / 0.39 \\
Q54 & 72.77 / 22.36 & 98.17 / 0.59 & 70.99 / 25.27 \\
Q55 & 20.49 / 5.84 & 99.00 / 0.58 & 21.77 / 4.32 \\
Q56 & 29.89 / 11.27 & 99.09 / 0.91 & 23.72 / 12.26 \\
\bottomrule
\end{tabular}
\par\smallskip
{\footnotesize Training uses 5\% of rows, or 10\% for Q54.
Error is relative aggregate error. Values are means from experiments repeated multiple times with paired seeds.
LR uses the original feature encoding.}

\end{table}

\subsubsection{Proxy Diagnostics of Sampling Imbalance}\label{sec:proxy_diagnostic}
A proxy can reveal sampling imbalance in the contribution it predicts. Reserve evaluation rows $V$ from the uniform sample of relation $R$ and fit the proxy on disjoint rows $T$. For $U=R\setminus T$ and predicted contribution $g_i$, the observable difference is
\begin{equation}\label{eq:proxy_sample_check}
 D_g=\sum_{i\in U}g_i-\frac{|U|}{|V|}\sum_{i\in V}g_i.
\end{equation}
The comparison needs no additional oracle labels. Our diagnostic uses two folds, exchanging training and evaluation roles and weighting each difference by its evaluation-fold size. COUNT uses predicted predicate membership, and SUM multiplies it by the known aggregate value. For AVG, the count and sum differences give the change in their ratio relative to the pooled sample estimate. The relative change provides an auxiliary diagnostic, with undefined ratios reported separately.

Prediction with sampled residual correction is an established statistical idea~\cite{rev_crossprediction}. We use its sample-versus-population comparison to inspect the sample, while retaining the original AQP output. A constant or inaccurate proxy can miss the same important rows as AQP, so a small difference is not an accuracy certificate. The following replay measures the diagnostic separately from label reuse and output ordering.

\myparagraph{Proxy diagnostics at fixed sample cost}
A separate replay is repeated multiple times with fresh seeds at ten and twenty-five AQP rounds. The unique samples cost 9.48\% to 9.55\% and 21.99\% to 22.16\% of oracle rows, respectively. Two-fold proxy diagnostics require zero additional oracle labels. Training and evaluation use disjoint folds, and the proxy check does not affect the original AQP output.

For Equation~\ref{eq:proxy_sample_check}, we evaluate both the balanced DT score and the unweighted label frequency in the same fitted leaf. The latter avoids treating class-weighted scores as population frequencies, but does not guarantee calibration. Its relative aggregate discrepancy has pooled AUC 0.659 and 0.671 for detecting AQP errors above 10\% at the two budgets. The AUC compares with the pooled unique-sample estimate and can reflect differences between queries. For example, Q52 has AUC 0.28 and 0.52. The diagnostic therefore provides an additional view of sampling imbalance, not a reliable per-query accuracy test.

Applying the frequency-based difference as an aggregate correction improves mean error over pooled AQP in only 3 of 16 query/budget settings. Balanced-score correction is undefined in 30\% of Q13 runs at ten rounds and 10\% at twenty-five rounds. We therefore retain the original AQP estimator and use the proxy signal as a separate diagnostic.

\subsubsection{Sensitivity to Errors in the Supplied Oracle}
\label{sec:upstream_error}
Sampling approximates the supplied oracle's aggregate, which can differ
from an aggregate based on correct task labels. To separate these effects,
we flip 5\% of the recorded predicate labels and run AQP on the flipped
labels. The original-label reference evaluates the query over every row
using recorded labels; the flipped-label reference evaluates every row
after the flips. These flips simulate label errors rather than estimate
the error rate of a particular model.

Table~\ref{tab:upstream_error} reports the effect of the flips and the
distance of AQP from each reference. On Q52, the two full-data reference
aggregates differ by 269.66\%. AQP's estimate is 3.26\% from the
flipped-label reference but 274.55\% from the original-label reference.
Sampling can therefore closely reproduce the supplied model's aggregate
even when model errors substantially change the desired result. Oracle
approximation error and upstream prediction error are distinct; reducing
the former does not correct the latter.
\begin{table}[!tp]

\centering\small
\setlength{\tabcolsep}{3pt}
\caption{Effect of flipping 5\% of oracle predicate labels on AQP estimates.}
\label{tab:upstream_error}
\begin{tabular}{lrrr}
\toprule
Query & \shortstack{Label-flip\\effect (\%)} & \shortstack{AQP vs.\\flipped (\%)} & \shortstack{AQP vs.\\original (\%)} \\
\midrule
Q7 & 1.14 & 1.12 & 1.26 \\
Q13 & 11.78 & 9.87 & 16.33 \\
Q50 & 9.38 & 4.22 & 10.78 \\
Q52 & 269.66 & 3.26 & 274.55 \\
Q53 & 110.50 & 0.78 & 110.99 \\
Q54 & 71.93 & 11.16 & 79.34 \\
Q55 & 283.62 & 1.82 & 279.90 \\
Q56 & 106.89 & 6.51 & 107.49 \\
\bottomrule
\end{tabular}
\par\smallskip
{\footnotesize The original and flipped references are query aggregates computed over all rows using the corresponding predicate labels. The label-flip effect is their absolute difference divided by the absolute original reference, expressed as a percentage; the AQP columns compare its sampled estimate with each reference.}

\end{table}

%% file: sections/evaluation_llm.tex
\subsection{Evaluating Reward-Model Selection on LLM Candidate Pools}
\label{sec:eval:llm}

\begin{figure*}[!t]
\centering
\begin{minipage}{\textwidth}
  \centering
  {\small
    \textcolor{RoyalBlue}{\rule{12pt}{2pt}}~{Fixed-budget sampling}
    \qquad
    {\large\textcolor{orange}{$\circ$}}~{Offline reference point}
    \qquad
    {\large\textcolor{BrickRed}{$\bigstar$}}~Strategy PM
    \qquad
    {\large\textcolor{ForestGreen}{$\blacklozenge$}}~Full-RM
  }
\end{minipage}
\vspace{4pt}
\begin{minipage}[t]{0.235\textwidth}
    \centering
    \includegraphics[width=\linewidth, height=3.5cm, keepaspectratio]{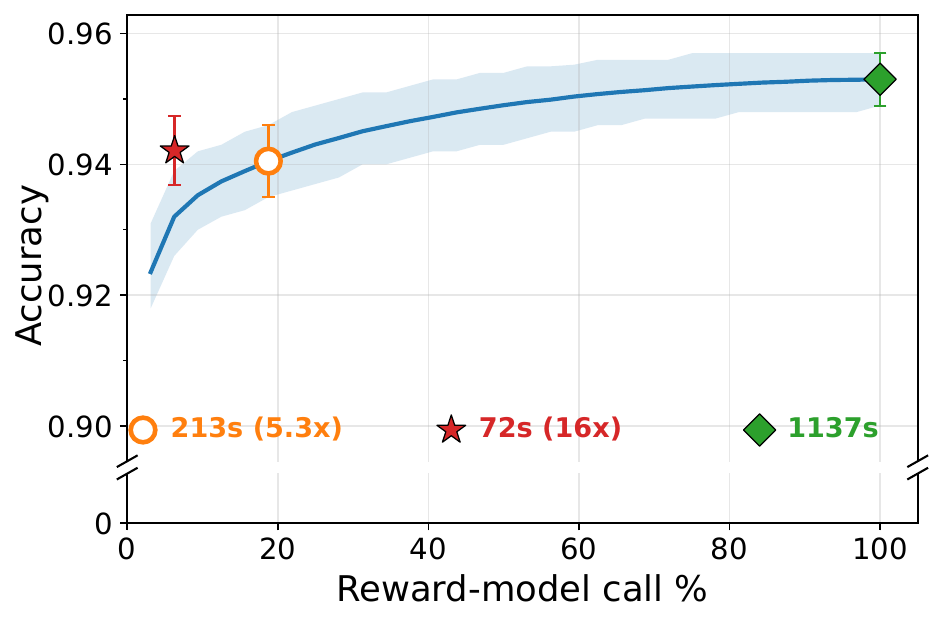}
    \subcaption{Qwen3-8B}
\end{minipage}\hfill
\begin{minipage}[t]{0.235\textwidth}
    \centering
    \includegraphics[width=\linewidth, height=3.5cm, keepaspectratio]
        {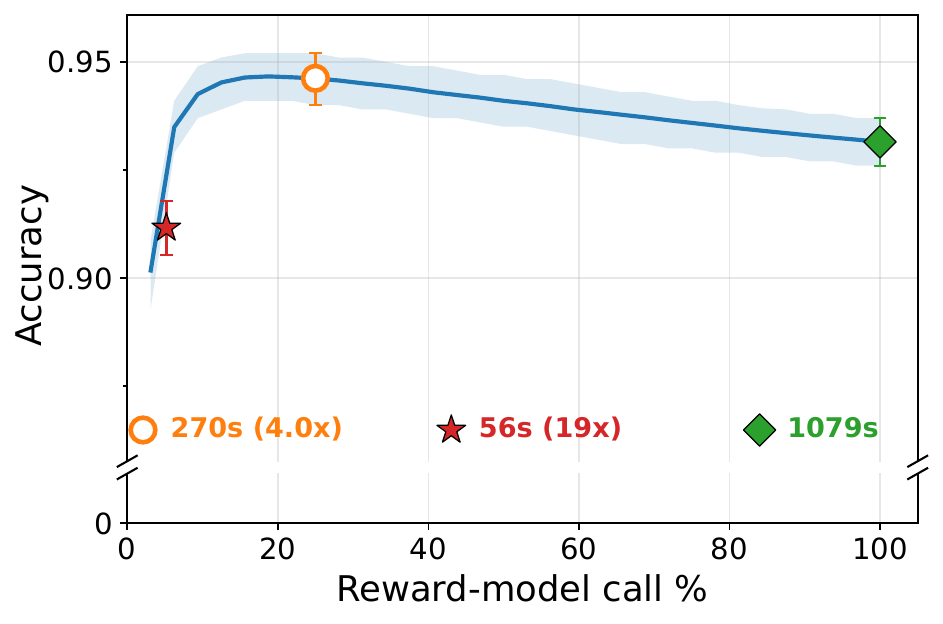}
    \subcaption{LLaMA-3.1-8B}
\end{minipage}\hfill
\begin{minipage}[t]{0.235\textwidth}
    \centering
    \includegraphics[width=\linewidth, height=3.5cm, keepaspectratio]{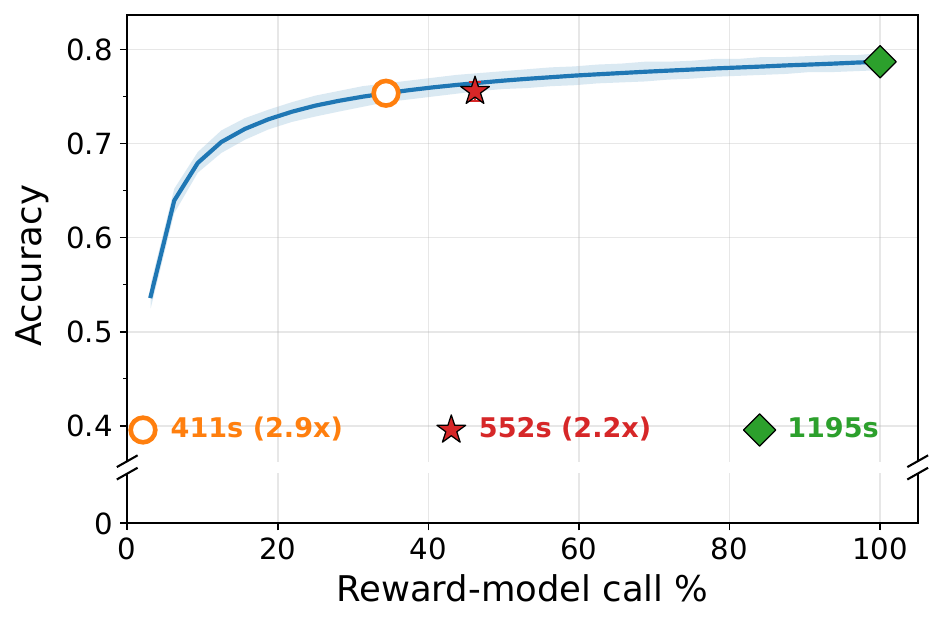}
    \subcaption{Mistral-7B-V0.3}
\end{minipage}
\begin{minipage}[t]{0.235\textwidth}
    \centering
    \includegraphics[width=\linewidth, height=3.5cm, keepaspectratio]{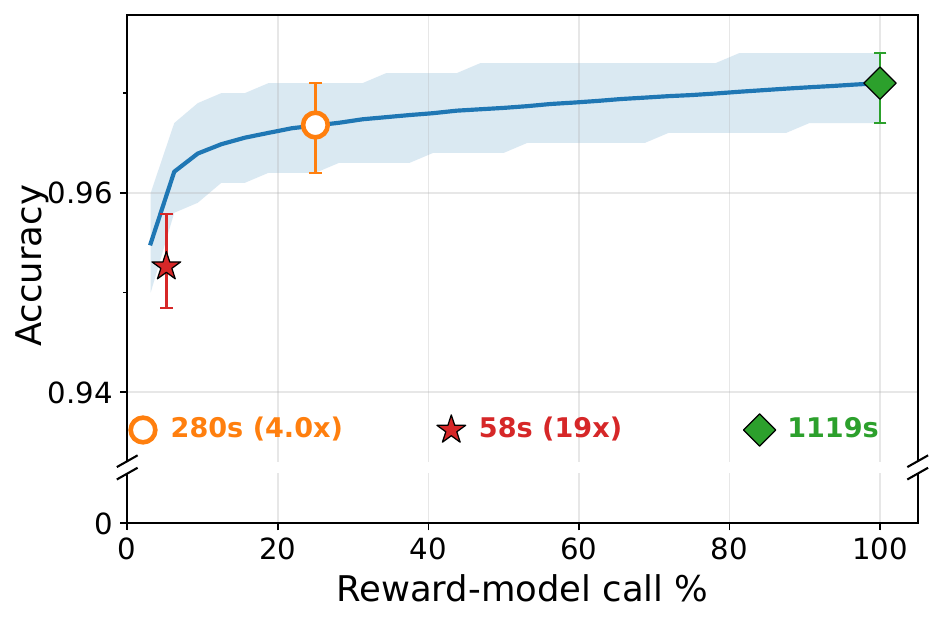}
    \subcaption{Qwen3-32B}
\end{minipage}

\par\vspace{2pt}
\parbox{\linewidth}{\small\textbf{(A) Math Reasoning (GSM8K).}
Accuracy is measured by exact-match correctness of the numerical answer.}
\vspace{6pt}

\begin{minipage}[t]{0.235\textwidth}
    \centering
    \includegraphics[width=\linewidth, height=3.5cm, keepaspectratio]{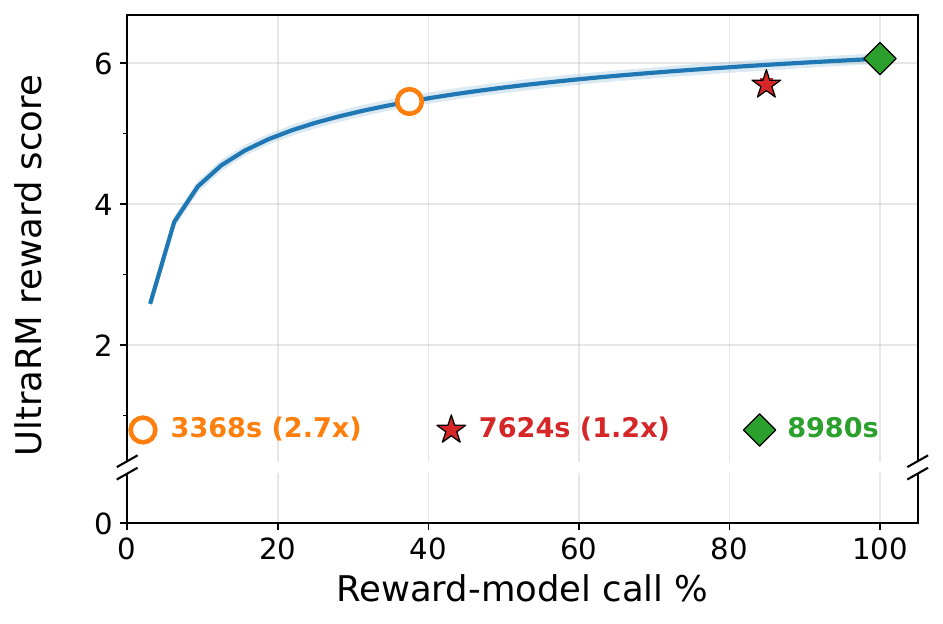}
    \subcaption{Qwen3-8B}
\end{minipage}\hfill
\begin{minipage}[t]{0.235\textwidth}
    \centering
    \includegraphics[width=\linewidth, height=3.5cm, keepaspectratio]
        {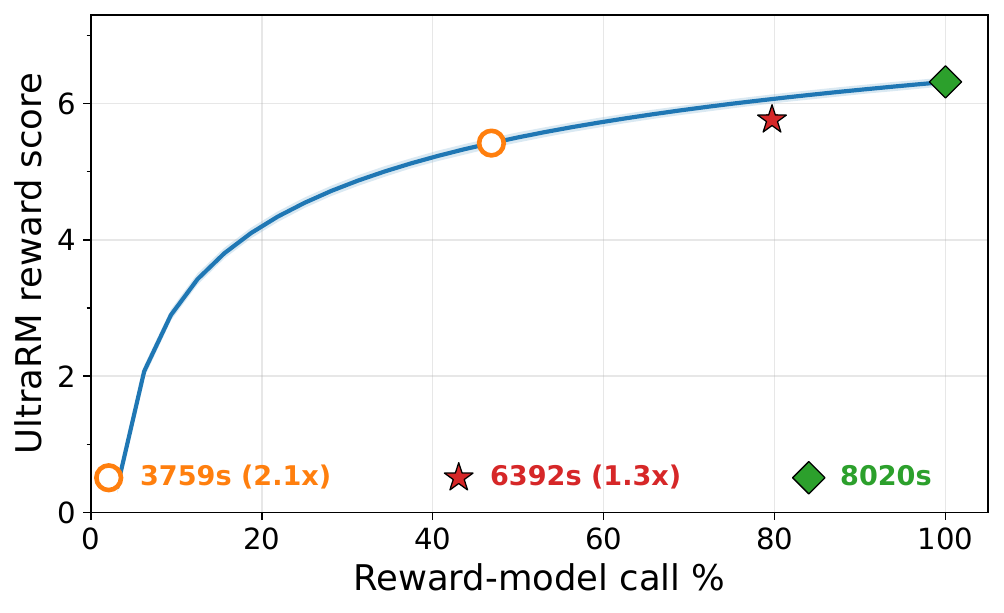}
    \subcaption{LLaMA-3.1-8B}
\end{minipage}\hfill
\begin{minipage}[t]{0.235\textwidth}
    \centering
    \includegraphics[width=\linewidth, height=3.5cm, keepaspectratio]
        {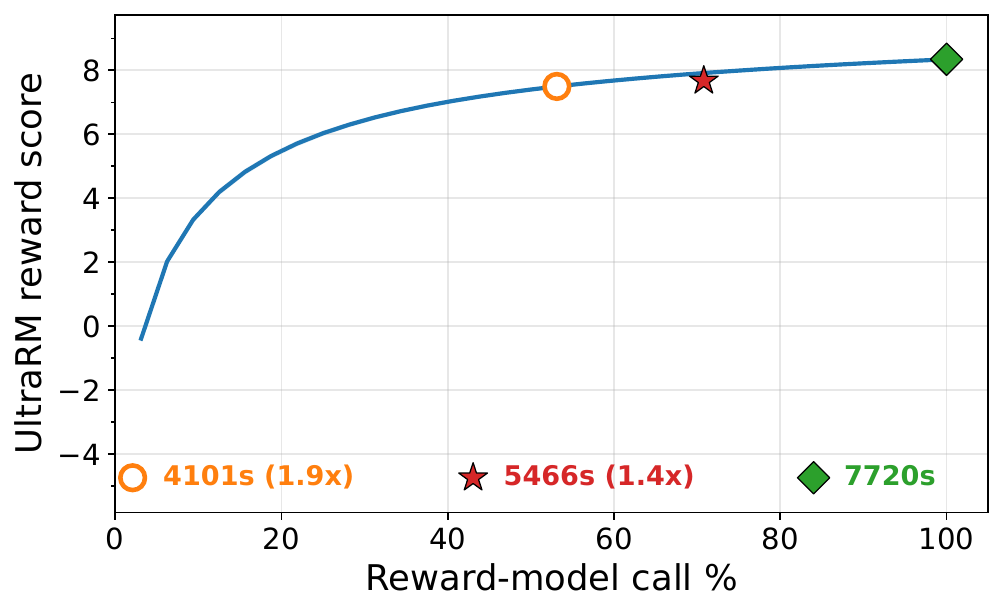}
    \subcaption{Mistral-7B-V0.3}
\end{minipage}\hfill
\begin{minipage}[t]{0.235\textwidth}
    \centering
    \includegraphics[width=\linewidth, height=3.5cm, keepaspectratio]{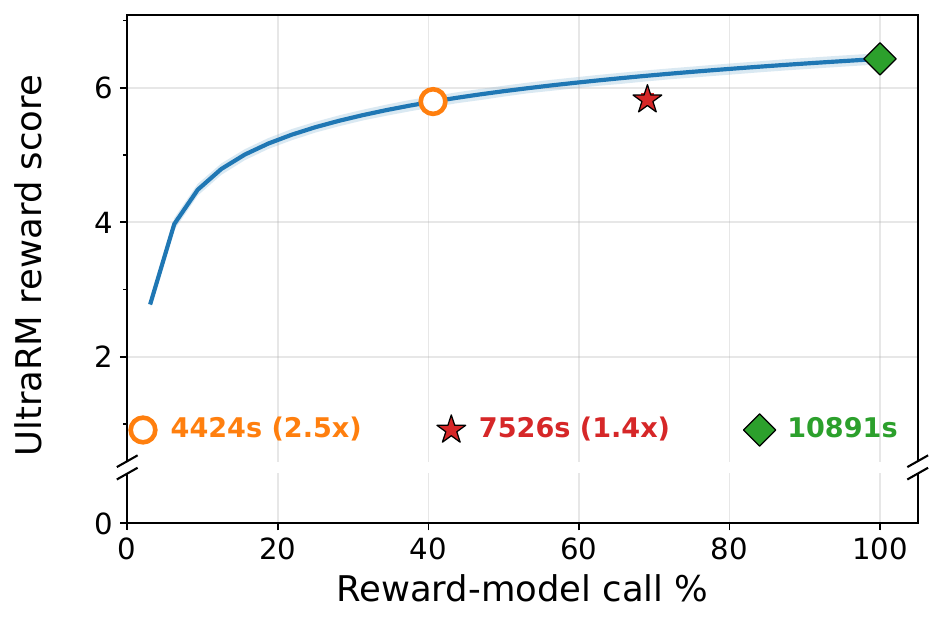}
    \subcaption{Qwen3-32B}
\end{minipage}

\par\vspace{2pt}
\parbox{\linewidth}{\small\textbf{(B) General Instruction Following (UltraFeedback).}
Quality is measured by the UltraRM-13B reward score of the selected
response, as no ground-truth labels exist for this open-ended domain.}
\vspace{6pt}

\begin{minipage}[t]{0.235\textwidth}
    \centering
    \includegraphics[width=\linewidth, height=3.5cm, keepaspectratio]{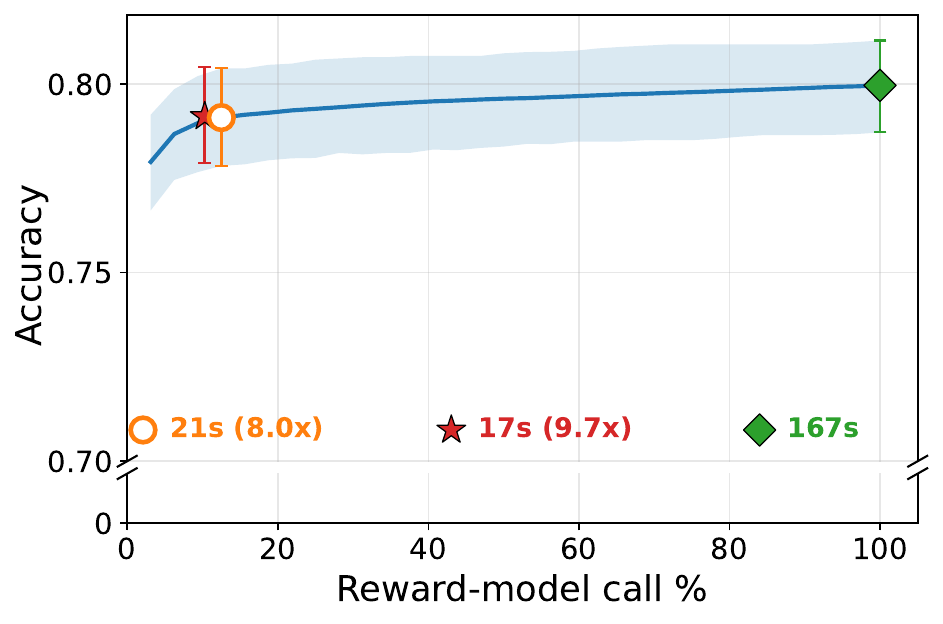}
    \subcaption{Qwen3-8B}
\end{minipage}\hfill
\begin{minipage}[t]{0.235\textwidth}
    \centering
    \includegraphics[width=\linewidth, height=3.5cm, keepaspectratio]{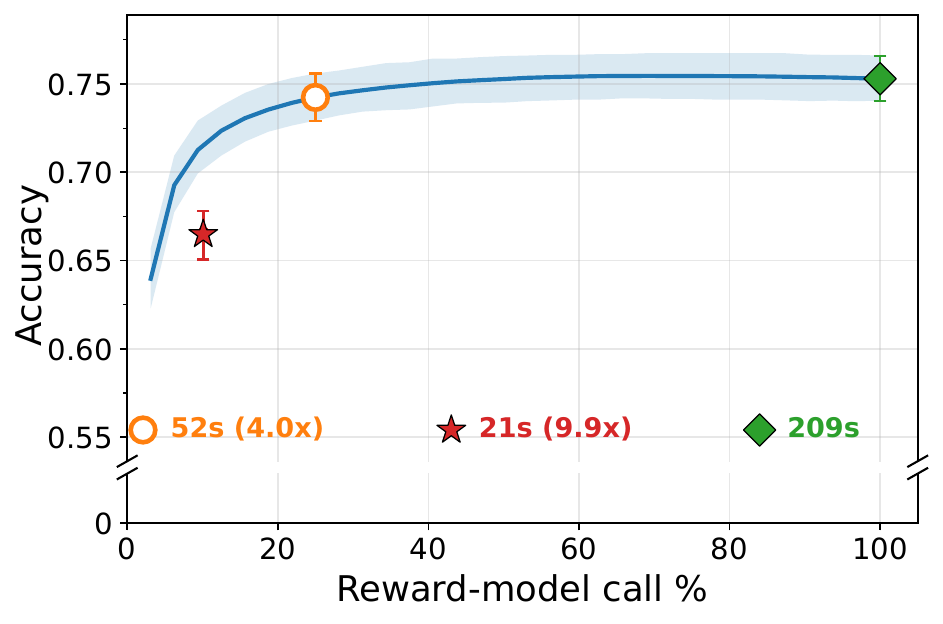}
    \subcaption{LLaMA-3.1-8B}
\end{minipage}\hfill
\begin{minipage}[t]{0.235\textwidth}
    \centering
    \includegraphics[width=\linewidth, height=3.5cm, keepaspectratio]{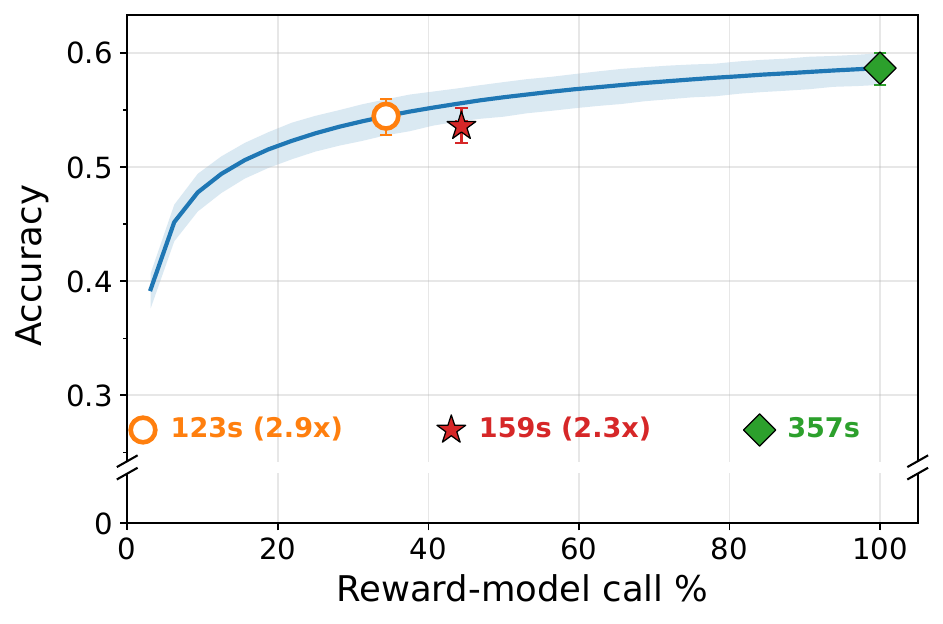}
    \subcaption{Mistral-7B-V0.3}
\end{minipage}\hfill
\begin{minipage}[t]{0.235\textwidth}
    \centering
    \includegraphics[width=\linewidth, height=3.5cm, keepaspectratio]{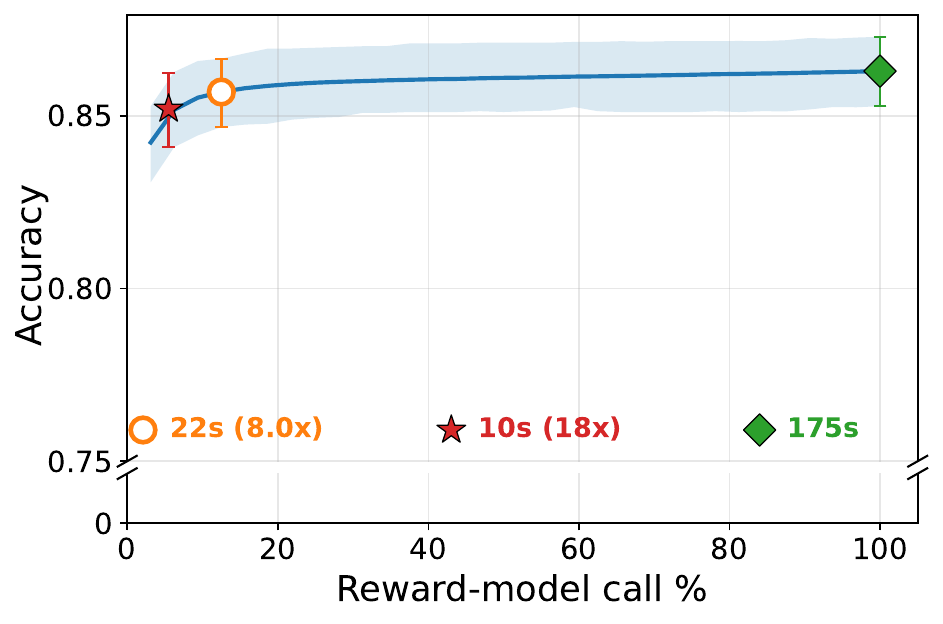}
    \subcaption{Qwen3-32B}
\end{minipage}

\par\vspace{2pt}
\parbox{\linewidth}{\small\textbf{(C) Code Generation (HumanEval+, MBPP).}
Accuracy is measured by pass@1.}

\caption{{Oracle cost and quality on three LLM tasks.}
{Shading and error bars show the 25th to 75th percentiles from 1,000 prompt resamples with the original proxy. Annotations show estimated RM time and speedup.} {Sampling, offline, and Full-RM quality use all prompts. PM quality uses test prompts. PM calls include training and are normalized by scoring all prompts.}}

\label{fig:llm_pareto}
\end{figure*}

\noindent \textbf{LLM pipeline workload.}
To evaluate \ours across diverse task domains in post-training pipelines, we select
three representative pipelines: a \emph{Math} fine-tuning task based
on GSM8K~\cite{gsm8k}, a \emph{General} task based on UltraFeedback~\cite{ultrafeedback}, and
a \emph{Code} task targeting code generation (a combination of two widely-used coding benchmarks HumanEval+~\cite{evalplus} and MBPP~\cite{mbpp}).
Each generator contributes 1,000 Math, 591 Code, and 2,000 General records. For each prompt from the pipeline, we generate $N=32$ candidate responses 
following standard practice in RLHF rejection 
sampling~\cite{grattafiori2024llama}.

Each pipeline uses {a task-specific Reward Model (RM), with Math and Code scored by Skywork-Reward-V2-Qwen3-8B~\cite{rev_skywork} and General scored by UltraRM-13B~\cite{rev_ultrarm},} as the oracle to score the generated candidates from LLMs on the questions in the dataset and select the top ones for fine-tuning in the next iteration. Thus, our goal is to reduce reward-model scoring calls while maintaining answer quality; candidate generation and fine-tuning costs are orthogonal to this evaluation.
For each domain we evaluate four generator models with varying numbers of parameters to understand our improvements on fine-tuning with different model scales: Qwen3-8B~\cite{qwen3}, LLaMA-3.1-8B~\cite{grattafiori2024llama},
Mistral-7B-V0.3~\cite{mistral7b_v03} and Qwen3-32B~\cite{qwen3}.

\myparagraph{Compared methods}
\textbf{Full RM} scores all 32 candidates and returns the highest-scoring response. \textbf{Fixed-budget sampling} scores a uniform subset and returns its best response. \textbf{PM} applies the gate and filter in Section~\ref{sec:strategy_pm}.

As discussed in Section~\ref{subsec:llm_opt}, we compare \textbf{Surface}, our PM's decision tree using 23 surface features, with \textbf{MiniLM}~\cite{rev_minilm} and \textbf{Nomic}~\cite{rev_nomic}, which use the same decision tree with embeddings. MiniLM uses a 256-token limit, while Nomic encodes the complete response. All three predict RM scores using the same training data and cascade rules. Table~\ref{tab:feature_timing} measures CPU feature-extraction time on matched batches of 128 responses using four threads.

\noindent\textbf{Accuracy metrics.} For Math and Code tasks, accuracy is defined as exact-match 
correctness---whether the numerical answer matches the ground 
truth and whether test cases pass, respectively.
{For the General task, we use the UltraRM-13B score of the selected response as the quality metric. Comparing with full RM measures how well each method preserves the pipeline's selection objective while reducing RM calls; it is not an independent measure of helpfulness.}
{In Table~\ref{tab:proxy_compact}, quality is accuracy (\%) for Math and Code and mean RM score for General.}

\noindent\textbf{RM calls.} {The sampling reference evaluates $\mathrm{RM}@m$ for predetermined $m=1,\ldots,32$. The orange point marks the original offline location where the increase in expected maximum RM score first falls below 0.1. It is not an execution of Equation~\ref{eq:adaptive_stop}. PM uses the original 5\% or 10\% prompt-training budgets and the presets in Table~\ref{tab:cascade_presets}.}
{Calls include scoring the training candidates and are normalized by full-workload scoring. The times in Figure~\ref{fig:llm_pareto} are estimated RM scoring times.}

\subsubsection{Selection Quality and Oracle Cost} We report the results in Figure~\ref{fig:llm_pareto}.
{The fixed-budget curves show that scoring a subset of candidates preserves quality close to full RM, and PM further reduces scoring on Math and Code. For LLaMA-3.1-8B Math, PM uses 5.23\% of full scoring, including training calls, or 19.1 times fewer calls than full RM. Accuracy on the test prompts is 91.16\%, compared with 92.95\% for full RM, a gap of 1.79 percentage points.}

{The General task gives more modest savings, because structural features are less informative for free-form responses. PM uses 69.07\% to 84.90\% of full scoring across generators, with RM-score gaps of 0.40 to 0.64. Math and Code often have several correct candidates, allowing PM to avoid most scoring calls. General requires more scoring to preserve the quality of the selected response.}

\subsubsection{Surface Features versus Embeddings}\phantomsection\label{sec:proxy_ablation}
We compare our PM's surface features (Surface) with MiniLM and Nomic embeddings on quality and RM calls (Table~\ref{tab:proxy_compact}) and CPU preparation time (Table~\ref{tab:feature_timing}). The comparison tests whether embeddings improve the quality and scoring savings enough to justify their preparation cost.

\emph{Quality and RM calls.} Table~\ref{tab:proxy_compact} shows that Surface achieves competitive quality and RM-call savings without embeddings. Surface gives higher quality in 11 of 12 cases against each embedding proxy. For Qwen3-32B Code, Surface gives 85.20\% accuracy at 5.60\% of calls, compared with 84.85\% at 13.60\% for MiniLM and 83.96\% at 22.30\% for Nomic. Embeddings sometimes reduce calls further, particularly on General, but with lower mean RM scores. MiniLM improves LLaMA-3.1-8B Math accuracy from 91.16\% to 91.68\%, while increasing calls from 5.23\% to 8.97\%. Even the smaller MiniLM encoder therefore offers no consistent improvement over Surface.

\emph{CPU preparation time.} Table~\ref{tab:feature_timing} shows that Surface's feature extraction takes 0.009 to 0.052 seconds per 128 responses, compared with 1.498 to 2.814 seconds for MiniLM and 16.391 to 84.770 seconds for Nomic. On matched batches, Surface's extraction is 54 to 173 times faster than MiniLM extraction and 1,469 to 2,029 times faster than Nomic extraction. Together, Tables~\ref{tab:proxy_compact} and~\ref{tab:feature_timing} show that Surface retains competitive quality and RM-call savings at substantially lower preparation cost. A decision tree with inexpensive structural features can therefore exploit the workflow's selection requirement without the cost of generating embeddings.

\begin{table}[!tp]

\centering\small
\caption{PM versus embedding proxies. Each cell shows quality / RM calls (\%). Quality is accuracy (\%) for Math and Code and mean RM score for General; higher quality and fewer calls are better. Bold marks the best quality and the fewest calls in each row.}
\label{tab:proxy_compact}\label{tab:llm_quality}
\setlength{\tabcolsep}{2.5pt}
\begin{tabular}{@{}llrrr@{}}
\toprule
Task & Generator & \textbf{PM} & \textbf{MiniLM} & \textbf{Nomic} \\
\midrule
\multirow{4}{*}{Math} & Qwen3-8B & \textbf{94.21} / \textbf{6.26} & 92.11 / 7.00 & 93.26 / 6.75 \\
 & LLaMA-3.1-8B & 91.16 / \textbf{5.23} & \textbf{91.68} / 8.97 & 90.53 / 5.84 \\
 & Mistral-7B-V0.3 & \textbf{75.56} / 46.19 & 72.00 / 44.30 & 72.56 / \textbf{43.57} \\
 & Qwen3-32B & 95.26 / 5.18 & 95.05 / \textbf{5.02} & \textbf{95.79} / 6.92 \\
\midrule
\multirow{4}{*}{General} & Qwen3-8B & \textbf{5.69} / 84.90 & 5.43 / \textbf{76.17} & 5.64 / 82.10 \\
 & LLaMA-3.1-8B & \textbf{5.76} / 79.67 & 4.42 / \textbf{65.21} & 4.78 / 71.53 \\
 & Mistral-7B-V0.3 & \textbf{7.67} / 70.81 & 7.54 / \textbf{67.81} & 7.55 / 67.82 \\
 & Qwen3-32B & \textbf{5.82} / 69.07 & 5.23 / \textbf{57.46} & 5.55 / 69.33 \\
\midrule
\multirow{4}{*}{Code} & Qwen3-8B & \textbf{79.14} / \textbf{10.26} & 78.57 / 10.99 & 77.82 / 11.98 \\
 & LLaMA-3.1-8B & \textbf{66.49} / \textbf{10.18} & 65.24 / 13.97 & 65.78 / 11.79 \\
 & Mistral-7B-V0.3 & \textbf{53.57} / 44.35 & 50.56 / 47.23 & 49.25 / \textbf{43.00} \\
 & Qwen3-32B & \textbf{85.20} / \textbf{5.60} & 84.85 / 13.60 & 83.96 / 22.30 \\
\bottomrule
\end{tabular}

\par\smallskip
{\footnotesize Quality is accuracy (\%) for Math and Code, and mean RM score for General.}

\end{table}

\begin{table}[!tp]

\centering\small
\caption{CPU feature-extraction time for 128 responses.}
\label{tab:feature_timing}
\begin{tabular}{lrrr}
\toprule
Method & Min (s) & Median (s) & Max (s) \\
\midrule
PM & 0.009 & 0.016 & 0.052 \\
MiniLM & 1.498 & 2.293 & 2.814 \\
Nomic & 16.391 & 25.732 & 84.770 \\
\bottomrule
\end{tabular}

\end{table}

\subsubsection{Feature and Predictor Ablation}
\label{sec:llm_predictors}
Table~\ref{tab:proxy_comparison} extends the feature comparison to DT,
Ridge regression, and XGBoost. Ridge is the linear baseline because these
targets are continuous RM scores, whereas the TPC-DS comparison uses LR
for binary labels. Candidate pools, training splits, and cascade rules
remain fixed. The predictors run on CPU using four threads. MiniLM
uses a 256-token input limit; Nomic covers complete responses, with
chunking for the six responses exceeding its context window.

With surface features across the 12 settings, Ridge improves quality in
three cases and XGBoost in four. Each quality gain also uses more RM
calls than DT. Table~\ref{tab:proxy_timing} reports CPU fitting and
prediction costs separately from feature extraction. XGBoost has median
fitting and prediction times of 0.0659 and 0.0093 seconds, compared with
0.0116 and 0.0010 seconds for DT. Ridge fits faster at 0.0011 seconds
but does not consistently improve both quality and call savings.
These results support a small decision tree as the default proxy under
our current cascade rules; they do not establish that DT is universally
the most accurate predictor.
\begin{table}[!tp]

\centering\small
\setlength{\tabcolsep}{3pt}
\caption{CPU cost of fitting and applying each proxy.}
\label{tab:proxy_timing}
\begin{tabular}{llrr}
\toprule
Features & Predictor & \shortstack{Fit time (s)\\median / max} & \shortstack{Prediction time (s)\\median / max} \\
\midrule
\multirow{3}{*}{Surface} & DT & 0.0116 / 0.0703 & 0.0010 / 0.0045 \\
 & Ridge & 0.0011 / 0.0021 & 0.0015 / 0.0034 \\
 & XGBoost & 0.0659 / 0.0851 & 0.0093 / 0.0222 \\
\midrule
\multirow{3}{*}{MiniLM} & DT & 0.3350 / 2.0122 & 0.0038 / 0.0128 \\
 & Ridge & 0.0056 / 0.0158 & 0.0258 / 0.0492 \\
 & XGBoost & 1.3174 / 1.8455 & 0.0137 / 0.0293 \\
\midrule
\multirow{3}{*}{Nomic} & DT & 0.6774 / 4.0581 & 0.0061 / 0.0190 \\
 & Ridge & 0.0144 / 0.0428 & 0.0502 / 0.0948 \\
 & XGBoost & 2.6724 / 3.9150 & 0.0164 / 0.0304 \\
\bottomrule
\end{tabular}

\par\smallskip
{\footnotesize Each timing result is the median of three runs.
Median and maximum are across the 12 generator/task pools.
Surface with DT is the original PM.}

\end{table}

\begin{table*}[!tp]

\centering\small
\caption{Feature and predictor choices for RM selection.}
\label{tab:proxy_comparison}
\setlength{\tabcolsep}{3pt}
\renewcommand{\arraystretch}{0.95}
\begin{tabular}{@{}llrrrrrr@{}}
\toprule
 & & \multicolumn{2}{c}{Surface} & \multicolumn{2}{c}{MiniLM (256 tokens)} & \multicolumn{2}{c}{Nomic (complete text)} \\
\cmidrule(lr){3-4}\cmidrule(lr){5-6}\cmidrule(lr){7-8}
Generator & Predictor & Quality & RM calls (\%) & Quality & RM calls (\%) & Quality & RM calls (\%) \\
\midrule
\multicolumn{8}{l}{\textbf{Math} (quality: accuracy (\%))} \\
\multirow{3}{*}{Qwen3-8B} & DT & 94.21 & 6.26 & 92.11 & 7.00 & 93.26 & 6.75 \\
 & Ridge & 94.21 & 5.76 & 93.58 & 12.30 & 92.84 & 11.74 \\
 & XGBoost & 92.63 & 5.75 & 92.74 & 5.00 & 92.74 & 5.00 \\
\multirow{3}{*}{LLaMA-3.1-8B} & DT & 91.16 & 5.23 & 91.68 & 8.97 & 90.53 & 5.84 \\
 & Ridge & 90.53 & 5.19 & 90.63 & 5.54 & 91.37 & 6.18 \\
 & XGBoost & 90.00 & 5.08 & 90.63 & 5.00 & 90.63 & 5.00 \\
\multirow{3}{*}{Mistral-7B-V0.3} & DT & 75.56 & 46.19 & 72.00 & 44.30 & 72.56 & 43.57 \\
 & Ridge & 74.56 & 49.74 & 74.78 & 68.67 & 75.22 & 67.21 \\
 & XGBoost & 72.22 & 43.02 & 75.22 & 69.59 & 74.56 & 64.05 \\
\multirow{3}{*}{Qwen3-32B} & DT & 95.26 & 5.18 & 95.05 & 5.02 & 95.79 & 6.92 \\
 & Ridge & 95.26 & 5.48 & 95.16 & 5.37 & 95.37 & 5.52 \\
 & XGBoost & 95.05 & 5.00 & 95.05 & 5.00 & 95.05 & 5.00 \\
\midrule
\multicolumn{8}{l}{\textbf{General} (quality: mean RM score)} \\
\multirow{3}{*}{Qwen3-8B} & DT & 5.69 & 84.90 & 5.43 & 76.17 & 5.64 & 82.10 \\
 & Ridge & 5.90 & 85.17 & 5.70 & 84.90 & 5.82 & 87.41 \\
 & XGBoost & 5.89 & 86.76 & 6.00 & 96.27 & 6.02 & 96.47 \\
\multirow{3}{*}{LLaMA-3.1-8B} & DT & 5.76 & 79.67 & 4.42 & 65.21 & 4.78 & 71.53 \\
 & Ridge & 6.20 & 91.47 & 6.04 & 90.46 & 6.03 & 89.31 \\
 & XGBoost & 6.23 & 90.91 & 6.37 & 99.90 & 6.37 & 99.81 \\
\multirow{3}{*}{Mistral-7B-V0.3} & DT & 7.67 & 70.81 & 7.54 & 67.81 & 7.55 & 67.82 \\
 & Ridge & 7.28 & 73.76 & 7.17 & 68.70 & 7.17 & 68.20 \\
 & XGBoost & 7.95 & 81.04 & 8.14 & 88.61 & 8.10 & 87.64 \\
\multirow{3}{*}{Qwen3-32B} & DT & 5.82 & 69.07 & 5.23 & 57.46 & 5.55 & 69.33 \\
 & Ridge & 6.28 & 86.16 & 6.05 & 83.60 & 6.19 & 85.86 \\
 & XGBoost & 6.22 & 85.35 & 6.31 & 95.62 & 6.32 & 95.88 \\
\midrule
\multicolumn{8}{l}{\textbf{Code} (quality: accuracy (\%))} \\
\multirow{3}{*}{Qwen3-8B} & DT & 79.14 & 10.26 & 78.57 & 10.99 & 77.82 & 11.98 \\
 & Ridge & 78.76 & 10.71 & 77.07 & 12.35 & 77.82 & 12.25 \\
 & XGBoost & 78.38 & 13.28 & 76.88 & 10.77 & 77.44 & 11.03 \\
\multirow{3}{*}{LLaMA-3.1-8B} & DT & 66.49 & 10.18 & 65.24 & 13.97 & 65.78 & 11.79 \\
 & Ridge & 65.06 & 5.82 & 67.74 & 22.82 & 68.81 & 23.67 \\
 & XGBoost & 64.35 & 6.32 & 64.71 & 5.27 & 62.92 & 5.19 \\
\multirow{3}{*}{Mistral-7B-V0.3} & DT & 53.57 & 44.35 & 50.56 & 47.23 & 49.25 & 43.00 \\
 & Ridge & 51.88 & 61.29 & 51.50 & 67.63 & 52.07 & 60.32 \\
 & XGBoost & 52.82 & 51.78 & 54.89 & 75.72 & 53.57 & 77.70 \\
\multirow{3}{*}{Qwen3-32B} & DT & 85.20 & 5.60 & 84.85 & 13.60 & 83.96 & 22.30 \\
 & Ridge & 85.03 & 7.14 & 85.56 & 9.35 & 83.60 & 10.50 \\
 & XGBoost & 85.03 & 8.78 & 84.85 & 11.52 & 83.60 & 9.78 \\
\bottomrule
\end{tabular}

\par\smallskip
{\footnotesize Higher quality and fewer calls are preferred. RM calls include training.
Surface with DT is the original PM; MiniLM uses 256 tokens and Nomic covers complete responses.}

\end{table*}

\subsubsection{Paired Quality and Training-Sample Variability}
\label{sec:llm_variability}
The cost-quality figure uses the original proxy fit and saved candidate
pools. To compare PM and full RM on identical data, Table~\ref{tab:paired_detail}
reports their quality on the same held-out prompts, with paired 95\%
bootstrap intervals for the quality gap.
\myparagraph{Paired quality comparison}
Table~\ref{tab:paired_detail} uses the original splits. Math has 50 training and 950 test records, except Mistral with 100 and 900. General has 200 and 1,800, except Mistral with 100 and 1,900. Code has 59 and 532 for Qwen3-8B and Mistral, and 30 and 561 for the other generators. The Code pool contains 589 distinct normalized prompts among 591 records, including one prompt shared between training and testing per generator. We retain that original split.

The paired intervals resample the same held-out prompts for PM and full RM. Both selectors use candidate identifiers to break tied scores. The figure additionally resamples candidate order for its sampling curves, retaining their first-drawn tie rule and fixed offline budgets. 

\begin{table*}[!tp]

\centering\small
\caption{Original DT versus full RM on the same held-out prompts.}
\label{tab:paired_detail}
\setlength{\tabcolsep}{5pt}
\begin{tabular}{llrrrr}
\toprule
Task & Generator & \shortstack{Original DT\\(PM) quality} & \shortstack{Full RM\\quality} & \shortstack{Quality gap\\PM $-$ Full RM} & \shortstack{95\% paired interval\\for the gap} \\
\midrule
\multirow{4}{*}{Math} & Qwen3-8B & 94.21 & 94.95 & -0.74 & [-2.00, 0.42] \\
 & LLaMA-3.1-8B & 91.16 & 92.95 & -1.79 & [-3.89, 0.21] \\
 & Mistral-7B-V0.3 & 75.56 & 78.78 & -3.22 & [-5.34, -1.22] \\
 & Qwen3-32B & 95.26 & 96.95 & -1.68 & [-2.95, -0.52] \\
\midrule
\multirow{4}{*}{General} & Qwen3-8B & 5.69 & 6.09 & -0.40 & [-0.47, -0.33] \\
 & LLaMA-3.1-8B & 5.76 & 6.37 & -0.61 & [-0.70, -0.54] \\
 & Mistral-7B-V0.3 & 7.67 & 8.31 & -0.64 & [-0.74, -0.55] \\
 & Qwen3-32B & 5.82 & 6.43 & -0.61 & [-0.69, -0.54] \\
\midrule
\multirow{4}{*}{Code} & Qwen3-8B & 79.14 & 80.45 & -1.32 & [-3.37, 0.94] \\
 & LLaMA-3.1-8B & 66.49 & 75.04 & -8.56 & [-11.76, -5.70] \\
 & Mistral-7B-V0.3 & 53.57 & 58.65 & -5.08 & [-7.87, -2.63] \\
 & Qwen3-32B & 85.20 & 86.45 & -1.25 & [-3.21, 0.71] \\
\bottomrule
\end{tabular}

\par\smallskip
{\footnotesize Quality is accuracy (\%) for Math and Code and mean RM score for General.
Gaps are PM minus full RM, in percentage points or RM-score units, respectively.}

\end{table*}

For LLaMA-3.1-8B Math, the gap is $-1.79$ percentage points, with a 95\%
interval of $[-3.89,0.21]$. Since this interval includes zero, it does not
resolve a quality difference for that setting. Other settings have larger
losses: LLaMA-3.1-8B Code has an $-8.56$ percentage-point gap with interval
$[-11.76,-5.70]$. General has consistently negative intervals, reflecting
lower preservation of the RM objective. The observed 19-fold call saving
and the uncertainty in task quality are separate measurements.
\myparagraph{Retraining the proxy}
For Table~\ref{tab:proxy_refits}, we retrain PM multiple times on bootstrap samples of the original training prompts and recalibrate thresholds using only the resampled training observations. Tree and random-selection seeds vary; the held-out set, responses, and RM scores remain fixed. The experiment measures sensitivity to the available training sample without generating new responses or rerunning the RM. It retains the disclosed Code prompt overlap.

\begin{table*}[!tp]

\centering\small
\caption{Original DT quality and RM calls after retraining.}
\label{tab:proxy_refits}
\begin{tabular}{llrrr}
\toprule
Task & Generator & \shortstack{Original fit\\quality} & \shortstack{Refit quality\\mean $\pm$ SD} & \shortstack{Refit RM calls (\%)\\mean $\pm$ SD} \\
\midrule
\multirow{4}{*}{Math} & Qwen3-8B & 94.21 & $93.41 \pm 0.59$ & $7.41 \pm 0.99$ \\
 & LLaMA-3.1-8B & 91.16 & $90.69 \pm 0.89$ & $5.81 \pm 0.46$ \\
 & Mistral-7B-V0.3 & 75.56 & $75.41 \pm 0.81$ & $45.94 \pm 1.38$ \\
 & Qwen3-32B & 95.26 & $95.40 \pm 0.48$ & $5.29 \pm 0.36$ \\
\midrule
\multirow{4}{*}{General} & Qwen3-8B & 5.69 & $5.79 \pm 0.10$ & $84.38 \pm 2.77$ \\
 & LLaMA-3.1-8B & 5.76 & $4.95 \pm 0.55$ & $70.91 \pm 6.74$ \\
 & Mistral-7B-V0.3 & 7.67 & $7.62 \pm 0.09$ & $70.35 \pm 2.26$ \\
 & Qwen3-32B & 5.82 & $5.82 \pm 0.18$ & $70.08 \pm 4.31$ \\
\midrule
\multirow{4}{*}{Code} & Qwen3-8B & 79.14 & $78.37 \pm 0.70$ & $12.19 \pm 1.72$ \\
 & LLaMA-3.1-8B & 66.49 & $65.67 \pm 1.27$ & $10.49 \pm 3.07$ \\
 & Mistral-7B-V0.3 & 53.57 & $50.84 \pm 1.06$ & $42.91 \pm 2.10$ \\
 & Qwen3-32B & 85.20 & $84.80 \pm 0.63$ & $13.65 \pm 3.10$ \\
\bottomrule
\end{tabular}

\par\smallskip
{\footnotesize Quality is accuracy (\%) for Math and Code and mean RM score for General.
Refit columns report mean and standard deviation across fits. Calls include training.}

\end{table*}

Retraining retains strong Math savings but exposes more sensitivity on
General. For LLaMA-3.1-8B General, the original fit has mean RM score 5.76,
whereas refits average $4.95\pm0.55$ with $70.91\pm6.74\%$ of full-workload
calls. Qwen3-32B Code also shows call-budget sensitivity: the original fit
uses 5.60\% of calls, compared with $13.65\pm3.10\%$ across refits.
Prompt resampling and refitting reuse saved responses and RM scores;
they do not represent independent candidate generation or new RM evaluations.

\subsubsection{Selection Agreement and Variance-Gate Ablation}
\label{sec:selection_agreement}
Task quality does not imply recovery of the full-RM winner. The variance
gate can return a random response when predicted scores have low dispersion,
and the candidate filter can discard the highest-scoring response even
when its survivor set is nonempty. The empty-set fallback cannot recover
such a discarded winner. We therefore measure agreement separately and
test an alternative that spends additional RM calls on gated prompts.

For each prompt that the original gate would skip, the alternative samples
five candidates uniformly and returns the one with the highest RM score.
Proxy training and the candidate-filter branch remain unchanged. Both
methods return one response from a pool of 32; the five-candidate alternative
does not return a top-five set.

Table~\ref{tab:agreement} reports agreement for all 12 generator/task
settings. Agreement accepts the same candidate or an equal stored RM
score. F1 measures word overlap with the full-RM response on General,
including identical selections. Table~\ref{tab:gate_retention} measures
correct answer retention: among test prompts where full RM selects a
correct answer, it is the percentage where PM also selects a correct
answer. Task labels evaluate correctness, while response selection uses
RM scores.
\begin{table*}[!tp]

\centering\small
\caption{Agreement and F1 with full-RM selections for the original PM and the five-candidate alternative.}
\label{tab:agreement}
\begin{tabular}{llrrrr}
\toprule
Task & Generator & \shortstack{Original PM\\agreement (\%)} & \shortstack{Five-candidate\\agreement (\%)} & \shortstack{Original PM\\F1} & \shortstack{Five-candidate\\F1} \\
\midrule
\multirow{4}{*}{Math} & Qwen3-8B & 4.84 & 18.21 & n/a & n/a \\
 & LLaMA-3.1-8B & 3.16 & 16.00 & n/a & n/a \\
 & Mistral-7B-V0.3 & 54.56 & 54.56 & n/a & n/a \\
 & Qwen3-32B & 3.37 & 16.53 & n/a & n/a \\
\midrule
\multirow{4}{*}{General} & Qwen3-8B & 85.44 & 85.44 & 0.9314 & 0.9314 \\
 & LLaMA-3.1-8B & 79.00 & 79.00 & 0.8950 & 0.8950 \\
 & Mistral-7B-V0.3 & 80.63 & 80.63 & 0.8024 & 0.8024 \\
 & Qwen3-32B & 67.78 & 67.78 & 0.8498 & 0.8498 \\
\midrule
\multirow{4}{*}{Code} & Qwen3-8B & 49.06 & 70.68 & n/a & n/a \\
 & LLaMA-3.1-8B & 19.25 & 35.83 & n/a & n/a \\
 & Mistral-7B-V0.3 & 36.47 & 36.47 & n/a & n/a \\
 & Qwen3-32B & 43.32 & 64.17 & n/a & n/a \\
\bottomrule
\end{tabular}

\par\smallskip
{\footnotesize The five-candidate alternative scores five random candidates only on prompts skipped by the variance gate.
Both return one response. Agreement accepts the same candidate or an equal stored RM score.
F1 measures word overlap with the full-RM response on General, including identical selections.
General has no skipped prompts, so its F1 values are unchanged.}

\end{table*}

\begin{table*}[!tp]

\centering\small
\setlength{\tabcolsep}{5pt}
\caption{Correct answer retention and RM calls after scoring five random candidates on prompts skipped by the variance gate.}
\label{tab:gate_retention}
\begin{tabular}{llrrrr}
\toprule
Task & Generator & \multicolumn{2}{c}{Correct answer retention (\%)} & \multicolumn{2}{c}{RM calls (\%)} \\
\cmidrule(lr){3-4}\cmidrule(lr){5-6}
 & & Original PM & Five RM calls & Original PM & Five RM calls \\
\midrule
\multirow{4}{*}{Math} & Qwen3-8B & 97.56 & 97.56 & 6.26 & 20.57 \\
 & LLaMA-3.1-8B & 93.32 & 97.73 & 5.23 & 19.98 \\
 & Mistral-7B-V0.3 & 91.40 & 91.40 & 46.19 & 46.19 \\
 & Qwen3-32B & 97.39 & 99.13 & 5.18 & 19.95 \\
\midrule
\multirow{4}{*}{Code} & Qwen3-8B & 95.56 & 97.90 & 10.26 & 24.24 \\
 & LLaMA-3.1-8B & 84.56 & 94.30 & 10.18 & 23.34 \\
 & Mistral-7B-V0.3 & 87.50 & 87.50 & 44.35 & 44.35 \\
 & Qwen3-32B & 96.08 & 98.35 & 5.60 & 20.28 \\
\bottomrule
\end{tabular}
\par\smallskip
{\footnotesize Retention is measured only on test prompts where full RM selects a correct answer.
It is the percentage for which PM also selects a correct answer.
Both call percentages include training and are normalized by full-workload scoring.
Five RM calls replace only the gate's random return; all other decisions remain unchanged.}

\end{table*}

Scoring five candidates improves agreement in all six settings with
skipped prompts and correct answer retention in five of them. On
LLaMA-3.1-8B Code, agreement rises from 19.25\% to 35.83\% and retention
from 84.56\% to 94.30\%, while total RM calls rise from 10.18\% to
23.34\%. On LLaMA-3.1-8B Math, agreement rises from 3.16\% to 16.00\%
and retention from 93.32\% to 97.73\%, while calls rise from 5.23\% to
19.98\%. Across the six affected settings, the alternative uses
19.95\%--24.24\% of full-workload calls, including training. It reduces
lost correct answers while retaining substantial call savings.

Mistral and General have no skipped prompts, so their results are
unchanged. General agreement ranges from 67.78\% to 85.44\%, with F1
from 0.8024 to 0.9314. High correct answer retention despite low agreement
shows that distinct Math or Code responses can both be correct. Scoring
five candidates improves the tradeoff but still does not guarantee
recovery of the full-RM winner.

%% file: sections/conclusion.tex
\section{Conclusion}
\label{sec:conclusion}
We presented a query-centric formulation of AI workflows and showed that
two classical database techniques---approximate query processing and
proxy-based predicate filtering---directly reduce the number of expensive
model invocations without modifying the underlying models or pipelines.
Experiments on TPC-DS confirm that
the two strategies are complementary: Strategy~AQP provides reliable
reduction under balanced label distributions, while Strategy~PM provides
additional gains when structural features support a reliable proxy.
{PM uses up to 19 times fewer RM calls than full scoring in the evaluated structured tasks, including training. For LLaMA-3.1-8B Math, accuracy is 91.16\%, compared with 92.95\% for full RM.}
{Shared execution avoids duplicate oracle calls across estimation and proxy training, and returns the plan with fewer remaining requests first. The results demonstrate how query-result requirements make existing database techniques effective for reducing AI workflow cost.}

%% file: references.bib
@article{abacus2025,
  title={Abacus: A cost-based optimizer for Semantic Operator Systems},
  author={Russo, Matthew and Liu, Chunwei and Sudhir, Sivaprasad and Vitagliano, Gerardo and Cafarella, Michael and Kraska, Tim and Madden, Samuel},
  journal={Proceedings of the VLDB Endowment},
  volume={19}, number={5}, pages={1060--1073},
  doi={10.14778/3796195.3796215},
  url={https://arxiv.org/abs/2505.14661},
  year={2026}
}

@article{aero2025,
author = {Kakkar, Gaurav Tarlok and Cao, Jiashen and Sengupta, Aubhro and Arulraj, Joy and Kim, Hyesoon},
title = {Aero: Adaptive Query Processing of ML Queries},
year = {2025},
issue_date = {June 2025},
publisher = {Association for Computing Machinery},
address = {New York, NY, USA},
volume = {3},
number = {3},
url = {https://doi.org/10.1145/3725408},
doi = {10.1145/3725408},
journal = {Proc. ACM Manag. Data},
month = jun,
articleno = {174},
numpages = {27}
}

@misc{biswal2025tag,
      title={Text2SQL is Not Enough: Unifying AI and Databases with TAG}, 
      author={Asim Biswal and Liana Patel and Siddarth Jha and Amog Kamsetty and Shu Liu and Joseph E. Gonzalez and Carlos Guestrin and Matei Zaharia},
      year={2024},
      eprint={2408.14717},
      archivePrefix={arXiv},
      primaryClass={cs.DB},
      url={https://arxiv.org/abs/2408.14717}, 
}

@article{chen2023frugalgpt,
  title={FrugalGPT: How to Use Large Language Models While Reducing Cost and Improving Performance},
  author={Chen, Lingjiao and Zaharia, Matei and Zou, James},
  journal={Transactions on Machine Learning Research},
  year={2024},
  url={https://openreview.net/forum?id=cSimKw5p6R}
}

@inproceedings{evalplus,
  title = {Is Your Code Generated by Chat{GPT} Really Correct? Rigorous Evaluation of Large Language Models for Code Generation},
  author = {Liu, Jiawei and Xia, Chunqiu Steven and Wang, Yuyao and Zhang, Lingming},
  booktitle = {Thirty-seventh Conference on Neural Information Processing Systems},
  year = {2023},
  url = {https://proceedings.neurips.cc/paper_files/paper/2023/hash/43e9d647ccd3e4b7b5baab53f0368686-Abstract.html},
}

@misc{grattafiori2024llama,
      title={The Llama 3 Herd of Models}, 
      author={Grattafiori, Aaron and Dubey, Abhimanyu and Jauhri, Abhinav and Pandey, Abhinav and Kadian, Abhishek and Al-Dahle, Ahmad and Letman, Aiesha and Mathur, Akhil and Schelten, Alan and Vaughan, Alex and others},
      year={2024},
      eprint={2407.21783},
      archivePrefix={arXiv},
      primaryClass={cs.AI},
      url={https://arxiv.org/abs/2407.21783}, 
}

@article{gsm8k,
  title={Training Verifiers to Solve Math Word Problems},
  author={Cobbe, Karl and Kosaraju, Vineet and Bavarian, Mohammad and Chen, Mark and Jun, Heewoo and Kaiser, Lukasz and Plappert, Matthias and Tworek, Jerry and Hilton, Jacob and Nakano, Reiichiro and Hesse, Christopher and Schulman, John},
  journal={arXiv preprint arXiv:2110.14168},
  year={2021}
}

@article{hellerstein1997online,
author = {Hellerstein, Joseph M. and Haas, Peter J. and Wang, Helen J.},
title = {Online aggregation},
year = {1997},
issue_date = {June 1997},
publisher = {Association for Computing Machinery},
address = {New York, NY, USA},
volume = {26},
number = {2},
issn = {0163-5808},
url = {https://doi.org/10.1145/253262.253291},
doi = {10.1145/253262.253291},
journal = {SIGMOD Rec.},
month = jun,
pages = {171–182},
numpages = {12}
}

@misc{kalayci2025pandora,
      title={Optimal Stopping vs Best-of-$N$ for Inference Time Optimization}, 
      author={Yusuf Kalayci and Vinod Raman and Shaddin Dughmi},
      year={2025},
      eprint={2510.01394},
      archivePrefix={arXiv},
      primaryClass={cs.LG},
      url={https://arxiv.org/abs/2510.01394}, 
}

@article{kang2017noscope,
author = {Kang, Daniel and Emmons, John and Abuzaid, Firas and Bailis, Peter and Zaharia, Matei},
title = {NoScope: optimizing neural network queries over video at scale},
year = {2017},
issue_date = {August 2017},
publisher = {VLDB Endowment},
volume = {10},
number = {11},
issn = {2150-8097},
url = {https://doi.org/10.14778/3137628.3137664},
doi = {10.14778/3137628.3137664},
journal = {Proc. VLDB Endow.},
month = aug,
pages = {1586–1597},
numpages = {12}
}

@article{kang2019blazeit,
author = {Kang, Daniel and Bailis, Peter and Zaharia, Matei},
title = {BlazeIt: optimizing declarative aggregation and limit queries for neural network-based video analytics},
year = {2019},
issue_date = {December 2019},
publisher = {VLDB Endowment},
volume = {13},
number = {4},
issn = {2150-8097},
url = {https://doi.org/10.14778/3372716.3372725},
doi = {10.14778/3372716.3372725},
journal = {Proc. VLDB Endow.},
month = dec,
pages = {533–546},
numpages = {14}
}

@article{kang2021abae,
author = {Kang, Daniel and Guibas, John and Bailis, Peter and Hashimoto, Tatsunori and Sun, Yi and Zaharia, Matei},
title = {Accelerating approximate aggregation queries with expensive predicates},
year = {2021},
issue_date = {July 2021},
publisher = {VLDB Endowment},
volume = {14},
number = {11},
issn = {2150-8097},
url = {https://doi.org/10.14778/3476249.3476285},
doi = {10.14778/3476249.3476285},
journal = {Proc. VLDB Endow.},
month = jul,
pages = {2341–2354},
numpages = {14}
}

@inproceedings{li2016wander,
  title={Wander join: Online aggregation via random walks},
  author={Li, Feifei and Wu, Bin and Yi, Ke and Zhao, Zhuoyue},
  booktitle={Proceedings of the 2016 International Conference on Management of Data},
  pages={615--629},
  year={2016}
}

@inproceedings{lu2018pp,
author = {Lu, Yao and Chowdhery, Aakanksha and Kandula, Srikanth and Chaudhuri, Surajit},
title = {Accelerating Machine Learning Inference with Probabilistic Predicates},
year = {2018},
isbn = {9781450347037},
publisher = {Association for Computing Machinery},
address = {New York, NY, USA},
url = {https://doi.org/10.1145/3183713.3183751},
doi = {10.1145/3183713.3183751},
booktitle = {Proceedings of the 2018 International Conference on Management of Data},
pages = {1493–1508},
numpages = {16},
location = {Houston, TX, USA},
series = {SIGMOD '18}
}

@article{mbpp,
  title={Program Synthesis with Large Language Models},
  author={Austin, Jacob and Odena, Augustus and Nye, Maxwell and Bosma, Maarten and Michalewski, Henryk and Dohan, David and Jiang, Ellen and Cai, Carrie and Terry, Michael and Le, Quoc and Sutton, Charles},
  journal={arXiv preprint arXiv:2108.07732},
  year={2021},
  }

@misc{mistral7b_v03,
  title        = {Mistral-7B-v0.3},
  author       = {{Mistral AI}},
  year         = {2024},
  howpublished = {\url{https://huggingface.co/mistralai/Mistral-7B-v0.3}},
  note         = {Apache 2.0 License}
}

@article{patel2025lotus,
author = {Patel, Liana and Jha, Siddharth and Pan, Melissa and Gupta, Harshit and Asawa, Parth and Guestrin, Carlos and Zaharia, Matei},
title = {Semantic Operators and Their Optimization: Enabling LLM-Based Data Processing with Accuracy Guarantees in LOTUS},
year = {2025},
issue_date = {July 2025},
publisher = {VLDB Endowment},
volume = {18},
number = {11},
issn = {2150-8097},
url = {https://doi.org/10.14778/3749646.3749685},
doi = {10.14778/3749646.3749685},
journal = {Proc. VLDB Endow.},
month = jul,
pages = {4171–4184},
numpages = {14}
}

@misc{qwen3,
      title={Qwen3 Technical Report}, 
      author={Yang, An and Li, Anfeng and Yang, Baosong and Zhang, Beichen and Hui, Binyuan and Zheng, Bo and Yu, Bowen and Gao, Chang and Huang, Chengen and Lv, Chenxu and others},
      year={2025},
      eprint={2505.09388},
      archivePrefix={arXiv},
      primaryClass={cs.CL},
      url={https://arxiv.org/abs/2505.09388}, 
}

@article{rev_bargain,
  author = {Sepanta Zeighami and Shreya Shankar and Aditya Parameswaran},
  title = {Cut Costs, Not Accuracy: {LLM}-Powered Data Processing with Guarantees},
  year = {2025},
  journal = {Proceedings of the ACM on Management of Data},
  volume = {3}, number = {6}, pages = {1--26},
  doi = {10.1145/3769776},
  url = {https://arxiv.org/abs/2509.02896}
}

@inproceedings{rev_blinkdb,
  author = {Sameer Agarwal and Barzan Mozafari and Aurojit Panda and Henry Milner and Samuel Madden and Ion Stoica},
  title = {{BlinkDB}: Queries with Bounded Errors and Bounded Response Times on Very Large Data},
  booktitle = {European Conference on Computer Systems}, year = {2013},
  pages = {29--42}, doi = {10.1145/2465351.2465355},
  url = {https://web.eecs.umich.edu/~mozafari/papers/eurosys_2013.pdf}
}

@article{rev_chung,
  author = {Yeounoh Chung and Rushabh Desai and Jian He and Yu Xiao and Thibaud Hottelier and Yves-Laurent {Kom Samo} and Pushkar Khadilkar and Xianshun Chen and Sam Idicula and Fatma {\"O}zcan and Alon Halevy and Yannis Papakonstantinou},
  title = {100x Cost \& Latency Reduction: Performance Analysis of {AI} Query Approximation using Lightweight Proxy Models: [Experiments \& Analysis]},
  year = {2026},
  journal = {Proceedings of the ACM on Management of Data},
  volume = {4}, number = {3}, articleno = {125}, numpages = {23},
  doi = {10.1145/3802002},
  url = {https://dl.acm.org/doi/10.1145/3802002}
}

@article{rev_crossprediction,
  author = {Tijana Zrnic and Emmanuel J. Cand{\`e}s},
  title = {Cross-Prediction-Powered Inference},
  year = {2024},
  journal = {Proceedings of the National Academy of Sciences},
  volume = {121}, number = {15}, pages = {e2322083121},
  doi = {10.1073/pnas.2322083121},
  url = {https://arxiv.org/abs/2309.16598}
}

@misc{rev_minilm,
  author = {{Sentence Transformers}},
  title = {all-MiniLM-L6-v2 Model Card},
  year = {n.d.},
  howpublished = {Hugging Face},
  url = {https://huggingface.co/sentence-transformers/all-MiniLM-L6-v2},
  note = {Accessed 26 September 2026}
}

@misc{rev_nomic,
  author = {{Nomic AI}},
  title = {nomic-embed-text-v1.5 Model Card},
  year = {2024}, howpublished = {Hugging Face},
  url = {https://huggingface.co/nomic-ai/nomic-embed-text-v1.5},
  note = {Accessed 26 September 2026}
}

@inproceedings{rev_palimpzest,
  author = {Chunwei Liu and Matthew Russo and Michael Cafarella and Lei Cao and Peter Baile Chen and Zui Chen and Michael Franklin and Tim Kraska and Samuel Madden and Rana Shahout and Gerardo Vitagliano},
  title = {{Palimpzest}: Optimizing {AI}-Powered Analytics with Declarative Query Processing},
  booktitle = {Conference on Innovative Data Systems Research}, year = {2025},
  url = {https://www.vldb.org/cidrdb/papers/2025/p12-liu.pdf}
}

@article{rev_quest,
  author = {Zhaoze Sun and Chengliang Chai and Qiyan Deng and Kaisen Jin and Xinyu Guo and Han Han and Ye Yuan and Guoren Wang and Lei Cao},
  title = {{QUEST}: Query Optimization in Unstructured Document Analysis},
  journal = {Proceedings of the VLDB Endowment}, volume = {18}, number = {11},
  pages = {4560--4573}, year = {2025}, doi = {10.14778/3749646.3749713},
  url = {https://www.vldb.org/pvldb/vol18/p4560-chai.pdf}
}

@misc{rev_seed,
  author = {Zui Chen and Lei Cao and Sam Madden and Tim Kraska and Zeyuan Shang and Ju Fan and Nan Tang and Zihui Gu and Chunwei Liu and Michael Cafarella},
  title = {{SEED}: Domain-Specific Data Curation With Large Language Models},
  howpublished = {arXiv:2310.00749}, year = {2023},
  url = {https://arxiv.org/abs/2310.00749}
}

@misc{rev_skywork,
  author = {{Skywork}},
  title = {Skywork-Reward-V2-Qwen3-8B Model Card},
  year = {2025}, howpublished = {Hugging Face},
  url = {https://huggingface.co/Skywork/Skywork-Reward-V2-Qwen3-8B},
  note = {Accessed 26 September 2026}
}

@article{rev_thalamus,
  author = {Saehan Jo and Immanuel Trummer},
  title = {{ThalamusDB}: Approximate Query Processing on Multi-Modal Data},
  journal = {Proceedings of the ACM on Management of Data},
  volume = {2}, number = {3}, articleno = {186}, year = {2024},
  doi = {10.1145/3654989},
  url = {https://saehanjo.github.io/assets/files/thalamusdb.pdf}
}

@misc{rev_ultrarm,
  author = {{OpenBMB}},
  title = {UltraRM-13b Model Card},
  year = {2023}, howpublished = {Hugging Face},
  url = {https://huggingface.co/openbmb/UltraRM-13b},
  note = {Accessed 26 September 2026}
}

@inproceedings{rev_uqe,
  author = {Hanjun Dai and Bethany Yixin Wang and Xingchen Wan and Bo Dai and Sherry Yang and Azade Nova and Pengcheng Yin and Phitchaya Mangpo Phothilimthana and Charles Sutton and Dale Schuurmans},
  title = {{UQE}: A Query Engine for Unstructured Databases},
  booktitle = {Advances in Neural Information Processing Systems},
  volume = {37}, year = {2024},
  url = {https://proceedings.neurips.cc/paper_files/paper/2024/hash/34b3a40ec9752c1ae48fe85fef8fe8dc-Abstract-Conference.html}
}

@inproceedings{rev_verdictdb,
  author = {Yongjoo Park and Barzan Mozafari and Joseph Sorenson and Junhao Wang},
  title = {{VerdictDB}: Universalizing Approximate Query Processing},
  booktitle = {Proceedings of the 2018 International Conference on Management of Data},
  pages = {1461--1476}, year = {2018}, doi = {10.1145/3183713.3196905},
  url = {https://arxiv.org/abs/1804.00770}
}

@inproceedings{rev_zendb,
  author = {Yiming Lin and Madelon Hulsebos and Ruiying Ma and Shreya Shankar and Sepanta Zeighami and Aditya G. Parameswaran and Eugene Wu},
  title = {Querying Templatized Document Collections with Large Language Models},
  booktitle = {IEEE International Conference on Data Engineering}, year = {2025},
  pages = {2422--2435}, doi = {10.1109/ICDE65448.2025.00183},
  url = {https://ir.cwi.nl/pub/35884}
}

@article{shankar2025docetl,
author = {Shankar, Shreya and Chambers, Tristan and Shah, Tarak and Parameswaran, Aditya G. and Wu, Eugene},
title = {DocETL: Agentic Query Rewriting and Evaluation for Complex Document Processing},
year = {2025},
issue_date = {May 2025},
publisher = {VLDB Endowment},
volume = {18},
number = {9},
issn = {2150-8097},
url = {https://doi.org/10.14778/3746405.3746426},
doi = {10.14778/3746405.3746426},
journal = {Proc. VLDB Endow.},
month = may,
pages = {3035–3048},
numpages = {14}
}

@article{shankar2026cascades,
author = {Shankar, Shreya and Zeighami, Sepanta and Parameswaran, Aditya},
title = {Task Cascades for Efficient Unstructured Data Processing},
year = {2026},
issue_date = {April 2026},
publisher = {Association for Computing Machinery},
address = {New York, NY, USA},
volume = {4},
number = {1},
url = {https://doi.org/10.1145/3786702},
doi = {10.1145/3786702},
journal = {Proc. ACM Manag. Data},
month = apr,
articleno = {88},
numpages = {26}
}

@inproceedings{tasti2021,
author = {Kang, Daniel and Guibas, John and Bailis, Peter D. and Hashimoto, Tatsunori and Zaharia, Matei},
title = {TASTI: Semantic Indexes for Machine Learning-Based Queries over Unstructured Data},
year = {2022},
isbn = {9781450392495},
publisher = {Association for Computing Machinery},
address = {New York, NY, USA},
url = {https://doi.org/10.1145/3514221.3517897},
doi = {10.1145/3514221.3517897},
booktitle = {Proceedings of the 2022 International Conference on Management of Data},
pages = {1934–1947},
numpages = {14},
location = {Philadelphia, PA, USA},
series = {SIGMOD '22}
}

@misc{touvron2023llama2,
      title={Llama 2: Open Foundation and Fine-Tuned Chat Models}, 
      author={Touvron, Hugo and Martin, Louis and Stone, Kevin and Albert, Peter and Almahairi, Amjad and Babaei, Yasmine and Bashlykov, Nikolay and Batra, Soumya and Bhargava, Prajjwal and Bhosale, Shruti and others},
      year={2023},
      eprint={2307.09288},
      archivePrefix={arXiv},
      primaryClass={cs.CL},
      url={https://arxiv.org/abs/2307.09288}, 
}

@misc{ultrafeedback,
      title={UltraFeedback: Boosting Language Models with Scaled AI Feedback},
      author={Ganqu Cui and Lifan Yuan and Ning Ding and Guanming Yao and Bingxiang He and Wei Zhu and Yuan Ni and Guotong Xie and Ruobing Xie and Yankai Lin and Zhiyuan Liu and Maosong Sun},
      year={2023},
      eprint={2310.01377},
      archivePrefix={arXiv},
      primaryClass={cs.CL},
      url={https://arxiv.org/abs/2310.01377v2}
}

@article{yang2022core,
author = {Yang, Zhihui and Wang, Zuozhi and Huang, Yicong and Lu, Yao and Li, Chen and Wang, X. Sean},
title = {Optimizing machine learning inference queries with correlative proxy models},
year = {2022},
issue_date = {June 2022},
publisher = {VLDB Endowment},
volume = {15},
number = {10},
issn = {2150-8097},
url = {https://doi.org/10.14778/3547305.3547310},
doi = {10.14778/3547305.3547310},
journal = {Proc. VLDB Endow.},
month = jun,
pages = {2032–2044},
numpages = {13}
}

@inproceedings{zanette2024specrejection,
author = {Sun, Hanshi and Haider, Momin and Zhang, Ruiqi and Yang, Huitao and Qiu, Jiahao and Yin, Ming and Wang, Mengdi and Bartlett, Peter L. and Zanette, Andrea},
title = {Fast best-of-N decoding via speculative rejection},
year = {2024},
isbn = {9798331314385},
publisher = {Curran Associates Inc.},
address = {Red Hook, NY, USA},
booktitle = {Proceedings of the 38th International Conference on Neural Information Processing Systems},
articleno = {1025},
numpages = {23},
location = {Vancouver, BC, Canada},
series = {NIPS '24}
}
